\documentclass[aps,prd,groupaddress,floatfix,tighten,nofootinbib,showpacs,amsfonts,superscriptaddress]{revtex4}
\usepackage{gensymb}

\usepackage{url}
\usepackage{xcolor}
\usepackage{hyperref}
\usepackage{amssymb,amsfonts}
\usepackage{amsmath,color}
\usepackage{epsfig}
\usepackage{graphicx,epsf,epsfig}
\usepackage{footnote}
\usepackage{multirow}
\usepackage{graphicx}
\usepackage{subfigure}
\usepackage{lipsum}
\usepackage{multirow} 
\usepackage{tabulary}
\usepackage{rotating}
\usepackage{array}
\usepackage[normalem]{ulem}
\usepackage{color}
\colorlet{shadecolor}{gray!15}
\definecolor{greenLinks}{rgb}{0,0.6,0}
\definecolor{blueLinks}{rgb}{0,0,0.6}
\definecolor{redLinks}{rgb}{0.6,0,0}
\definecolor{tempText}{rgb}{0.55,0.10,0.67}
\definecolor{eprintLinks}{rgb}{0.4,0.4,0.4}
\definecolor{journalLinks}{rgb}{0.6,0,0}
\def\slc#1{
 \setbox0=\hbox{$#1$}                      
 \dimen0=\wd0                              
 \setbox1=\hbox{/} \dimen1=\wd1            
 \ifdim\dimen0>\dimen1                     
 \rlap{\hbox to \dimen0{\hfil/\hfil}}      
 #1                                        
 \else                                     
 \rlap{\hbox to \dimen1{\hfil$#1$\hfil}}   
 /                                         
\fi}

\def\be{\begin{equation}}
\def\ee{\end{equation}}
\def\gs{\mathrel{\rlap{\raise 0.511ex \hbox{$>$}}{\lower 0.511ex \hbox{$\sim$}}}}
\def\ls{\mathrel{\rlap{\raise 0.511ex \hbox{$<$}}{\lower 0.511ex \hbox{$\sim$}}}}

\newcommand{\ba}{\begin{array}{c}}
\newcommand{\baz}{\begin{array}{cc}}
\newcommand{\barrr}{\begin{array}{rrr}}
\newcommand{\bad}{\begin{array}{ccc}}
\newcommand{\bav}{\begin{array}{cccc}}
\newcommand{\baf}{\begin{array}{ccccc}}
\newcommand{\bea}{\begin{equation} \begin{array}{c}}
\newcommand{\eea}{ \end{array} \end{equation}}
\newcommand{\ea}{\end{array}}

\usepackage[normalem]{ulem}

\def\21{$\mathrm{SU(2)_L \otimes U(1)_Y}$ }

\newcommand{\ignore}[1]{}

\usepackage{soul}
\usepackage{ulem}
\usepackage{cancel}

\usepackage{subfigure}
\usepackage{bm}
\usepackage{slashed}

\newcommand{\soul}[1]{}
\allowdisplaybreaks \allowdisplaybreaks[2]
 \newcommand{\AddrFCFMBUAP}{Facultad de Ciencias F\'{\i}sico Matem\'aticas, 
  Benem\'erita Universidad Aut\'onoma de Puebla,\\
  Apdo. Postal 1152, Puebla, Pue.  72000, M\'exico.}
 \newcommand{\AddrFCEBUAP}{Facultad de Ciencias de la Electr\'onica, 
  Benem\'erita Universidad Aut\'onoma de Puebla,\\
  Apdo. Postal 542, Puebla, Pue. 72000, M\'exico.}
 
\begin{document}
%
\title{Deviation to the Tri-Bi-Maximal flavor pattern and equivalent classes}
%
%
\author{E. Barradas-Guevara}
 \email{barradas@fcfm.buap.mx}
 \affiliation{\AddrFCFMBUAP}
%
%
\author{O. F\'elix-Beltr\'an}
 \email{olga.felix@correo.buap.mx}
 \affiliation{\AddrFCEBUAP}
%
%
\author{F. Gonzalez-Canales}
 \email{felix.gonzalezc@correo.buap.mx}
 \affiliation{\AddrFCEBUAP}
%
\date{\today}
%
%
\begin{abstract}
 In the model-independent context, where the neutrino mass matrix is assumed to be diagonalized by means of a unitary matrix that possess the Tri-Bi-Maximal (TBM) 
 flavor mixing pattern. We present an analysis where the TBM deviation is explored by considering different forms, with texture zeros, for the charged lepton mass 
 matrix. These last mass matrices are classified into equivalent classes.   
 We are interested in the charged lepton mass matrices with the minimum free parameter number, $i.e.$ the maximum number of texture zeros, that allows us to correctly 
 reproduce the reactor mixing angle value.  
 We show a deviation from the TBM pattern in terms of the charged lepton masses as well as the theoretical expressions and their parameter space for the mixing angles. 
 Finally, we present the phenomenological implications of numerical values of the ``Majorana-like'' phase factors on the neutrinoless double-beta decay. 
\end{abstract}
%
\maketitle
%
%
%

\section{Introduction}\label{sec:introduction}
Although the discovery of masses and flavor mixing of the neutrino can be regarded as one of the biggest breakthroughts in the understanding of particle physics, since it was 
the first evidence in favor of a beyond the Standard Model physics (BSM)~\cite{OHLSSON20161}. The experimental discovery of a nonzero reactor mixing angle ~\cite{DoubleChooz:2011ymz,DayaBay:2012fng,RENO:2012mkc} marked the beginning of a new era in particle physics. For these experimental results have expanded the 
flavor structure in the lepton sector by providing us the first indication for a new CP violation source~\cite{global-fit-2021}.


However, neutrino oscillation experiments do not resolve the question of whether neutrinos are Majorana o Dirac particles, or give us information about the absolute 
neutrino mass scale, nor about  Majorana phase factors. Majorana phases enter in decay amplitudes that violate the leptonic number, such as neutrinoless double 
beta decay~\cite{Zyla:2020zbs}. Therefore, an experimental observation of neutrinoless double-beta decay can test the absolute scale of neutrinos masses and the 
nature of the neutrino mass term, {\it i.e.,} the neutrinos would be Majorana particles. Now it is well known that neutrinos have a small value for their masses, less than eV, 
which naturally can be explained by considering neutrinos as Majorana particles~\cite{Athar:2021xsd}.  

The Tri-bi-maximal mixing pattern (TBM)~\cite{Harrison:2002er} considers a maximal atmospheric mixing angle $\theta_{23}= 45^{\circ}$ and solar angle 
$\theta_{12} \approx 35.26^{\circ}$, while the reactor angle is postulated as zero. Also, in the TBM framework, the charged lepton mass matrix is considered a diagonal 
matrix. The TBM flavor pattern was ruled out by the experimental measurement of the reactor angle, which reports a reactor mixing angle of the order of eight 
degrees~\cite{DoubleChooz:2011ymz,DayaBay:2012fng,RENO:2012mkc}. However, all is not lost with respect to the TBM pattern, if we remember that the 
leptonic mixing matrix, PMNS, arises from the mismatch between diagonalization of the mass matrices of charged leptons and the  left-handed neutrinos.
Realistic Tri-bi-maximal-like Neutrino Mixing Matrix~\cite{Ahn:2011ep,Chen:2018eou}.

A generalization can be proposed in which the unitary matrix diagonalizing to the neutrino mass matrix is represented by the TBM flavor pattern, while the unitary matrix that diagonalizes to the charged lepton mass matrix corresponds to corrections to the reactor, solar and atmospheric mixing angles. 

This work is structured as follows. In section~\ref{sec:Preliminares} we make a brief discussion on the leptonic flavor mixing matrix and its parametrization, the TBM lepton 
flavor pattern, and neutrinoless double beta decay. In section~\ref{Sec:Deviations} we present a framework where the corrections to the TBM pattern come from the charged 
lepton mass matrix. These last mass matrices are classified into equivalent classes and parameterized in terms of the charged lepton masses. In addition, the allowed regions for the 
leptonic flavor mixing angles and neutrinoless double beta decay are shown. Finally, in the section~\ref{sec:summary} we provide a brief sum-up discussion at the end. 

%
\section{Preliminaries}\label{sec:Preliminares}
%

In the particular case of considering neutrinos as Majorana particles, the low energy neutrino oscillation phenomenon is described by the Lagrangian~\cite{Hochmuth:2007wq}
\begin{equation}\label{Eq:Lagrangiano}
 \begin{array}{l}
  {\cal L} =
  - \frac{ g }{ \sqrt{2} } \overline{ \ell }_{L}  \gamma^{\mu} \nu_{L} W_{\mu} 
  - \frac{1}{2} \overline{\nu_{R}^{c}} \, {\bf M}_{\nu} \nu_{L}
  - \overline{\ell}_{R} {\bf M}_{\ell} \ell_{L} 
  + \textrm{H. c.} \, ,
 \end{array}
\end{equation}
which is written at the base of the flavor eigenstates. In this Lagrangian the first term corresponds to the charged currents, the second is a Majorana mass term for 
neutrinos, and the third part corresponds to the charged lepton mass term. Thus, the ${\bf M}_{\nu}$ is the neutrino mass matrix, while ${\bf M}_{\ell}$ is the charged 
lepton mass matrix. As well, the ${\bf M}_{\nu}$ is the neutrino mass matrix which is a $3 \times 3$ complex symmetric matrix, while ${\bf M}_{\ell}$ is the charged 
lepton mass matrix which, in general, is a $3 \times 3$ complex mass matrix. 
The matrices in eq.~(\ref{Eq:Lagrangiano}) can be rotated to the mass eigenstates basis by means of the unitary transformations
\begin{equation}\label{Eq:Transf-Unitarias}
 \begin{array}{l}
  {\bf M}_{\nu} = {\bf U}_{\nu}^{*} {\bf \Delta}_{\nu} {\bf U}_{\nu}^{\dagger}
  \quad \textrm{and} \quad
  {\bf M}_{\ell} = {\bf V}_{\ell} {\bf \Delta}_{\ell} {\bf U}_{\ell}^{\dagger}.
 \end{array}
\end{equation}
On this basis the mass matrices have the diagonal form, ${\bf \Delta}_{\nu} = \mathrm{diag} \left( m_{\nu 1}, m_{\nu 2}, m_{\nu 3} \right)$ and 
${\bf \Delta}_{\ell} = \mathrm{diag} \left( m_{e}, m_{\mu}, m_{\tau} \right)$. 
The unitary matrices in eq.~(\ref{Eq:Transf-Unitarias}) are obtained from the singular value decomposition theorem. 
From eqs.~(\ref{Eq:Lagrangiano}) 
and~(\ref{Eq:Transf-Unitarias}) the charged currents term takes the form 
\begin{equation}
 \begin{array}{l}
 {\cal L}_{cc} = \overline{\ell'}_{L} \gamma^{\mu} {\bf U}_{\mathrm{PMNS}} \nu_{L}', 
 \end{array}
\end{equation}
where $\ell'_{L} = {\bf U}_{\ell} \ell_{L}$, $\nu_{L}'= {\bf U}_{\nu} \nu_{L}$,  and
\begin{equation}\label{Eq:Upmns-0}
 \begin{array}{l}
  {\bf U}_{\mathrm{PMNS}} = {\bf U}_{\ell}^{\dagger}  {\bf U}_{\nu} =
  \left( \begin{array}{ccc}
   U_{e1}     & U_{e2}     & U_{e3}    \\
   U_{\mu 1}  & U_{\mu 2}  & U_{\mu 3} \\
   U_{\tau 1} & U_{\tau 2} & U_{\tau 3}    
  \end{array} \right).
 \end{array}
\end{equation}
The last expression is the leptonic flavor mixing matrix, which is known as the PMNS matrix and governs the neutrinos and lepton couplings.
In the symmetric parametrization the leptonic flavor mixing matrix has the form~\cite{Rodejohann:2011vc}
\begin{equation}\label{Ec-PMNS-Symmetric}
 \left( \begin{array}{ccc}
  c_{12} c_{13} & 
  s_{12} c_{13} e^{ -i \phi_{12} } & 
  s_{13} e^{ -i \phi_{13} } \\
  - s_{12} c_{23} e^{ i \phi_{12} } - c_{12} s_{13} s_{23} e^{- i \left( \phi_{23} - \phi_{13} \right) } &
  c_{23} c_{12}  - s_{23} s_{12} s_{13} e^{- i \left( \phi_{12} + \phi_{23} - \phi_{13} \right) } &
  c_{13} s_{23} e^{ - i \phi_{23} } \\
  s_{12} s_{23} e^{ i \left( \phi_{12} + \phi_{23} \right) } - c_{12} s_{13} c_{23} e^{ i \phi_{13} } &
  - c_{12} s_{23} e^{ i \phi_{23} } - s_{12} s_{13} c_{23} e^{- i \left( \phi_{12} - \phi_{13} \right) } &
  c_{13} c_{23} 
 \end{array}  \right),
\end{equation}
where $c_{ij} = \cos \theta_{ij}$, $s_{ij} = \sin \theta_{ij}$, and $\phi_{12}, \phi_{13}, \phi_{23}$ are the physical phases.
The symmetric  and PDG~\cite{Zyla:2020zbs} parametrization are related each other by means 
${\bf U}_{\mathrm{PDG}} = {\bf K} {\bf U}_{sym}$, where ${\bf K} = \mathrm{diag \left(1, e^{i \beta_{1}/2}, e^{i \beta_{2}/2} \right)}$ with the
phase factors $\beta_{1} = - 2 \phi_{12}$, $\beta_{2} = - 2 \left( \phi_{12} + \phi_{23} \right)$, and 
$\delta_{\mathrm{CP}} = \phi_{13} - \phi_{23} - \phi_{12}$~\cite{Barradas-Guevara:2017iyt}. The mixing angles in terms of the PMNS matrix entries are
\begin{equation}\label{Eq:Mix-Angle-0}
 \begin{array}{l}
  \sin^{2} \theta_{12} = \frac{ |U_{e2}|^{2} }{ 1 - |U_{e3}|^{2} }, \qquad
  \sin^{2} \theta_{23} = \frac{ |U_{\mu 3}|^{2} }{ 1 - |U_{e3}|^{2} }, 
  \qquad \mathrm{and} \qquad
  \sin^{2} \theta_{13} = |U_{e3}|^{2} .
 \end{array}
\end{equation}
While the phase factors associated with the CP violation phases are:
\begin{equation}\label{Ec-Fases-Para-Simetrica}
 \begin{array}{l}
  \sin \delta_{\mathrm{CP}} = 
  \frac{ 
   \mathcal{J}_{\mathrm{CP}} \left( 1 -  |U_{e3}|^{2} \right)
  }{
   |U_{e1}| |U_{e2}| |U_{e3}| |U_{\mu 3}| |U_{\tau 3}|
  }, \qquad
  \sin \left( - 2 \phi_{12} \right) =
  \frac{ 
   \mathcal{I}_{1}
  }{
   |U_{e1}|^{2} |U_{e2}|^{2}
  }, 
  \qquad \mathrm{and} \qquad
  \sin \left( - 2 \phi_{13} \right) =
  \frac{ 
   \mathcal{I}_{2}
  }{
   |U_{e1}|^{2} |U_{e3}|^{2}
  } ,
 \end{array}
\end{equation}
where $\mathcal{J}_{\mathrm{CP}} = \mathbb{I}m \left \{ U_{e3} U_{\mu 3}^{*} U_{e1}^{*} U_{\mu 1} \right \}$ is the Jarlskog invariant which is associated with the CP
violation phase Dirac-like. Also, $\mathcal{I}_{1} = \mathbb{I}m \left \{ |U_{e1}|^{2} |U_{e2}|^{2} \right \}$ and 
$\mathcal{I}_{2} = \mathbb{I}m \left \{ |U_{e1}|^{2} |U_{e3}|^{2} \right \}$ are the invariants associated with the CP violation phase factors 
Majorana-like~\cite{Barradas-Guevara:2017iyt}.

%
\subsection{The TBM leptonic flavor pattern}\label{subsec:TBMpattern}
%
In the framework of TBM flavor mixing pattern the charged lepton mass matrix has a diagonal form, while the solar, atmospheric and reactor mixing angles have the values 
$\sin^{2} \theta_{12} = \frac{1}{2}$, $\theta_{23} = \frac{\pi}{4}$, and $\theta_{13} = 0$, respectively. Furthermore, the CP symmetry is preserved, this means that phase 
factors are null. Consequently, the unitary matrix ${\bf U}_{\ell}$ is equal to identity matrix, and the PMNS matrix has the form 
${\bf U}_{\mathrm{TBM}} = {\bf U}_{\mathrm{PMNS}} = {\bf U}_{\nu} $~\cite{Harrison:2002er,Rahat_2018},
\begin{equation}\label{Eq:Utbm-0}
 {\bf U}_{\mathrm{TBM}} = 
 \left( 
 \begin{array}{ccc}
  \sqrt{ \frac{ 2 }{ 3 } } & \frac{ 1 }{ \sqrt{3} } & 0 \\
  - \frac{ 1 }{ \sqrt{6} } & \frac{ 1 }{ \sqrt{3} } & \frac{ 1 }{ \sqrt{2} }  \\
   \frac{ 1 }{ \sqrt{6} }   & - \frac{ 1 }{ \sqrt{3} } & \frac{ 1 }{ \sqrt{2} } 
 \end{array} \right).
\end{equation}
In this scheme,  the neutrino mass matrix has the form
\begin{equation}\label{Eq:M_TBM}
 {\bf M}_{\nu} =
 \left( \begin{array}{ccc}
  b_{\nu}   & a_{\nu}           & - a_{\nu} \\
  a_{\nu}   & b_{\nu} + d_{\nu} & b_{\nu} + c_{\nu} \\
  - a_{\nu} & b_{\nu} + c_{\nu} & b_{\nu} + d_{\nu} 
 \end{array}  \right),
\end{equation}
where $a_{\nu} = \frac{1}{3} \left( m_{\nu 2} - m_{\nu 1} \right)$,
$b_{\nu} = \frac{1}{3} \left( 2 m_{\nu 1} + m_{\nu 2} \right)$,
$c_{\nu} = \frac{1}{2} \left( m_{\nu 3} - \frac{4}{3} m_{\nu 2} - \frac{5}{3}  m_{\nu 1} \right) $, and
$d_{\nu} = \frac{1}{2} \left( m_{\nu 3} - m_{\nu 1} \right)$.

Unfortunately, in agreement with the current experimental data on neutrino oscillations, the TBM leptonic flavor pattern can not done correct description of nature, since 
the  reactor mixing  angle is non null. Also, there is a mount of  evidences  for  CP  violation  in neutrino  oscillations. From the current results obtained in a 
global fit of neutrino oscillation data in the simplest three neutrino theoretical framework, for a normal and inverted hierarchy, we have  the following numerical 
values for the neutrino oscillation parameters at the best fit point $\pm 1\sigma$ and $3\sigma$~\cite{global-fit-2021}:  
\begin{equation}
 \begin{array}{ll}\vspace{2mm}
  \Delta m_{21}^{2} \left( 10^{-5} \, \textrm{eV}^{2} \right) = 7.50_{-0.20}^{+0.22}, \quad 
   6.94-8.14, \\ 
  | \Delta m_{31}^{2} | \left( 10^{-3} \, \textrm{eV}^{2} \right) = \left \{ \begin{array}{ll}
  	2.55_{-0.03}^{+0.02}, & 2.47-2.63, \quad \textrm{for NH}, \\ 
  	2.45_{-0.03}^{+0.02}, &
  	2.37-2.53, \quad \textrm{for IH},
  \end{array}  \right.
 \end{array}
\end{equation}
\begin{equation}\label{Ec:exp-mixing-angles}
 \begin{array}{l} 
  \sin^{2} \theta_{12} \left( 10^{-1} \right) = 3.18 \pm 0.16, \quad 2.71-3.69,  \\ \vspace{2mm} 
  \sin^{2} \theta_{23} \left( 10^{-1} \right) = 
   \left \{ \begin{array}{ll}
    5.74 \pm 0.14, & 4.34-6.10, \quad \textrm{for NH,} \\
    5.78_{-0.17}^{+0.10}, & 4.33-6.08, \quad \textrm{for IH,}
   \end{array}   \right.\\ 
  \sin^{2} \theta_{13} \left( 10^{-2} \right) = 
   \left \{ \begin{array}{ll}
    2.200_{-0.062}^{+0.069}, & 2.000-2.405 \quad \textrm{for NH,} \\
    2.225_{-0.070}^{+0.064}, & 2.018-2.424 \quad \textrm{for IH.}
   \end{array}   \right.
 \end{array}
\end{equation}
In the above expressions $\Delta m_{ij}^{2} = m_{\nu_{i}}^{2}  - m_{\nu_{j}}^{2}$ and NH and IH denote the normal and inverted hierarchy in the neutrino mass spectrum, 
respectively.
%
\subsection{Neutrinoless double-beta decay}
%
The neutrinoless double beta decay $(0\nu\beta\beta)$ is a second-order process in which a nucleus decays into another by the emission of two 
electrons~\cite{Furry:1939qr},
\begin{equation}
 \begin{array}{l}
  (A,Z) \rightarrow (A,Z+2) + e^{-} + e^{-} .
\end{array}  
\end{equation}
This hypothesized nuclear transition is forbidden in the theoretical framework of the Standard Model. Consequently, the study of all variations of $0\nu\beta\beta$ is 
equally interesting for investigating the so-called new particle physics (NPP). The experimental discovery of one of these processes could solve the open question about the 
absolute value of neutrino masses and their hierarchy on the mass spectrum. Moreover, the $0\nu\beta\beta$ could be a fundamental tool to study neutrino physics, since this 
nuclear transition only exists if neutrinos are Majorana particles, which means that it would be the first signal of the non-conservation of the lepton number. The 
amplitude for $(0\nu\beta\beta)$ is proportional to the Majorana effective mass~\cite{10.3389/fphy.2019.00086}
\begin{equation}
 \begin{array}{l}
  m_{ee} = \sum \limits_{i} m_{\nu_{i}} U_{ei}^{2}, \quad i=1,2,3,
 \end{array}
\end{equation}
where $m_{\nu_{i}}$ ($i=1,2,3$) are the Majorana neutrino masses and $U_{ei}$ the elements of the first row of leptonic flavor mixing matrix PMNS, eq.~(\ref{Eq:Upmns-0}).
In the symmetric parametrization of leptonic flavor mixing matrix, eq.~(\ref{Ec-PMNS-Symmetric}), the Majorana effective mass has the form
\begin{equation}\label{ec:mee-0}
 \begin{array}{l} 
  |m_{ee}| = |m_{\nu_{1}} c_{12}^{2} c_{13}^{2}  + m_{\nu_{2}} s_{12}^{2} c_{13}^{2} e^{-i 2 \phi_{12} } + m_{\nu_{3}} s_{13}^{2}  e^{-i 2 \phi_{13} } |,
 \end{array}
\end{equation}
where $\phi_{12}$ and $\phi_{13}$ are the Majorana phase factor given in eq.~(\ref{Ec-Fases-Para-Simetrica}). In the above expression the $m_{\nu_{i}}$ neutrino masses 
can be written in terms of the lightest neutrino mass through the expressions 
\begin{equation}
 \begin{array}{l}
  m_{\nu_{3[2]}} = \sqrt{ m_{\nu_{1[3]}} + \Delta m_{31[23]}^{2} }
  \quad \textrm{and} \quad
  m_{\nu_{2[1]}} = \sqrt{ m_{\nu_{1[3]}} + \Delta m_{21[31]}^{2} } \, ,
 \end{array}
\end{equation}
where $m_{\nu_{1[3]}}$ is the lightest neutrino mass for the normal[inverted] hierarchy in the neutrino mass spectrum. Additionally, the  mass $m_{\nu_{1[3]}}$ is 
considered as the only free parameter in the effective mass $m_{ee}$.
%
\section{Deviations from the TBM flavor pattern}\label{Sec:Deviations}
%
A possible modification to the TBM flavor pattern may come from the charged lepton sector to considering the mass matrix of these without a diagonal form.  
In this generalization of TBM pattern, the neutrino mass matrix is given by eq.~(\ref{Eq:M_TBM}), whereas to fix the  form of the charged lepton mass matrix, 
we propose several equivalence classes whose elements are  Hermitian matrices with two texture zeros. These Hermitian matrices may be  written as 
\begin{equation}
 {\bf M}_{\ell}^{i} = {\bf U}_{\ell}^{i} {\bf \Delta}_{\ell} {\bf U}_{\ell}^{i \dagger},
\end{equation}
where 
\begin{equation}\label{Eq:Ul-0}
 {\bf U}_{\ell}^{i} = {\bf T}_{i} {\bf P}_{\ell}^{\dagger} {\bf O}_{\ell}, 
 \qquad i=0,...,5\, .
\end{equation}
In this expression, the ${\bf T}_{i}$ are the elements of $S_{3}$ real representation, see eq.~(\ref{eq:A-1}), ${\bf P}_{\ell}$ is the diagonal matrix of phase factors, 
which is obtained when the charged lepton mass matrix is written in a polar form, and ${\bf O}_{\ell}$ is a real orthogonal matrix whose explicit form is different for 
each equivalent class. 

From eqs.~(\ref{Eq:Upmns-0}), (\ref{Eq:Utbm-0}) and (\ref{Eq:Ul-0}) the PMNS matrix takes the form
\begin{equation}
 \begin{array}{l}
  {\bf U}_{\mathrm{PMNS}}^{i} = {\bf U}_{\ell}^{i \dagger} {\bf U}_{\nu} = 
  {\bf O}_{\ell}^{\top} {\bf P}_{\ell} {\bf T}_{i} {\bf U}_{\mathrm{TBM}}.
 \end{array}
\end{equation}
The explicit form of the ${\bf O}_{\ell}$ and ${\bf P}_{\ell}$ matrices depends on the equivalence class and the number of texture zeros it contains.
Before moving on to define the rule under which texture zeros in a matrix are counted. The rule is; one texture-zero on the diagonal counts as one, while two 
off-diagonal counts as one texture zero~\cite{Fritzsch:1999ee,Canales:2012dr}.
%
\subsection{Equivalent class with two texture zeros type-I}
%
The equivalent class for Hermitian matrices with two texture zeros type-I have the form~\cite{Canales:2012dr}: 
\begin{equation}\label{Eq:EQ-Type-I}
 \begin{array}{ccc}
  {\bf M}_{\ell}^{0} = 
  \left( \begin{array}{ccc}
   0            & a_{\ell}     & 0 \\
   a_{\ell}^{*} & b_{\ell}     & c_{\ell} \\
   0            & c_{\ell}^{*} & d_{\ell}
  \end{array} \right), &
  {\bf M}_{\ell}^{1} = 
  \left( \begin{array}{ccc}
   b_{\ell}     & a_{\ell}^{*} & c_{\ell} \\
   a_{\ell}     & 0            & 0 \\
   c_{\ell}^{*} & 0            & d_{\ell}
  \end{array} \right), &
  {\bf M}_{\ell}^{2} = 
  \left( \begin{array}{ccc}
   d_{\ell} & c_{\ell}^{*} & 0 \\
   c_{\ell} & b_{\ell}     & a_{\ell}^{*} \\
   0        & a_{\ell}     & 0
  \end{array} \right), \\
  {\bf M}_{\ell}^{3} = 
  \left( \begin{array}{ccc}
   0            & 0        & a_{\ell}  \\
   0            & d_{\ell} & c_{\ell}^{*}  \\
   a_{\ell}^{*} & c_{\ell} & b_{\ell} 
  \end{array} \right), &
  {\bf M}_{\ell}^{4} = 
  \left( \begin{array}{ccc}
   d_{\ell} & 0            & c_{\ell}^{*} \\
   0        & 0            & a_{\ell} \\
   c_{\ell} & a_{\ell}^{*} & b_{\ell}
  \end{array} \right), &
  {\bf M}_{\ell}^{5} = 
  \left( \begin{array}{ccc}
   b_{\ell}     & c_{\ell} & a_{\ell}^{*} \\
   c_{\ell}^{*} & d_{\ell} & 0 \\
   a_{\ell}     & 0        & 0
  \end{array} \right),
 \end{array}
\end{equation}
where $d_{\ell} = 1 - \delta_{\ell}$,
\begin{equation}
 \begin{array}{lll}
  a_{\ell} = \sqrt{ \frac{ \widetilde{m}_{e} \widetilde{m}_{\mu} }{ 1 - \delta_{\ell} } } \, e^{ i \phi_{a} } , &
  b_{\ell} = \left( \texttt{s}_{3} - 1 \right) + \texttt{s}_{1} \widetilde{m}_{e} + \texttt{s}_{2} \widetilde{m}_{\mu} + \delta_{\ell} , &
  c_{\ell} = \sqrt{ \frac{ f_{\ell 1} f_{\ell 2} f_{\ell 3} }{ 1 - \delta_{\ell} } } \, e^{ i \phi_{c} } ,
 \end{array}
\end{equation}
with 
$f_{\ell 1} = 1 - \texttt{s}_{1} \widetilde{m}_{e} - \delta_{\ell}$, 
$f_{\ell 2} = \texttt{s}_{3} \left( 1 -  \texttt{s}_{2} \widetilde{m}_{\mu} - \delta_{\ell} \right)$, 
$f_{\ell 3} = 1 + \texttt{s}_{3} \left( \delta_{\ell} - 1 \right)$,
$\widetilde{m}_{e} = \frac{ m_{e} }{ m_{\tau} }$, $\widetilde{m}_{\mu} = \frac{ m_{\mu} }{ m_{\tau} }$, 
$\phi_{a} = \arg \left \{ a_{\ell} \right \}$, and $\phi_{c} = \arg \left \{ c_{\ell} \right \}$. These last two expressions correspond to the phase factors of the 
complex mass matrix elements, which are related to the CP violation and are defined in the open-close interval $\left(-\pi, \pi \right]$. 
In this case, the diagonal matrix of phase factors is
${\bf P}_{\ell} = \mathrm{diag} \left( 1,  e^{ i \phi_{a} } ,e^{ i \left( \phi_{a} + \phi_{c} \right) }  \right) $. 
The real orthogonal matrix ${\bf O}_{\ell}$ is constructed with the help of the general eigenvectors given in eq.~(\ref{Eq:General_Eigenvector}), which are the 
eigenvectors of the charged lepton mass matrix. The explicit form of  
${\bf O}_{\ell} = \left( | M_{1} \rangle, | M_{2} \rangle, | M_{3} \rangle \right)$  is
\begin{equation}
 \begin{array}{l}
  {\bf O}_{\ell} = 
  \left( \begin{array}{ccc}\vspace{2mm}
   \texttt{s}_{1} \sqrt{ \frac{ \widetilde{m}_{\mu} f_{\ell 1} }{ D_{\ell 1} } } &
   \texttt{s}_{2} \sqrt{ \frac{ \widetilde{m}_{e}   f_{\ell 2} }{ D_{\ell 2} } } &
   \texttt{s}_{3} \sqrt{ \frac{ \widetilde{m}_{e} \widetilde{m}_{\mu} f_{\ell 3} }{ D_{\ell 3} } } \\ \vspace{2mm}
   \sqrt{ \frac{ \widetilde{m}_{e}   \left( 1 - \delta_{\ell} \right) f_{\ell 1} }{ D_{\ell 1} } } &
   \sqrt{ \frac{ \widetilde{m}_{\mu} \left( 1 - \delta_{\ell} \right) f_{\ell 2} }{ D_{\ell 2} } } &
   \sqrt{ \frac{ \left( 1 - \delta_{\ell} \right) f_{\ell 3} }{ D_{\ell 3} } } \\ \vspace{2mm}
  -\sqrt{ \frac{ \widetilde{m}_{e} f_{\ell 2} f_{\ell 3} }{ D_{\ell 1} } } &
   \texttt{s}_{1} \texttt{s}_{2} 
   \sqrt{ \frac{ \widetilde{m}_{\mu} f_{\ell 1} f_{\ell 3} }{ D_{\ell 2} } } &
   \texttt{s}_{3} \sqrt{ \frac{ f_{\ell 1} f_{\ell 2} }{ D_{\ell 3} } }
  \end{array} \right),
 \end{array}
\end{equation}
 where
 \begin{equation}
  \begin{array}{l} \vspace{2mm}
   D_{\ell 1} =       
    \left( 1 - \delta_{\ell} \right) \left( \widetilde{m}_{\mu} + \texttt{s}_{3} \widetilde{m}_{e} \right) \left( 1 + \texttt{s}_{2} \widetilde{m}_{e} \right), \quad
   D_{\ell 2} =       
    \left( 1 - \delta_{\ell} \right) \left( \widetilde{m}_{\mu} + \texttt{s}_{3} \widetilde{m}_{e} \right) \left( 1 + \texttt{s}_{1} \widetilde{m}_{\mu} \right), 
    \\ \vspace{2mm}  
   D_{\ell 3} =       
    \left( 1 - \delta_{\ell} \right) \left( 1 + \texttt{s}_{2} \widetilde{m}_{e} \right) \left( 1 + \texttt{s}_{1} \widetilde{m}_{\mu} \right) .  
  \end{array}
 \end{equation}  
In the last expressions $\texttt{s}_{k} = \mathrm{sign} \left( m_{\ell k} \right)$ with $k=1,2,3$, which means that we must consider that
$m_{\ell k} = - m_{\ell k}$ to avoid a sign inconsistency in the expression of $a_{\ell}$, which could cause this to be a purely imaginary amount. For charged lepton 
fields the sign of the mass is irrelevant since the sign can be changed by means of the chiral transformations; 
$\ell_{R} \rightarrow \ell_{R}' = e^{i \gamma_{5} \frac{\pi}{2} } \ell_{R}$ and 
$\ell_{L} \rightarrow \ell_{L}' = e^{i \gamma_{5} \frac{\pi}{2} } \ell_{L}$. 
These transformations change the sign of the eigenvalues, however, the rest of the Lagrangian  keeps invariant.
	\begin{figure}[t]
	\begin{center}
		\begin{tabular}{cc}
			\subfigure{ \includegraphics[width=7cm, height=5cm]{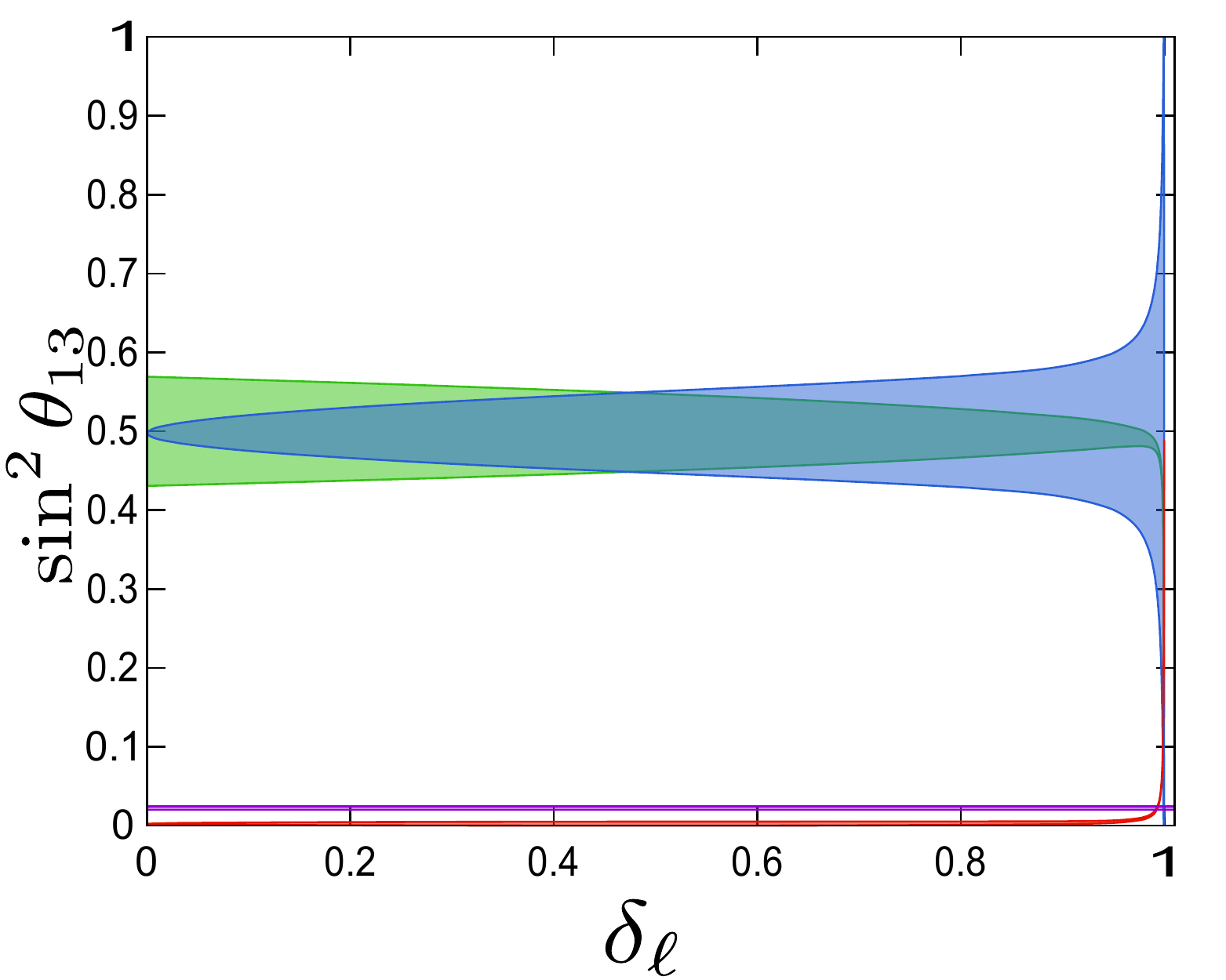}}
			\subfigure{ \includegraphics[width=7cm, height=5cm]{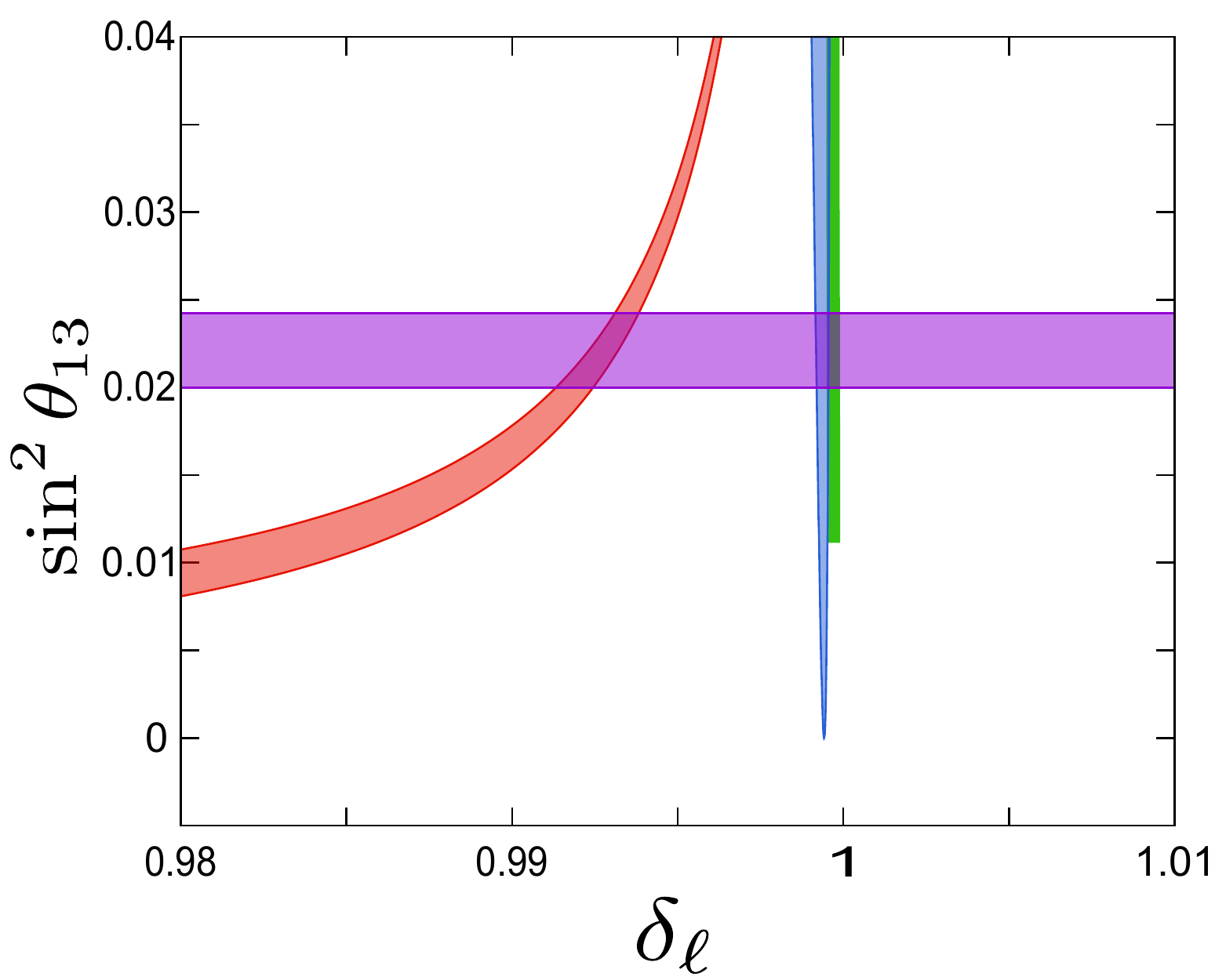}} \\
			\subfigure{ \includegraphics[width=7cm, height=5cm]{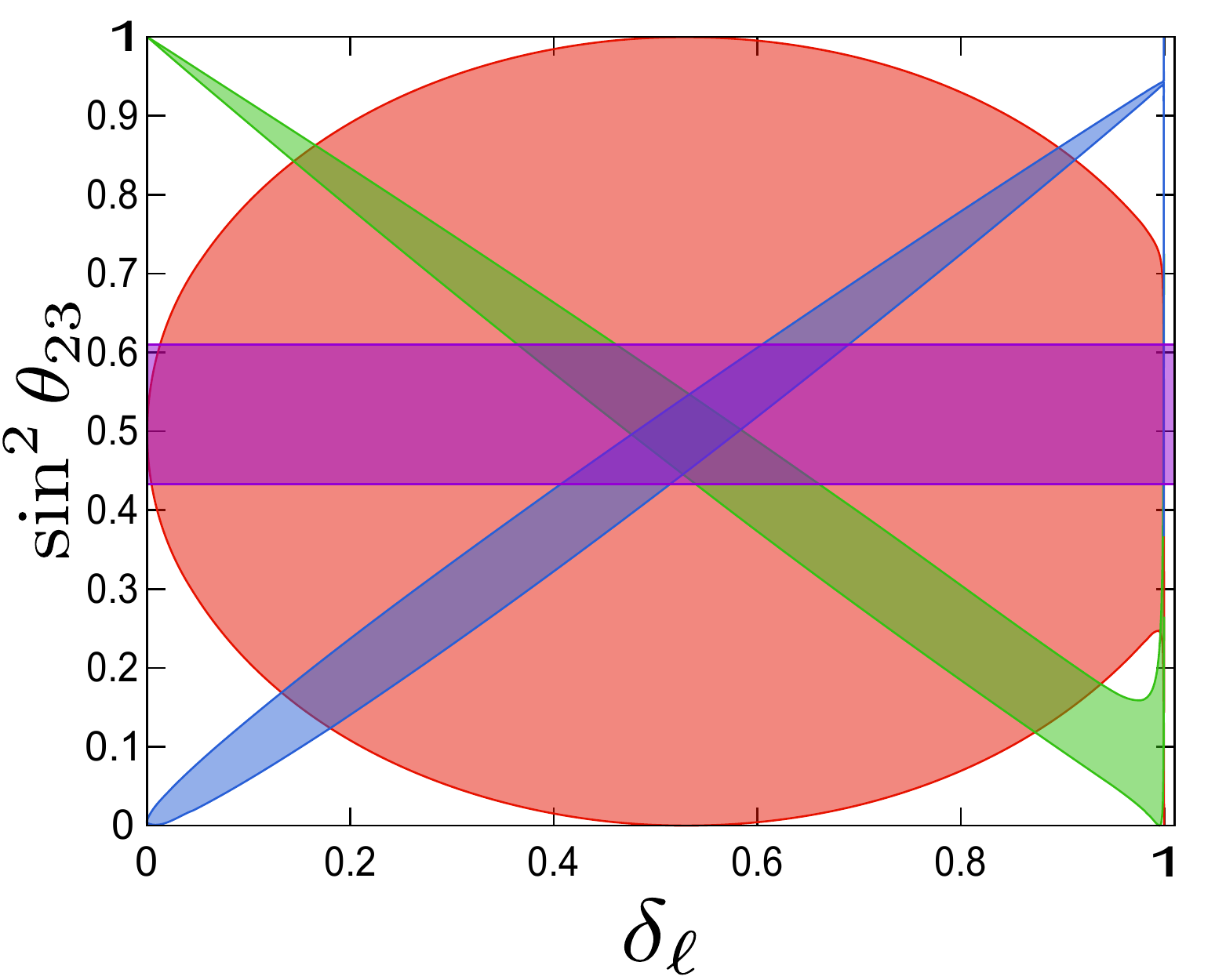}}
			\subfigure{ \includegraphics[width=7cm, height=5cm]{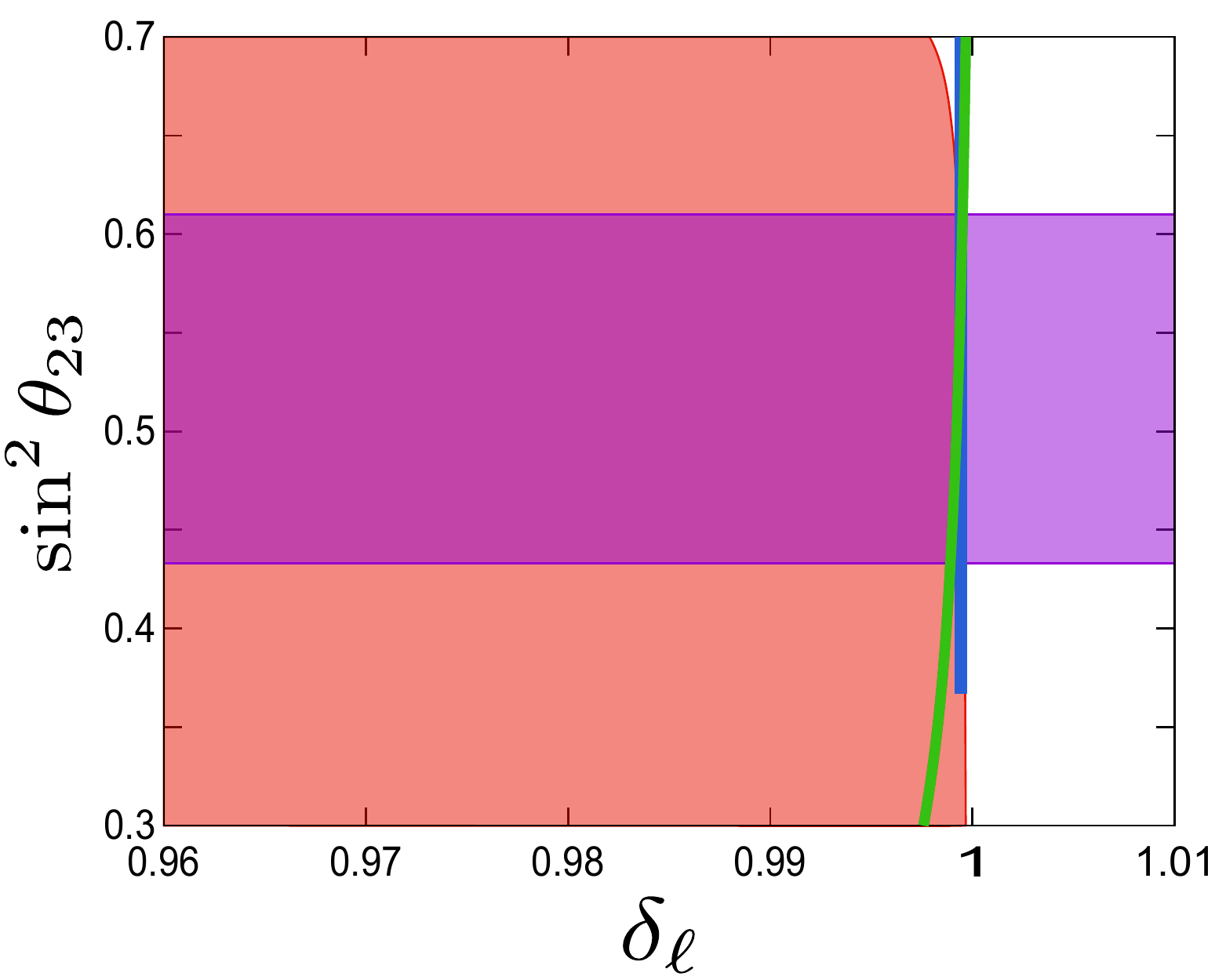}} \\    
			\subfigure{ \includegraphics[width=7cm, height=5cm]{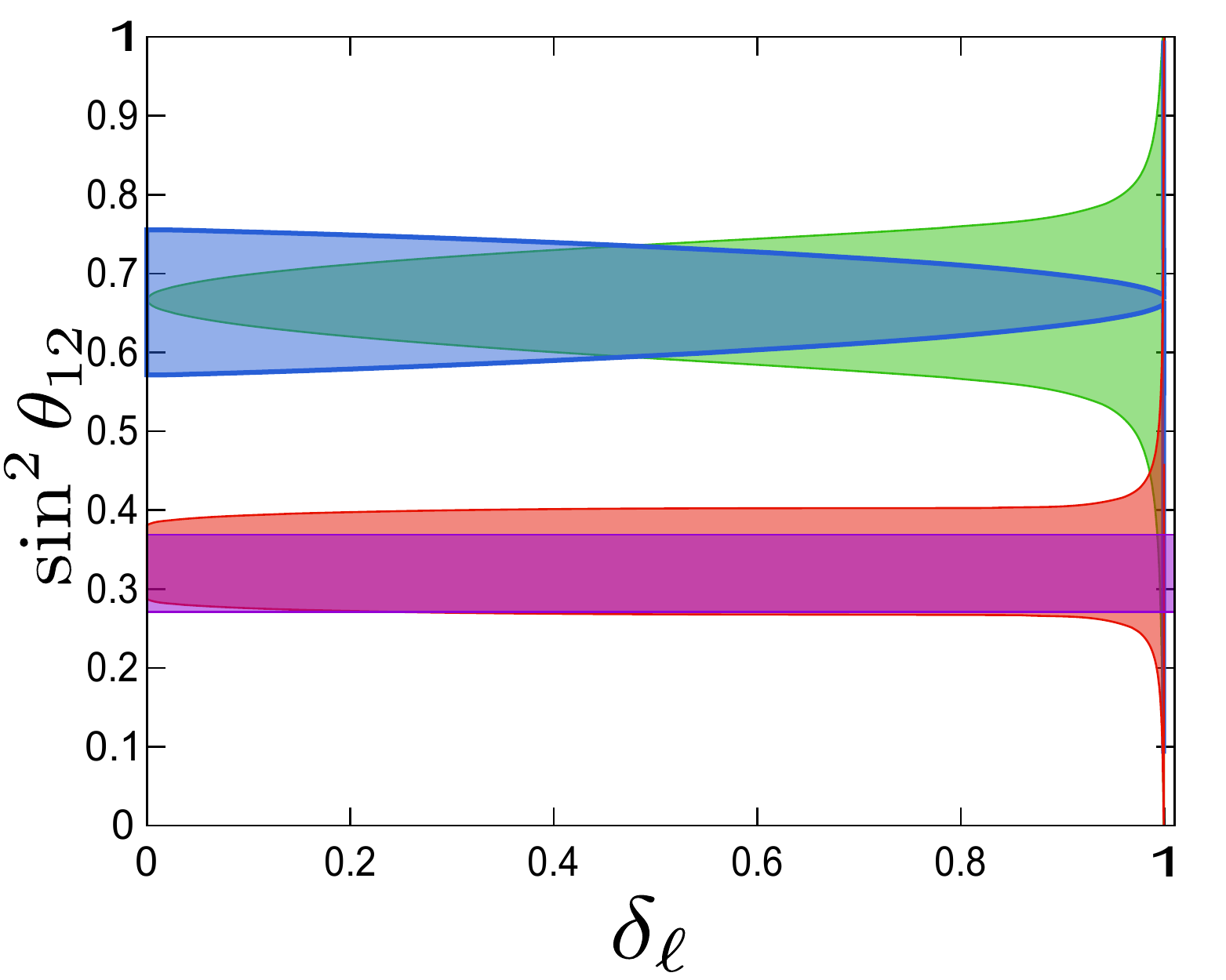}}
			\subfigure{ \includegraphics[width=7cm, height=5cm]{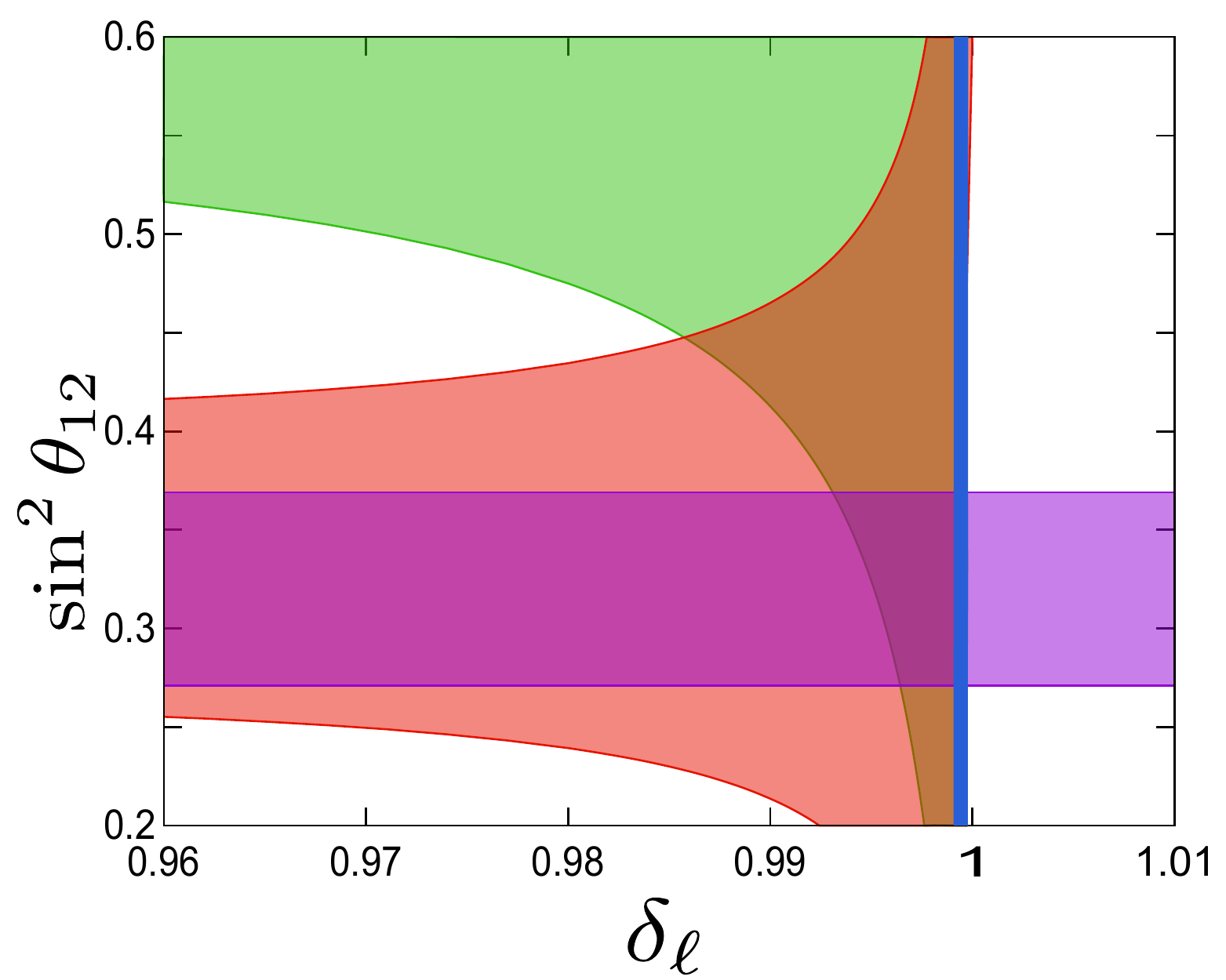}} 
		\end{tabular}
		\caption{The allowed regions for the reactor (upper panels), atmospheric (middle panels), and solar (lower panels) mixing angles and free parameter $\delta_{\ell}$ 
			for  equivalent class with two texture zeros type-I. 
			The purple stripe corresponds to the values at 3$\sigma$ for the reactor, atmospheric and solar mixing angles obtained from the global fit, for normal and inverted 
			hierarchy~\cite{global-fit-2021}.
			In these panels, the red area is for ${\bf M}_{\ell}^{0}$ and ${\bf M}_{\ell}^{3}$, blue area is for ${\bf M}_{\ell}^{1}$ and ${\bf M}_{\ell}^{5}$, 
			the green area is for ${\bf M}_{\ell}^{2}$ and ${\bf M}_{\ell}^{4}$. 
			The right panels show an amplification of the region in which the mixing angles theoretical expressions simultaneously reproduce the current experimental 
			data.}\label{Fig:Angulos-Clase-I-delta-l}
	\end{center}
\end{figure}
The parameter $\delta_{\ell}$ must satisfy the conditions 
\begin{equation}
 \begin{array}{ll}
  \textrm{\bf A.} \quad  0 < \delta_{\ell} < 1 - \widetilde{m}_{\mu}, & \mathrm{for } \qquad m_{e} = - m_{e} \quad 
   \left( \texttt{s}_{1} = -1, \; \texttt{s}_{2} = +1, \; \texttt{s}_{3} = +1 \right). \\
  \textrm{\bf B.} \quad  0 < \delta_{\ell} < 1 - \widetilde{m}_{e},\quad \delta_{\ell} \neq \widetilde{m}_{\mu} - \widetilde{m}_{e}, & \mathrm{for } \qquad m_{\mu} = - m_{\mu} \quad 
   \left( \texttt{s}_{1} = +1, \; \texttt{s}_{2} = -1, \; \texttt{s}_{3} = +1 \right). \\
 \textrm{\bf C.} \quad  1 - \widetilde{m}_{\mu} < \delta_{\ell} < 1 - \widetilde{m}_{e}, & \mathrm{for } \qquad  m_{\tau} = - m_{\tau} \quad 
   \left( \texttt{s}_{1} = +1, \; \texttt{s}_{2} = +1, \; \texttt{s}_{3} = -1 \right).
 \end{array}
\label{eq:conditionsdelta}
\end{equation}
In this case, the flavor mixing angles in eq.~(\ref{Eq:Mix-Angle-0}) have the form:
\begin{equation}
 \begin{array}{l}
  \sin^{2} \theta_{12} = \frac{1}{3} \frac{ \widetilde{m}_{e} }{ \widetilde{m}_{\mu} } \varepsilon_{12}, \qquad
  \sin^{2} \theta_{23} = \frac{1}{2} 
   \frac{
    \left( 1 + \texttt{s}_{2} \widetilde{m}_{e}   \right)
   }{
    \left( 1 + \texttt{s}_{1} \widetilde{m}_{\mu} \right) 
   } \varepsilon_{23}, \qquad
  \sin^{2} \theta_{13} = \frac{1}{2} \frac{ \widetilde{m}_{e} }{ \widetilde{m}_{\mu} } \varepsilon_{13}.
 \end{array}
\end{equation}
The explicit form of the $\varepsilon_{ij}$ parameters is given in the Appendix~\ref{Appendix-Para-Type-I}. 
\begin{figure}[t]
 \begin{center}
  \begin{tabular}{cc}
  \subfigure{ \includegraphics[width=7cm, height=5cm]{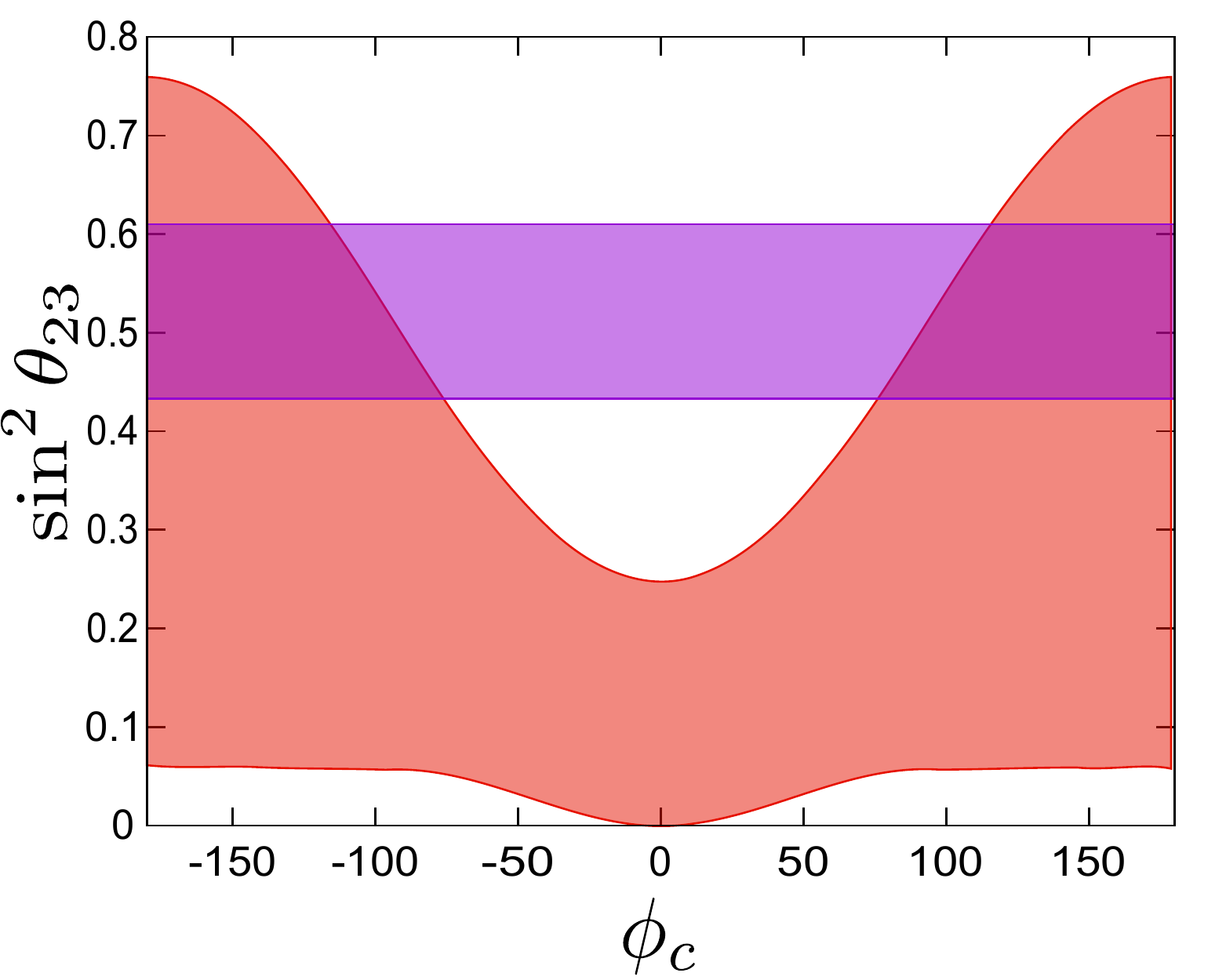}}
  \subfigure{ \includegraphics[width=7cm, height=5cm]{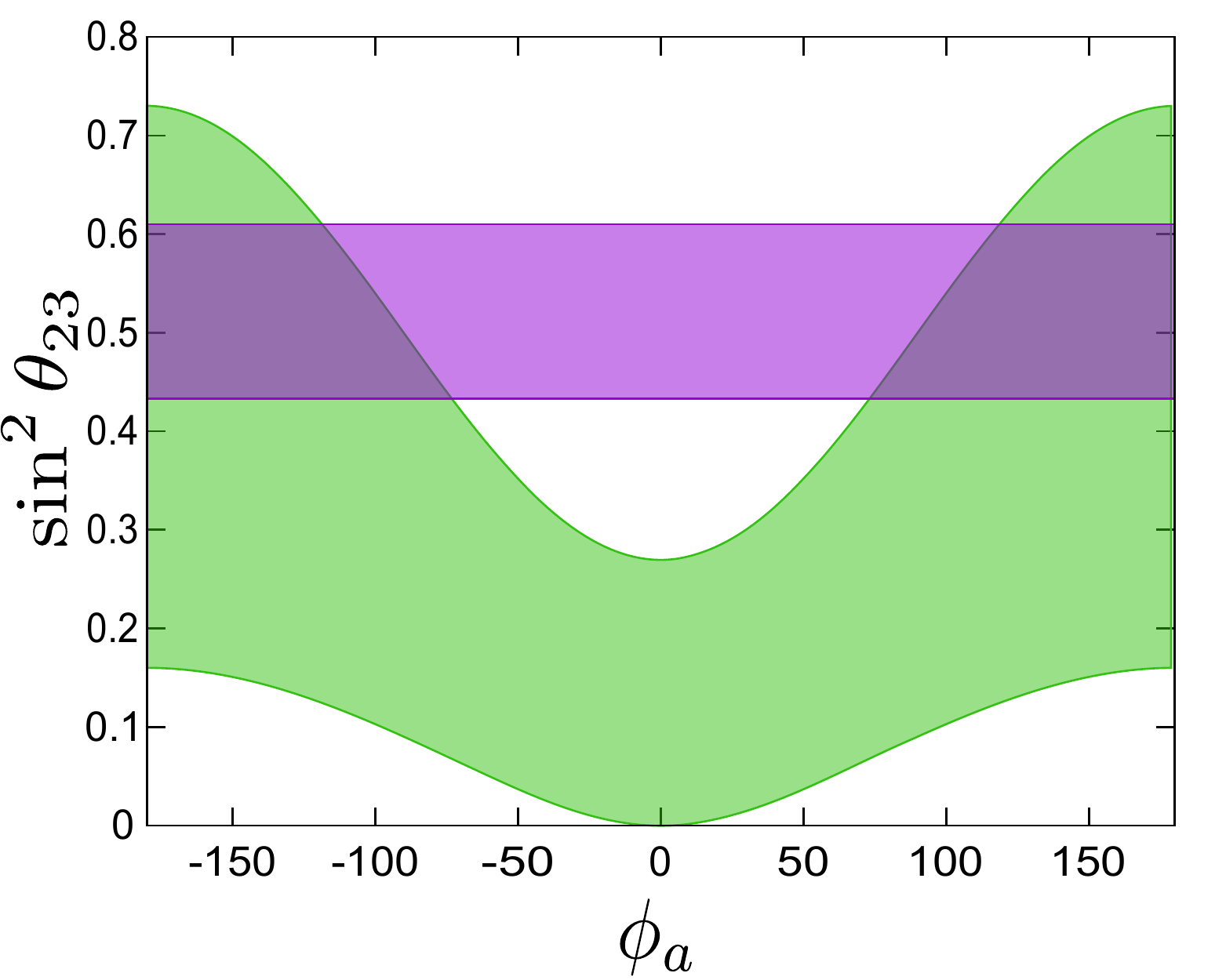}} 
  \end{tabular}
  \caption{The allowed regions for the atmospheric mixing angle and free parameters $\phi_{a}$ and $\phi_{c}$, for  
   equivalent class with two texture zeros type-I.
   The purple stripe corresponds to the values at 3$\sigma$ for atmospheric mixing angle obtained from the global fit, for normal and inverted 
   hierarchy~\cite{global-fit-2021}.
   In these panels, the red area is for ${\bf M}_{\ell}^{0}$ and ${\bf M}_{\ell}^{3}$, and 
   the green area is for ${\bf M}_{\ell}^{2}$ and ${\bf M}_{\ell}^{4}$.}\label{Fig:Angulos-Clase-I-phi-a-Phi-c}
 \end{center}
\end{figure}
From the allowed regions of flavor mixing angles shown in the figure~\ref{Fig:Angulos-Clase-I-delta-l}, which are computed taken into account the condition 
$\textrm{\bf B}$~\eqref{eq:conditionsdelta}  it is easy  to conclude that all charged lepton mass matrices are able to reproduce the current experimental values of 
reactor, solar and atmospheric angles. However, the numerical values interval of the free parameter $\delta_{\ell}$, for the ${\bf M}_{\ell}^{1}$, ${\bf M}_{\ell}^{2}$, 
${\bf M}_{\ell}^{4}$ and ${\bf M}_{\ell}^{5}$ mass matrices, is too small. 
Consequently, for this equivalent class, to reproduce the values for the leptonic flavor mixing angles, at $3\sigma$ obtained from the global fit 
eq.~(\ref{Ec:exp-mixing-angles}), for a normal (NH) and inverted (IH) hierarchy.  The free parameter $\delta_{\ell}$ should be in the following numerical interval:
\begin{equation}\label{Eq:Type-I:val-delta-l}
 \begin{array}{ll}\vspace{2mm}
  \delta_{\ell} \, \in \, \left[ 0.99132, 0.99382 \right] &\qquad \textrm{for} \, {\bf M}_{\ell}^{0} \, \textrm{and} \, {\bf M}_{\ell}^{3}, \\
  \delta_{\ell} \, \approx 0.9994 &\qquad  \textrm{for} \, {\bf M}_{\ell}^{1} \, \textrm{and} \, {\bf M}_{\ell}^{5},\\
  \delta_{\ell} \, \approx 0.9997 & \qquad \textrm{for} \, {\bf M}_{\ell}^{2} \, \textrm{and} \, {\bf M}_{\ell}^{4}.
 \end{array}
\end{equation}
In this equivalent class, from expressions in Appendix~\ref{Appendix-Para-Type-I} and figure~\ref{Fig:Angulos-Clase-I-phi-a-Phi-c} we conclude, shown:
\begin{enumerate}
 \item For the mass matrices ${\bf M}_{\ell}^{0}$ and ${\bf M}_{\ell}^{3}$, on the one hand the expression for solar mixing angle can reproduce the current experimental 
  data independently of numerical value of phase factors $\phi_{a}$ and $\phi_{c}$. 
  In other words, the mixing angle $\theta_{12}$  has a weak dependence on parameters $\phi_{a}$ and $\phi_{c}$. 
  On the other hand, the expression for the mixing angle $\theta_{13}$  does not has a explicit dependence on phase factor $\phi_{a}$, but if has a weak dependence on 
  phase factor $\phi_{c}$. 
  Finally, the expression for the mixing angle $\theta_{23}$ does not has a explicit dependence on phase factor $\phi_{a}$. However, to reproduce the current 
  experimental data at $3\sigma$ given in eq.~(\ref{Ec:exp-mixing-angles}) for  the $\theta_{23}$ angle, the phase factor $\phi_{c}$ must be on the following 
  numerical  interval; 
  \begin{equation}\label{Eq:Type-I:val-phi-c}
   |\phi_{c}| \in \left[ 76^{\circ}, 180^{\circ} \right] \quad \textrm{for} \, \, {\bf M}_{\ell}^{0} \, \textrm{and} \, {\bf M}_{\ell}^{3}.
  \end{equation}
 \item For the mass matrices ${\bf M}_{\ell}^{1}$ and ${\bf M}_{\ell}^{5}$, the reactor, solar and atmospheric mixing angles have a weak dependence on the phase factors 
  $\phi_{a}$ and $\phi_{c}$. 
 \item For the mass matrices ${\bf M}_{\ell}^{2}$ and ${\bf M}_{\ell}^{4}$, on the one hand the expression for the mixing angle $\theta_{12}$  has a weak dependence on 
  parameters $\phi_{a}$ and $\phi_{c}$. 
  On the other hand, the expression for the mixing angle $\theta_{13}$  does not has a explicit dependence on phase factor $\phi_{c}$, but if has a weak dependence on 
  phase factor $\phi_{a}$.
  Finally, the expression for the mixing angle  $\theta_{23}$ does not has a explicit dependence on phase factor $\phi_{c}$. 
  So, to  reproduce the current experimental data  eq.~(\ref{Ec:exp-mixing-angles}) at $3\sigma$ for  the $\theta_{23}$ angle, the phase factor $\phi_{a}$ runs over 
  the numerical range
  \begin{equation}\label{Eq:Type-I:val-phi-a}
   \begin{array}{ll}
    |\phi_{a}| \in \left[ 76^{\circ}, 180^{\circ} \right] & \textrm{for} \, \, {\bf M}_{\ell}^{2} \, \textrm{and} \, {\bf M}_{\ell}^{4}.
   \end{array}
  \end{equation}
\end{enumerate}
\begin{figure}[!htbp]
 \begin{center}
  \begin{tabular}{cc}
  \subfigure{ \includegraphics[width=8.5cm, height=7cm]{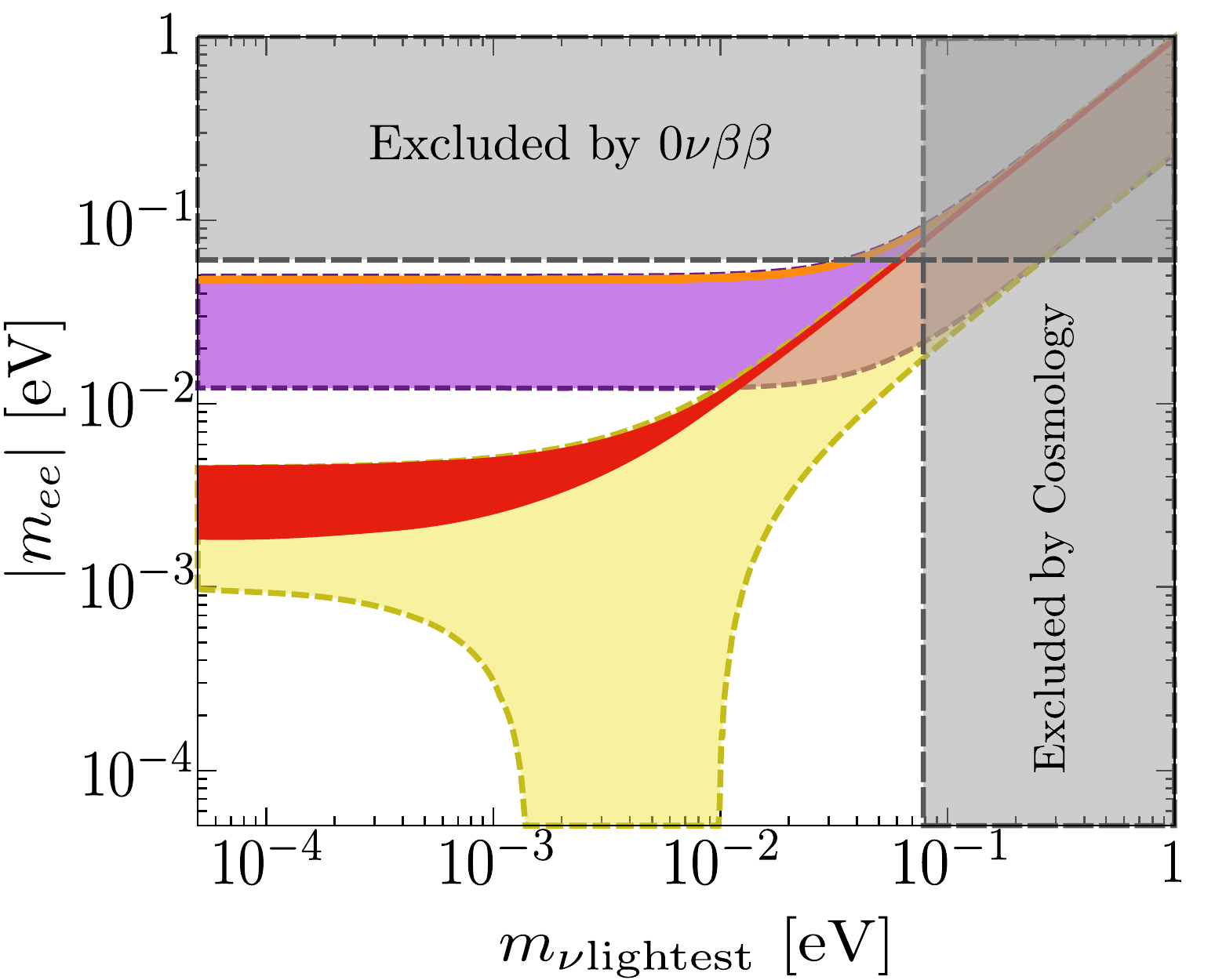}}
  \subfigure{ \includegraphics[width=8.5cm, height=7cm]{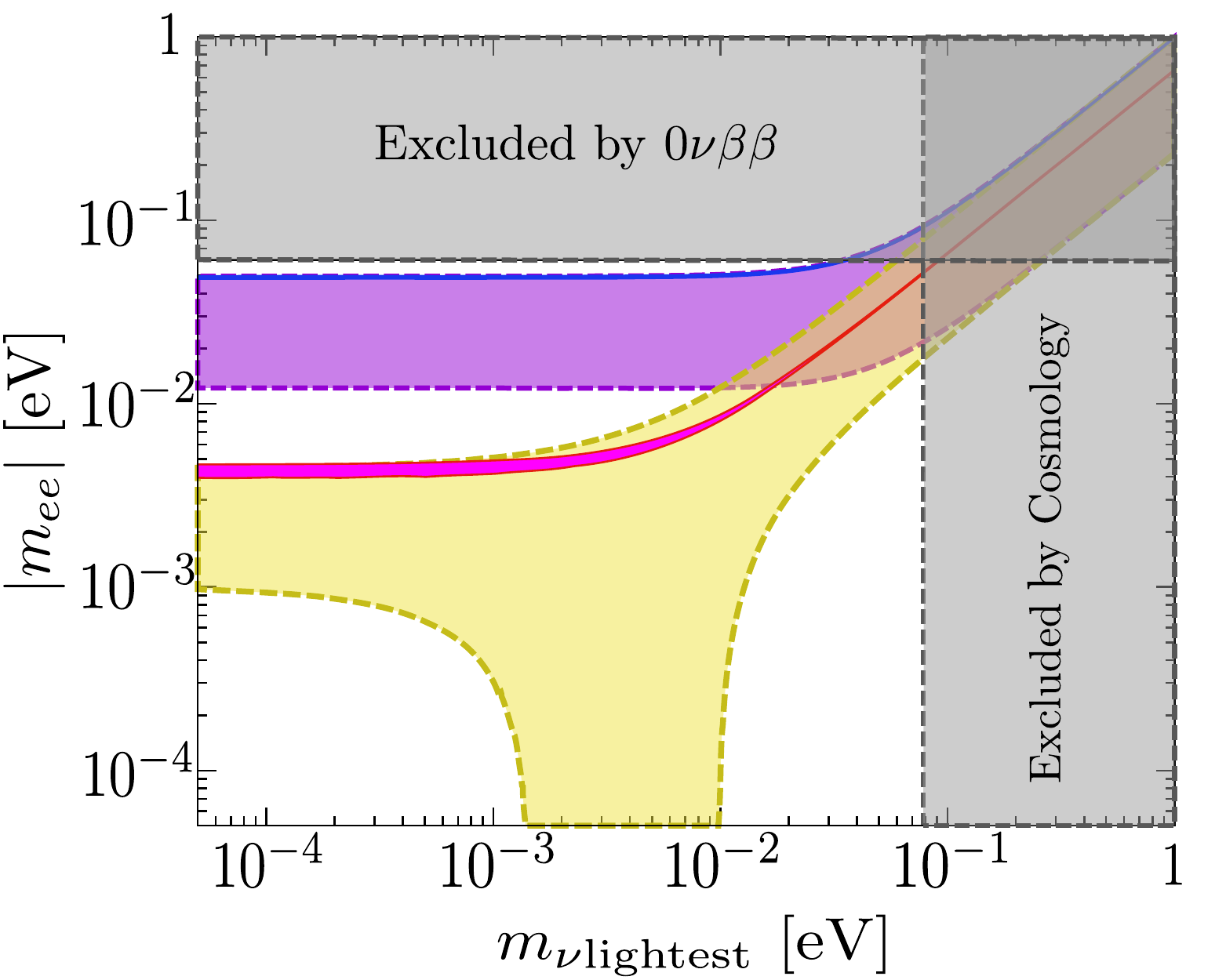}} \\
  \subfigure{ \includegraphics[width=8.5cm, height=7cm]{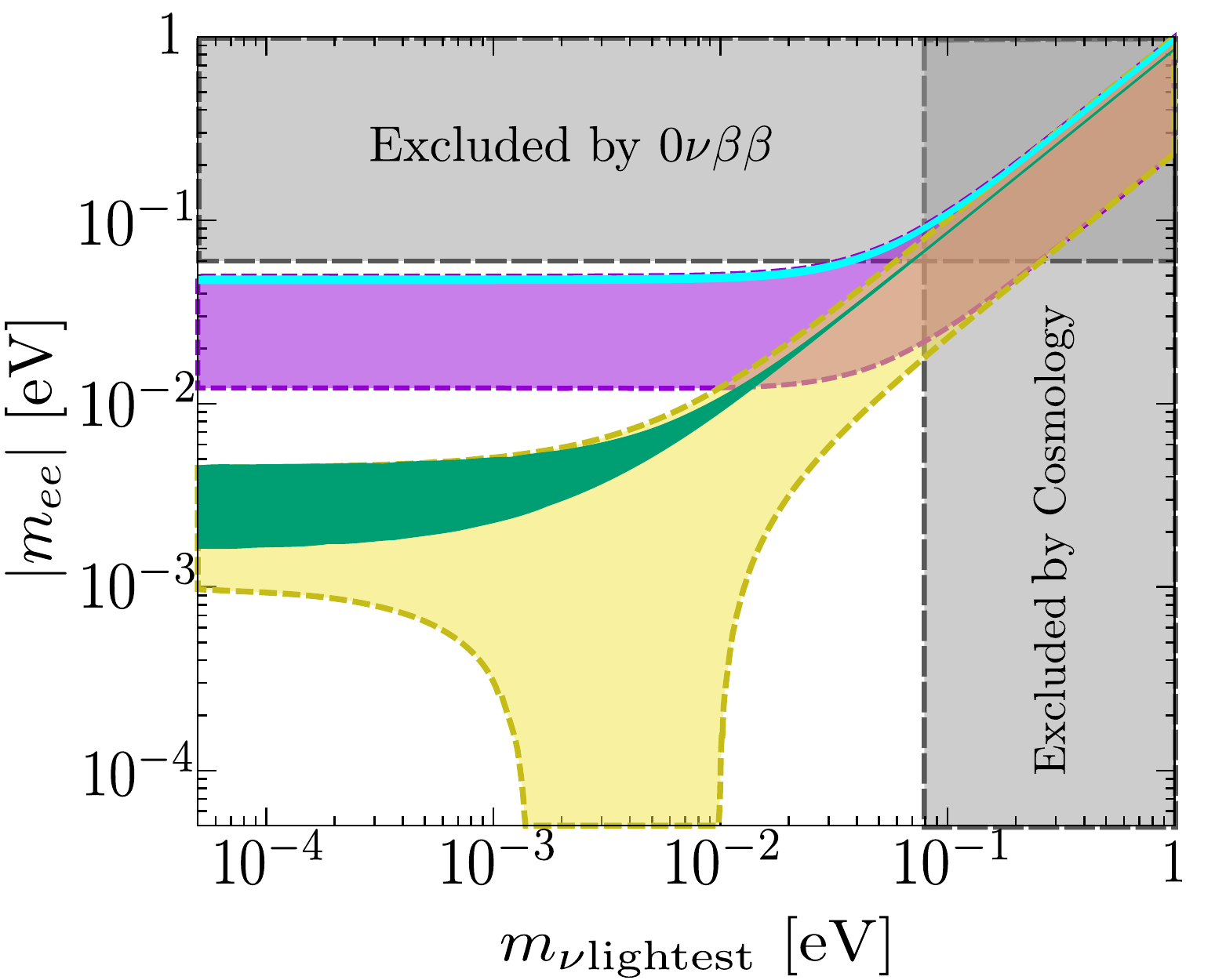}}
  \end{tabular}
  \caption{ In these panels, regions allowed for the magnitude of Majorana effective mass $|m_{ee}|$ are shown.
   Respectively, for an inverted and normal neutrino mass hierarchy, the yellow and purple stripes are obtained from the current experimental data on neutrino 
   oscillations at 3$\sigma$~\cite{global-fit-2021}.  The upper left panel, the red area is for a normal hierarchy while the orange area is for an inverted hierarchy, both areas are obtained from matrices  
   ${\bf M}_{\ell}^{0}$ and ${\bf M}_{\ell}^{3}$. The upper right panel, the magenta area is for a normal hierarchy while the blue area is for an inverted hierarchy, 
   both areas are obtained from ${\bf M}_{\ell}^{1}$ and ${\bf M}_{\ell}^{5}$. 
   In the lower panel, the green area is for a normal hierarchy while the cyan area is for an inverted hierarchy, both areas are obtained from  
   ${\bf M}_{\ell}^{2}$ and ${\bf M}_{\ell}^{4}$.
   From KamLAND-ZEN~\cite{KamLAND-Zen:2016pfg} and EXO-200~\cite{EXO:2017poz} the following upper limit $|mee| < 0.061$, which correspond to the 
   horizontal grey band. From the results reported by Planck collaboration, obtaining the vertical grey band~\cite{Planck:2015fie}.}\label{Fig:Mee:Type-I}
 \end{center}
\end{figure}
In figure~\ref{Fig:Mee:Type-I} we show the allowed regions for the magnitude of the Majorana effective mass $|m_{ee}|$, eq.~(\ref{ec:mee-0}), which were obtained in a 
model-independent context where the neutrino mass matrix has the form given in eq.~(\ref{Eq:M_TBM}), while the charged lepton matrix is represented for an element of 
the equivalent class with two texture zeros type-I. eq.~(\ref{Eq:EQ-Type-I}). Each one of these regions was obtained by considering the values given in 
eqs.~(\ref{Eq:Type-I:val-delta-l})-(\ref{Eq:Type-I:val-phi-a}) for the free parameter $\delta_{\ell}$ constrained by $\textrm{\bf B}$~\eqref{eq:conditionsdelta},  and 
the associated to the CP violation phases $\phi_{a}$ and $\phi_{c}$.
%
%
%
\subsection{Equivalent class with two texture zeros type-II}
%
The equivalent class for Hermitian matrices with two texture zeros type-II have the form~\cite{Canales:2012dr}: 
\begin{equation}\label{Eq:EQ-Type-II}
 \begin{array}{ccc}
  {\bf M}_{\ell}^{0} = 
  \left( \begin{array}{ccc}
   f_{\ell}     & a_{\ell}     & 0        \\
   a_{\ell}^{*} & 0            & c_{\ell} \\
   0            & c_{\ell}^{*} & d_{\ell}
  \end{array} \right), &
  {\bf M}_{\ell}^{1} = 
  \left( \begin{array}{ccc}
   0            & a_{\ell}^{*} & c_{\ell} \\
   a_{\ell}     & f_{\ell}     & 0 \\
   c_{\ell}^{*} & 0            & d_{\ell}
  \end{array} \right), &
  {\bf M}_{\ell}^{2} = 
  \left( \begin{array}{ccc}
   d_{\ell} & c_{\ell}^{*} & 0 \\
   c_{\ell} & 0            & a_{\ell}^{*} \\
   0        & a_{\ell}     & f_{\ell}
  \end{array} \right), \\
  {\bf M}_{\ell}^{3} = 
  \left( \begin{array}{ccc}
   f_{\ell}     & 0        & a_{\ell}  \\
   0            & d_{\ell} & c_{\ell}^{*}  \\
   a_{\ell}^{*} & c_{\ell} & 0 
  \end{array} \right), &
  {\bf M}_{\ell}^{4} = 
  \left( \begin{array}{ccc}
   d_{\ell} & 0            & c_{\ell}^{*} \\
   0        & f_{\ell}     & a_{\ell} \\
   c_{\ell} & a_{\ell}^{*} & 0
  \end{array} \right), &
  {\bf M}_{\ell}^{5} = 
  \left( \begin{array}{ccc}
   0     & c_{\ell} & a_{\ell}^{*} \\
   c_{\ell}^{*} & d_{\ell} & 0 \\
   a_{\ell}     & 0        & f_{\ell}
  \end{array} \right),
 \end{array}
\end{equation}
where $d_{\ell} = 1 - \delta_{\ell}$, 
\begin{equation}
 \begin{array}{l}
  a_{\ell} = 
   \sqrt{ \frac{ 
    \widetilde{\sigma}_{\ell 1} \widetilde{\sigma}_{\ell 2} \widetilde{\sigma}_{\ell 3} 
   }{ \widetilde{\mu}_{\ell} } }, \quad
  c_{\ell} = \sqrt{ \frac{ \delta_{\ell} f_{\ell 1} f_{\ell 2} }{ \widetilde{\mu}_{\ell} } } e^{i \phi_{c} }, \quad
  f_{\ell} = \widetilde{m}_{e} - \widetilde{m}_{\mu} + \delta_{\ell}, 
 \end{array}
\end{equation}
with 
$\widetilde{\sigma}_{\ell 1} = f_{\ell} - \widetilde{m}_{e}   = \delta_{\ell} - \widetilde{m}_{\mu}$,
$\widetilde{\sigma}_{\ell 2} = f_{\ell} + \widetilde{m}_{\mu} = \delta_{\ell} + \widetilde{m}_{e}  $, 
$\widetilde{\sigma}_{\ell 3} = 1 - f_{\ell} = 1 - \delta_{\ell} + \widetilde{m}_{\mu} - \widetilde{m}_{e}$, 
$\widetilde{\mu}_{\ell}      = \widetilde{m}_{e} - \widetilde{m}_{\mu} - 1 + 2 \delta_{\ell}$,
$f_{\ell 1} = - \left(1 - \widetilde{m}_{e} - \delta_{\ell} \right)$, 
$f_{\ell 2} = 1 + \widetilde{m}_{\mu} - \delta_{\ell} $, 
$f_{\ell 3} = \delta_{\ell}$, 
$\phi_{a} = \arg \left \{ a_{\ell} \right \}$, and $\phi_{c} = \arg \left \{ c_{\ell} \right \}$. In this case, the diagonal matrix of phase factors is
${\bf P}_{\ell} = \mathrm{diag} \left( 1,  e^{ i \phi_{a} } ,e^{ i \left( \phi_{a} + \phi_{c} \right) }  \right) $. 

\begin{figure}[t]
	\begin{center}
		\begin{tabular}{cc}  
			\subfigure{ \includegraphics[width=7cm, height=5cm]{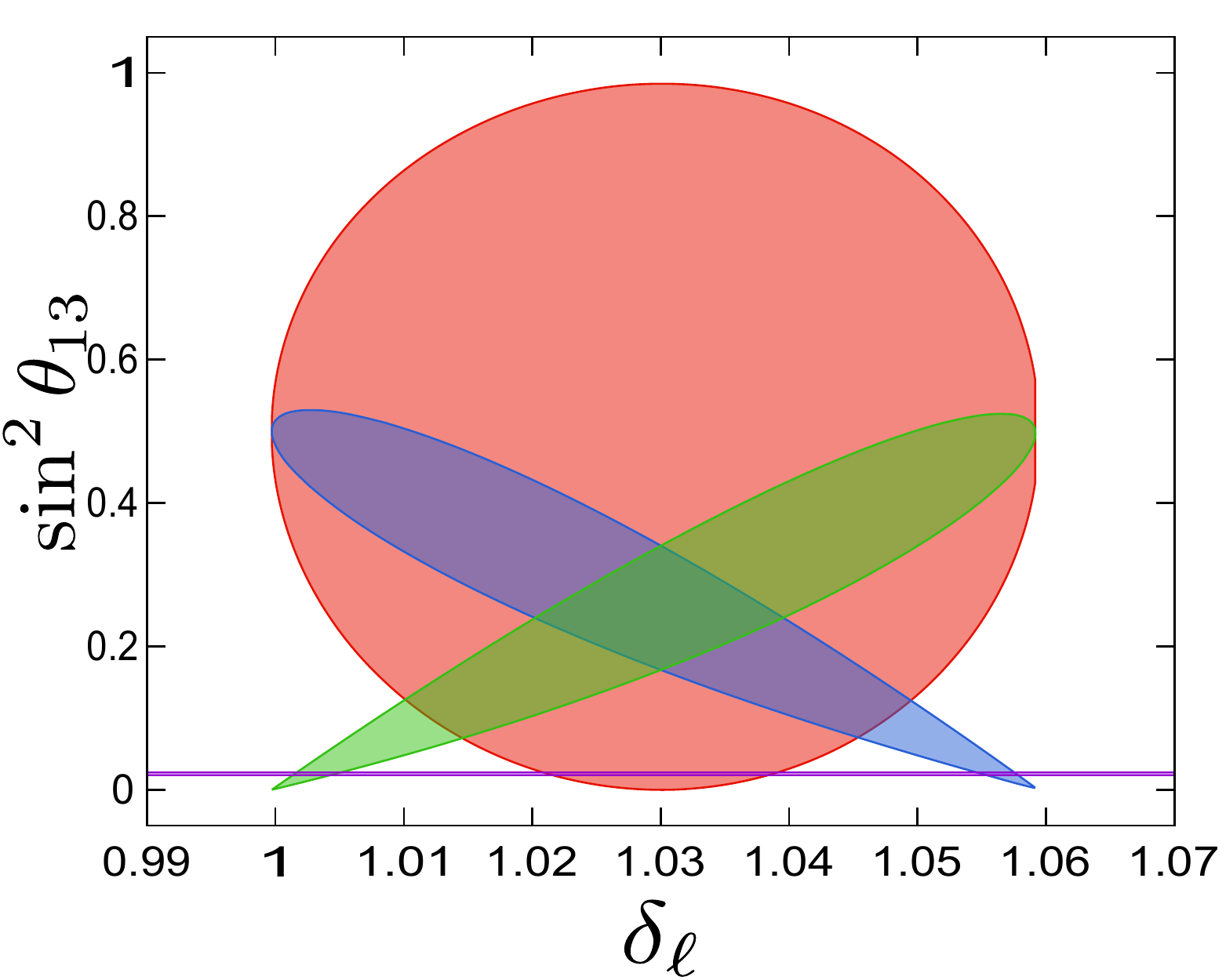}}  
			\subfigure{ \includegraphics[width=7cm, height=5cm]{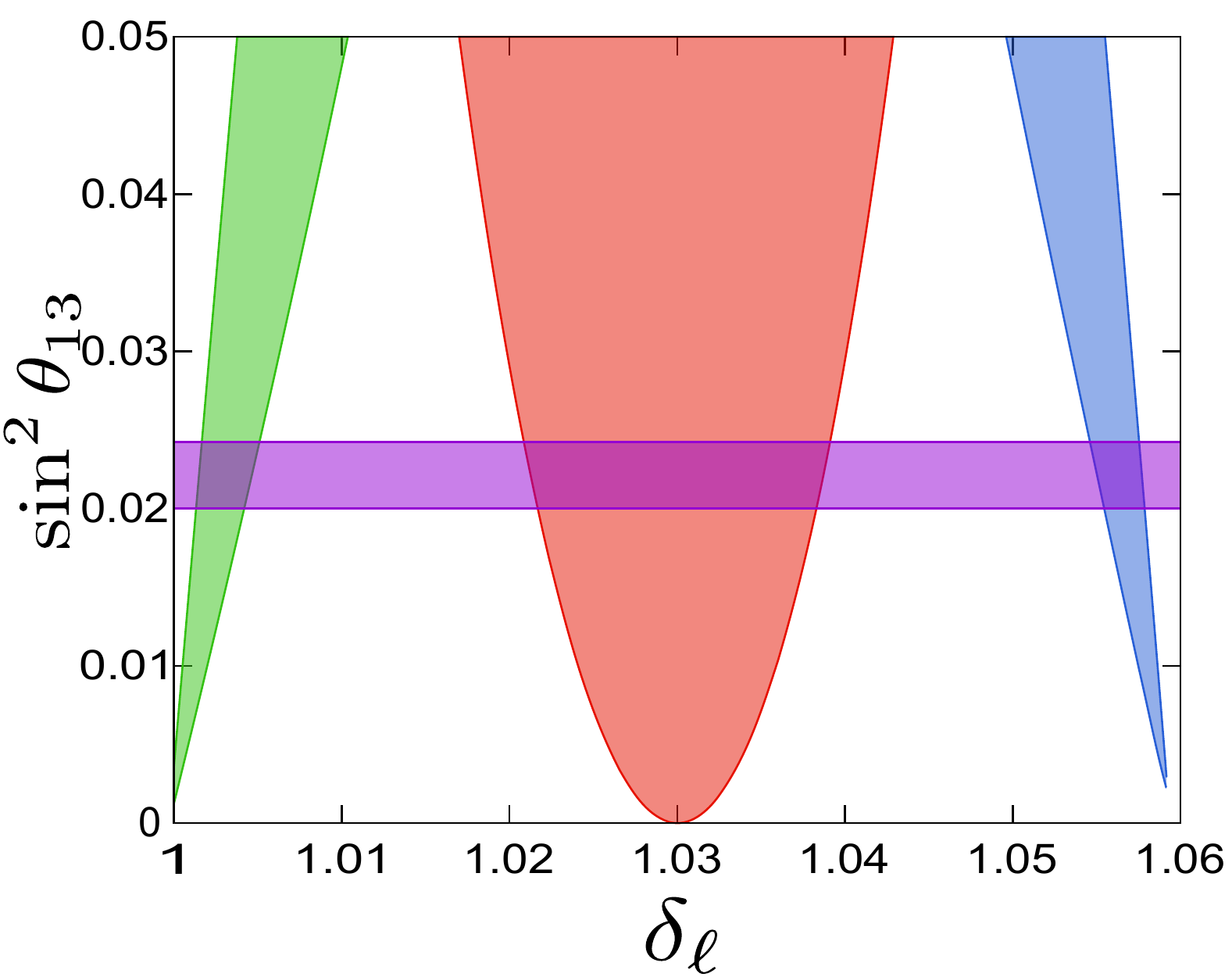}} \\   
			\subfigure{ \includegraphics[width=7cm, height=5cm]{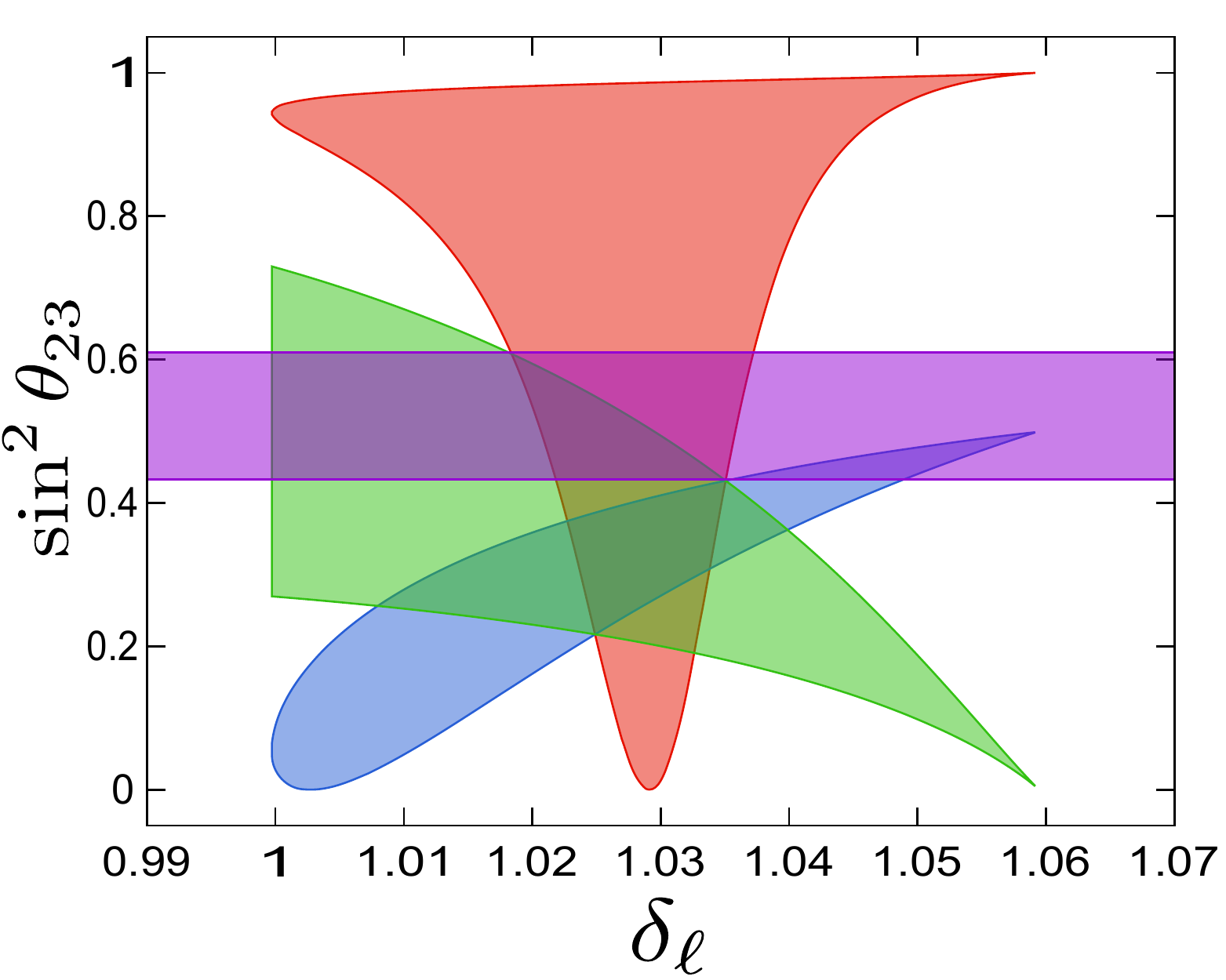}}
			\subfigure{ \includegraphics[width=7cm, height=5cm]{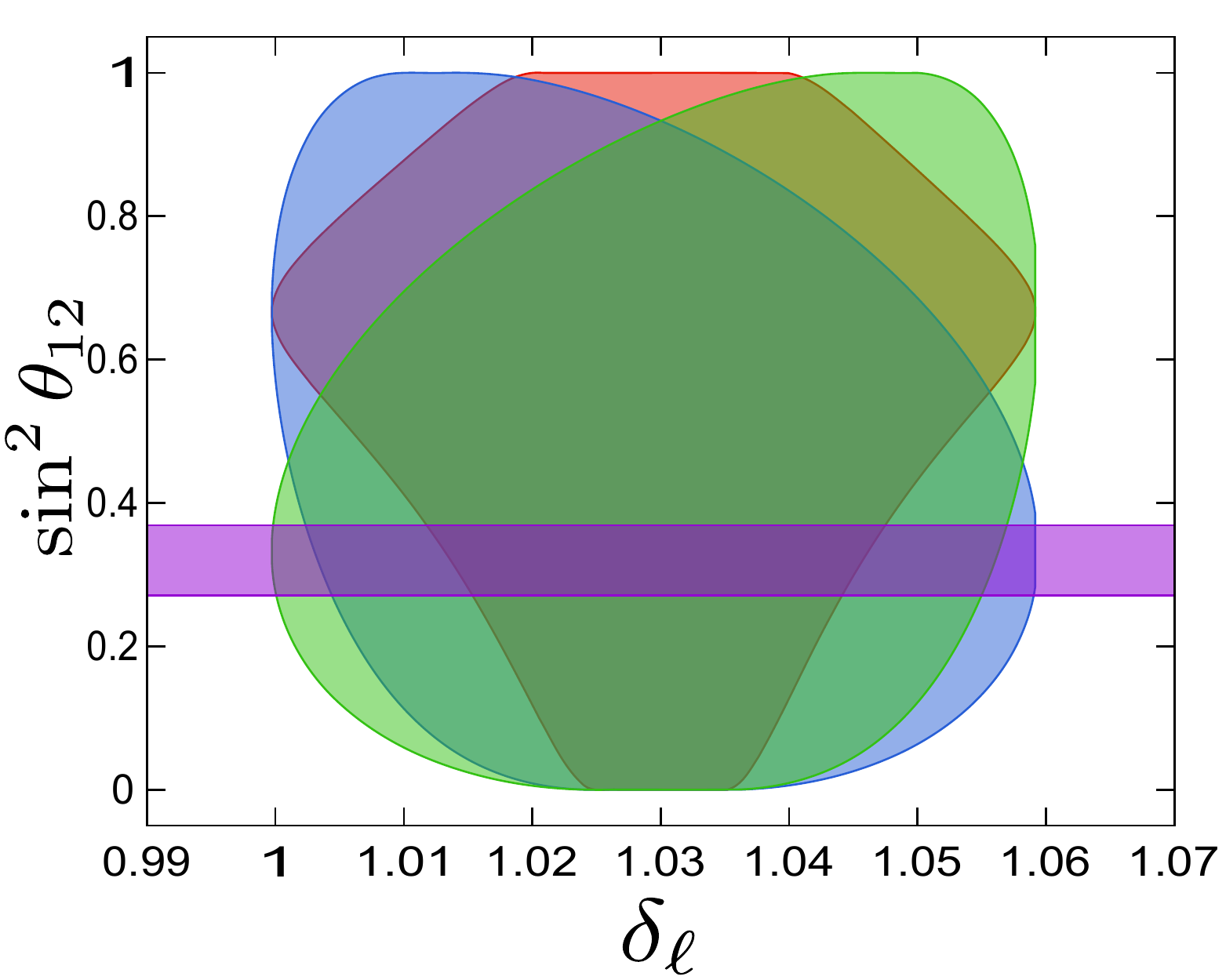}}
		\end{tabular}
		\caption{The allowed regions for the reactor (upper panels), atmospheric, and solar (lower panels) mixing angles and free parameter $\delta_{\ell}$, for  
			equivalent class with two texture zeros type-II. 
			The purple stripe corresponds to the values at 3$\sigma$ for the reactor, atmospheric and solar mixing angles obtained from the global fit, for normal and inverted 
			hierarchy~\cite{global-fit-2021}.
			In these panels, the red area is for ${\bf M}_{\ell}^{0}$ and ${\bf M}_{\ell}^{3}$, blue area is for ${\bf M}_{\ell}^{1}$ and ${\bf M}_{\ell}^{5}$, 
			the green area is for ${\bf M}_{\ell}^{2}$ and ${\bf M}_{\ell}^{4}$. 
			The right lower panel shows an amplification of the region in which the mixing angles theoretical expressions simultaneously reproduce the current experimental 
			data.}\label{Fig:Angulos-Clase-II-delta-l}
	\end{center}
\end{figure}
The real orthogonal matrix ${\bf O}_{\ell}$ is constructed with the help of the general eigenvectors given in eq.~(\ref{Eq:General_Eigenvector}), which are the 
eigenvectors of the charged lepton mass matrix. The explicit form of  
${\bf O}_{\ell} = \left( | M_{1} \rangle, | M_{2} \rangle, | M_{3} \rangle \right)$  is
\begin{equation}
 \begin{array}{l}
  {\bf O}_{\ell} = 
  \left( \begin{array}{ccc}\vspace{2mm}
     \sqrt{ \frac{ \widetilde{\sigma}_{\ell 2} \widetilde{\sigma}_{\ell 3} f_{\ell 1} }{ D_{\ell 1} } } &
   - \sqrt{ \frac{ \widetilde{\sigma}_{\ell 1} \widetilde{\sigma}_{\ell 3} f_{\ell 2} }{ D_{\ell 2} } } & 
     \sqrt{ \frac{ \widetilde{\sigma}_{\ell 1} \widetilde{\sigma}_{\ell 2} f_{\ell 3} }{ D_{\ell 3} } }  \\ \vspace{2mm}
   - \sqrt{ \frac{ \widetilde{\sigma}_{\ell 1} \widetilde{\mu}_{\ell} f_{\ell 1} }{  D_{\ell 1} } }  &
     \sqrt{ \frac{ \widetilde{\sigma}_{\ell 2} \widetilde{\mu}_{\ell} f_{\ell 2} }{  D_{\ell 2} } }  &
     \sqrt{ \frac{ \widetilde{\sigma}_{\ell 3} \widetilde{\mu}_{\ell} f_{\ell 3} }{  D_{\ell 3} } }  \\ \vspace{2mm}
   - \sqrt{ \frac{ \widetilde{\sigma}_{\ell 1} \delta_{\ell} f_{\ell 2} }{  D_{\ell 1} } }  &
   - \sqrt{ \frac{ \widetilde{\sigma}_{\ell 2} \delta_{\ell} f_{\ell 1} }{  D_{\ell 2} } }  &
     \sqrt{ \frac{ \widetilde{\sigma}_{\ell 3} f_{\ell 1} f_{\ell 2} }{  D_{\ell 3} } }     
  \end{array} \right),
 \end{array}
\end{equation}
where
 \begin{equation}
  \begin{array}{l} \vspace{2mm}
   {\cal D}_{\ell 1} =       
    \widetilde{\mu}_{\ell} \left( \widetilde{m}_{\mu} + \widetilde{m}_{e} \right) \left( 1 - \widetilde{m}_{e} \right), \quad
   {\cal D}_{\ell 2} =       
    \widetilde{\mu}_{\ell} \left( \widetilde{m}_{\mu} + \widetilde{m}_{e} \right) \left( 1 + \widetilde{m}_{\mu} \right), \quad
   {\cal D}_{\ell 3} =       
    \widetilde{\mu}_{\ell} \left( 1 - \widetilde{m}_{e} \right) \left( 1 + \widetilde{m}_{\mu} \right) .  
  \end{array}
 \end{equation} 
 The parameter $\delta_{\ell}$ must satisfy the conditions 
\begin{equation}
 \begin{array}{l}
  1 + \widetilde{m}_{\mu} - \widetilde{m}_{e} > \delta_{\ell} > 1 - \widetilde{m}_{e}, \quad \textrm{and} \quad
  \delta_{\ell} \neq \widetilde{m}_{\mu} - \widetilde{m}_{e}.
 \end{array}
\end{equation} 

\begin{figure}[!htbp]
	\begin{center}
		\begin{tabular}{cc} 
			\subfigure{ \includegraphics[width=7cm, height=5cm]{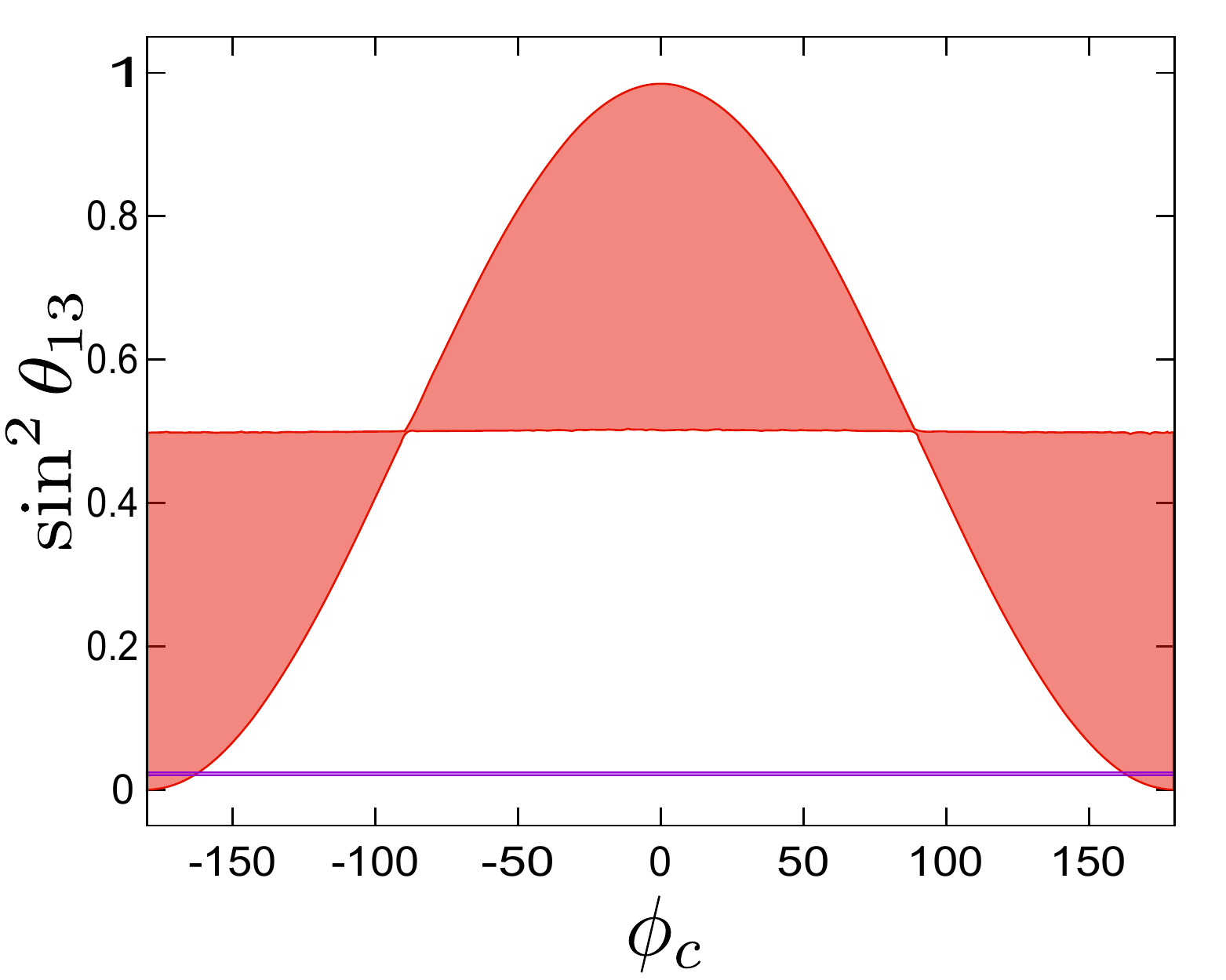}}
			\subfigure{ \includegraphics[width=7cm, height=5cm]{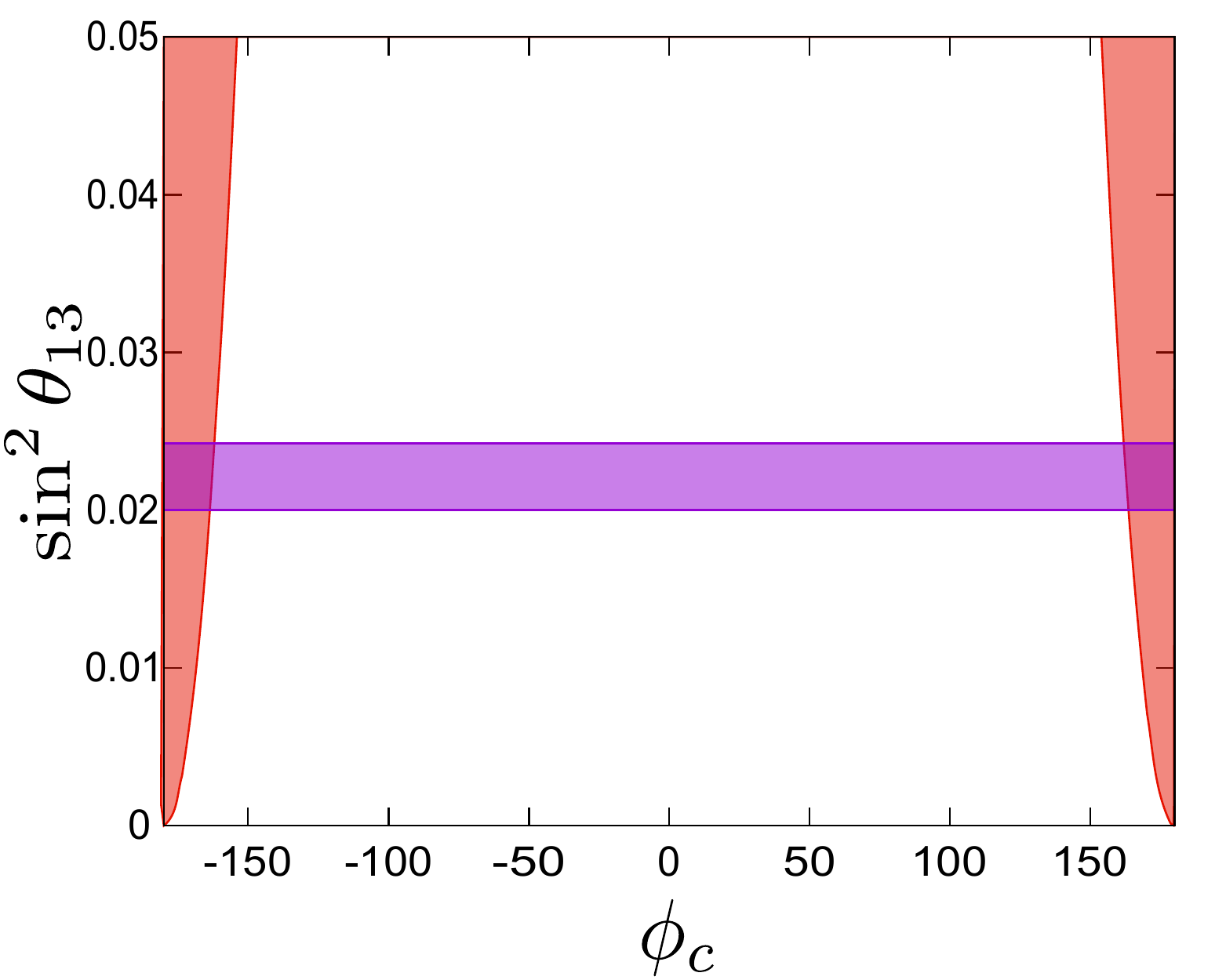}}\\
			\subfigure{ \includegraphics[width=7cm, height=5cm]{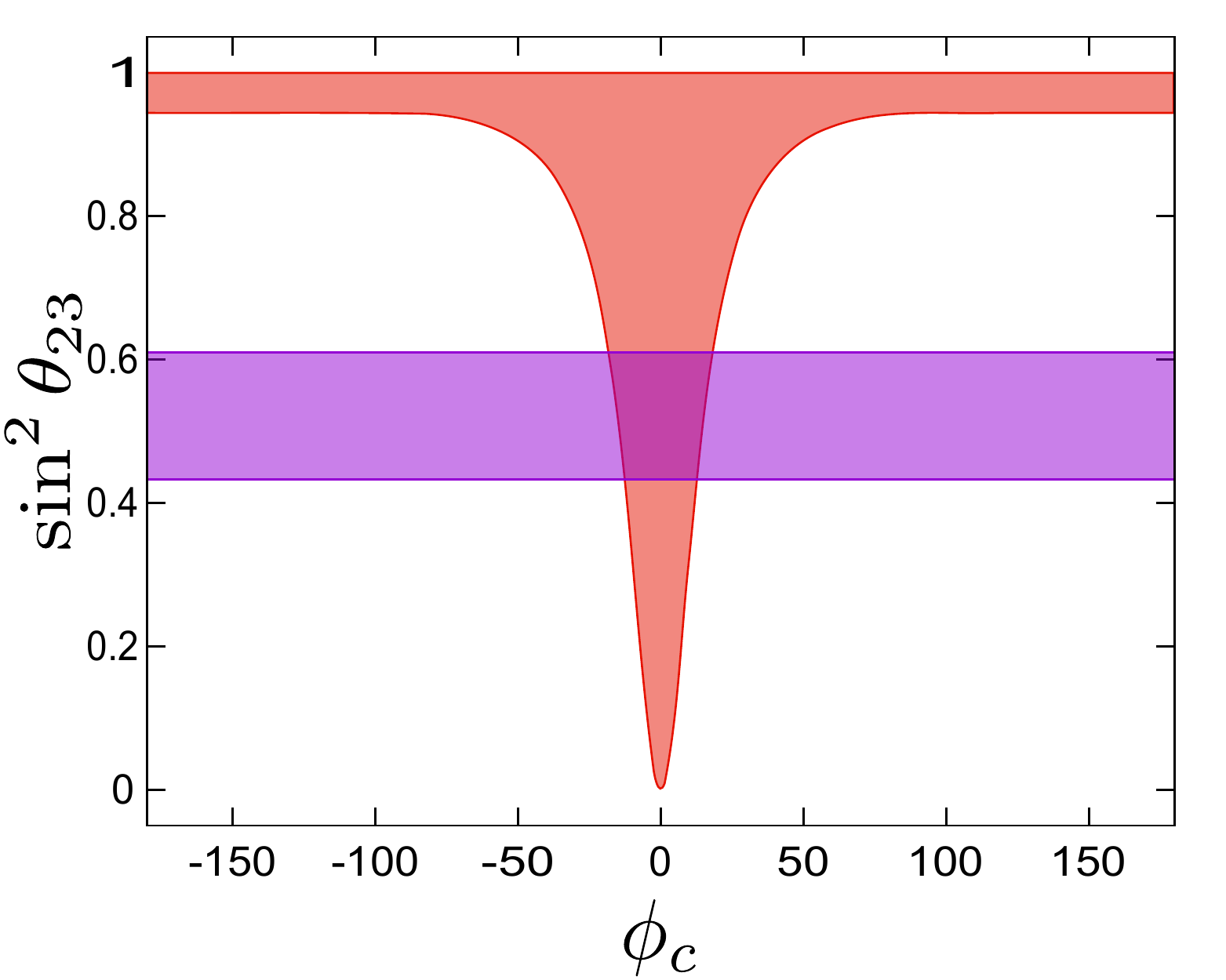}}
			\subfigure{ \includegraphics[width=7cm, height=5cm]{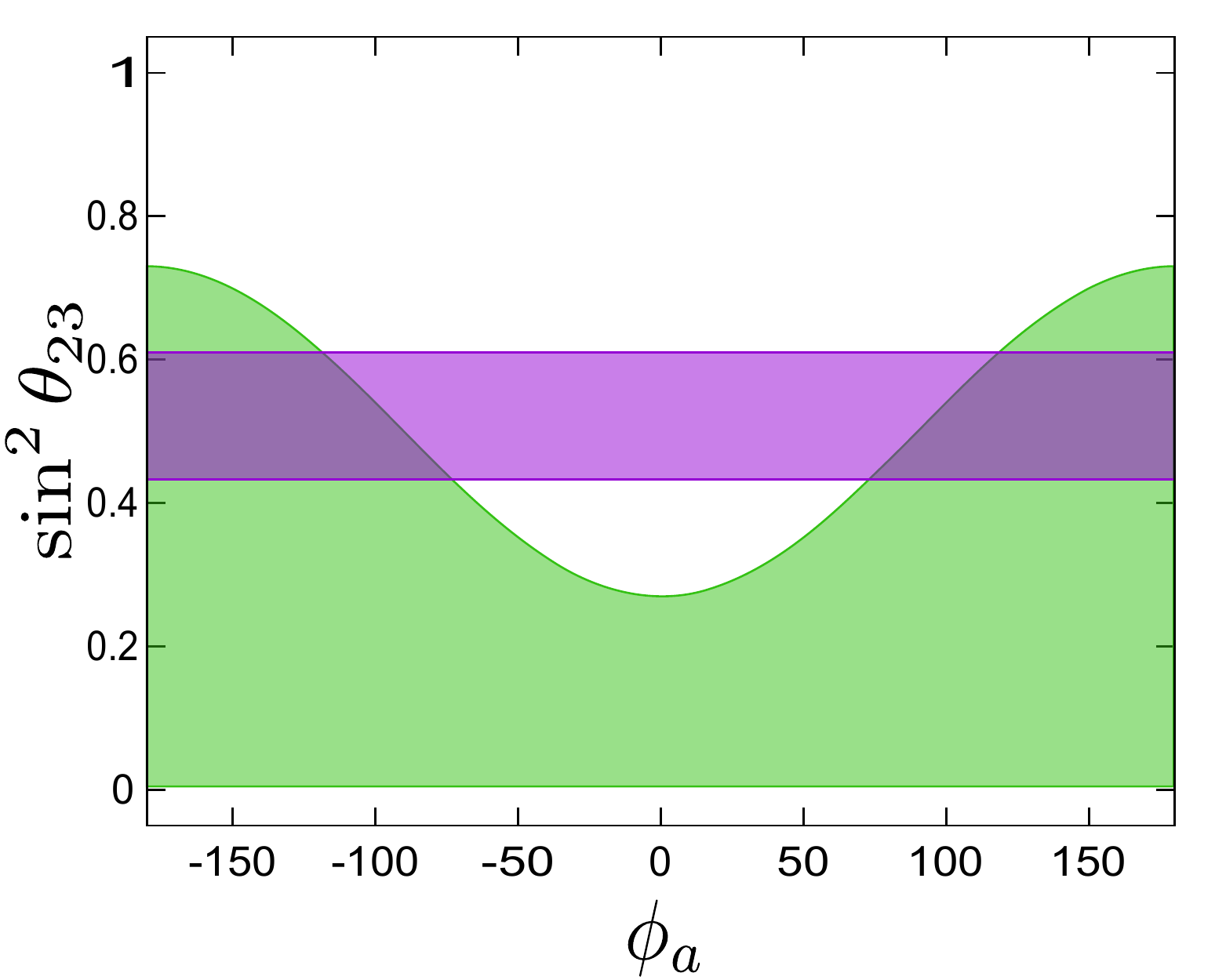}} \\
			\subfigure{ \includegraphics[width=7cm, height=5cm]{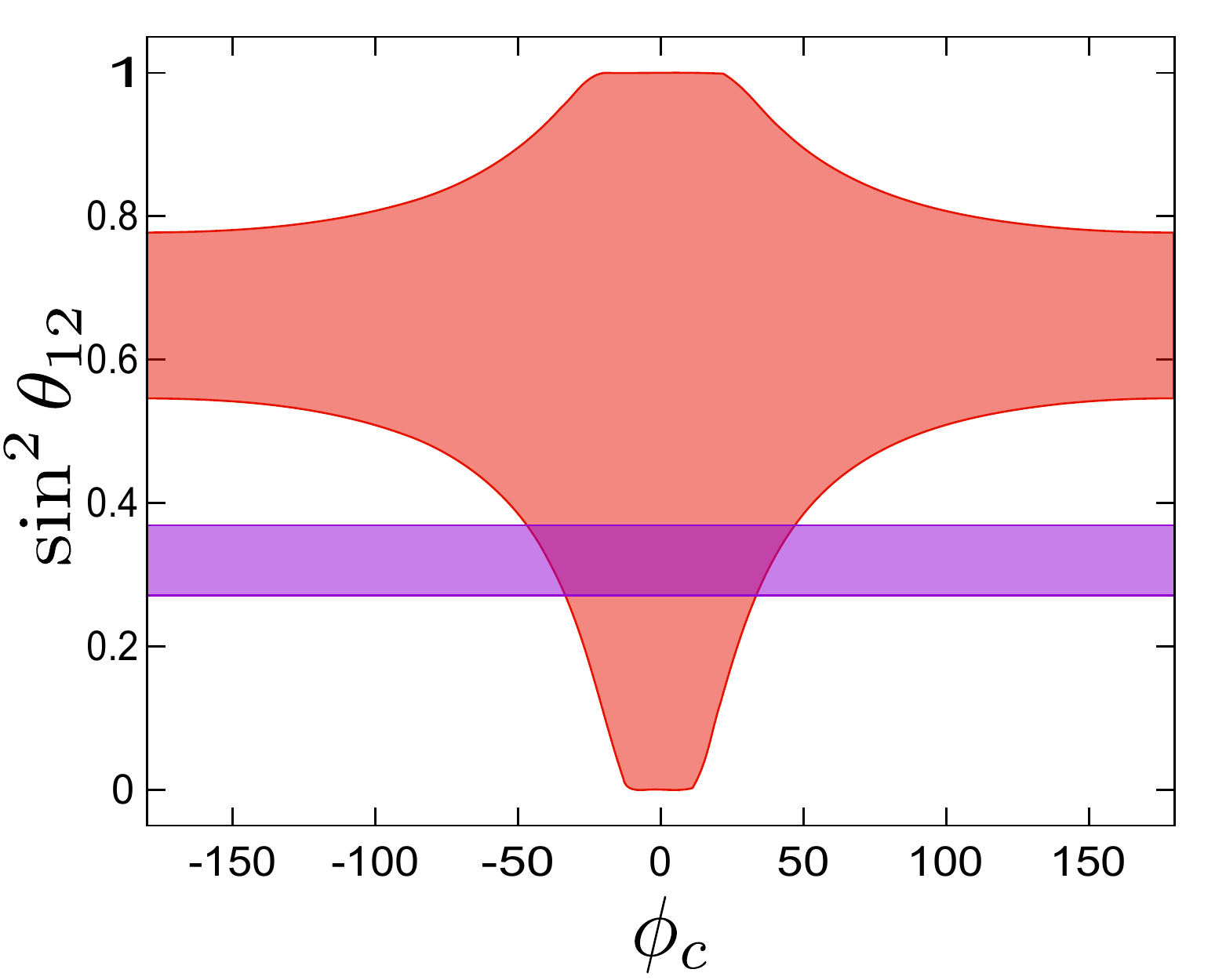}}
			\subfigure{ \includegraphics[width=7cm, height=5cm]{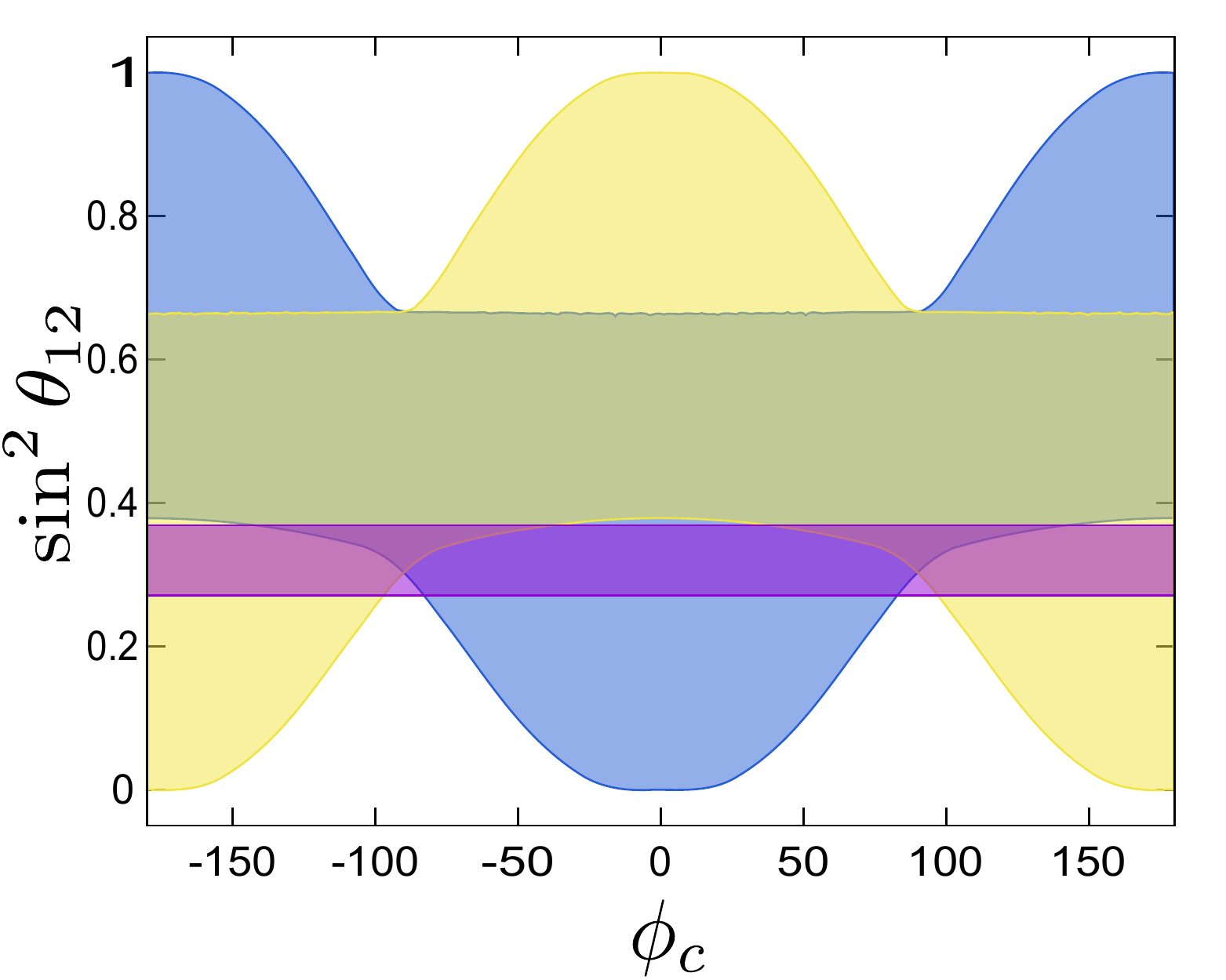}} 
		\end{tabular}
		\caption{The allowed regions for the reactor (upper panels), atmospheric (middle panels), and  solar (lower panels) mixing angles and free parameters 
			$\phi_{a}$ and $\phi_{c}$.
			The purple stripe corresponds to the values at 3$\sigma$ for the reactor, solar and atmospheric mixing angles obtained from the global fit, 
			for normal and inverted hierarchy~\cite{global-fit-2021}.
			In these panels, the red area is for ${\bf M}_{\ell}^{0}$ and ${\bf M}_{\ell}^{3}$,
			the green area is for ${\bf M}_{\ell}^{2}$ and ${\bf M}_{\ell}^{4}$, while the blue area is for ${\bf M}^{1}$ and yellow area is for ${\bf M}^{5}$.  
			The right upper panel shows an amplification of the region in which the reactor mixing angle theoretical expression reproduce the current experimental 
			data.}\label{Fig:Angulos-Clase-II-phi-a-Phi-c}
	\end{center}
\end{figure}
The flavor mixing angles in eq.~(\ref{Eq:Mix-Angle-0}) take the form:
\begin{equation}
 \begin{array}{l}
  \sin^{2} \theta_{12} = \frac{1}{3} \frac{ \widetilde{\sigma}_{\ell 1} }{ \widetilde{m}_{\mu} }  \varepsilon_{12}, \quad
  \sin^{2} \theta_{23} = \frac{1}{2} 
   \frac{ \left( 1 - \widetilde{m}_{e} \right) }{ \left( 1 + \widetilde{m}_{\mu} \right) } 
   \frac{ \widetilde{\sigma}_{\ell 2} }{ \widetilde{m}_{\mu} } \varepsilon_{23}, \quad
  \sin^{2} \theta_{13} = \frac{1}{2} \frac{ \widetilde{\sigma}_{\ell 1} }{ \widetilde{m}_{\mu} } \varepsilon_{13},   
 \end{array}
\end{equation}
The explicit form of the $\varepsilon_{ij}$ parameters is given in the Appendix~\ref{Appendix-Para-Type-II}.
From the allowed regions of flavor mixing angles shown in figure~\ref{Fig:Angulos-Clase-II-delta-l}, we obtain that all charged lepton mass matrices in this equivalent 
class are able to reproduce the current experimental data of the reactor, solar and atmospheric angles. However of the right panels in 
figure~\ref{Fig:Angulos-Clase-II-phi-a-Phi-c}, we can conclude that when considering the same numerical values interval for the phase factor $\phi_{c}$, in this 
equivalent class the mass matrices ${\bf M}_{\ell}^{1}$, ${\bf M}_{\ell}^{2}$, ${\bf M}_{\ell}^{4}$ and ${\bf M}_{\ell}^{5}$ are the only ones that can 
simultaneously reproduce the experimental data of the three mixing angles. 
%
%
	\begin{figure}[t]
	\begin{center}
		\begin{tabular}{cc}
			\subfigure{ \includegraphics[width=8.5cm, height=7cm]{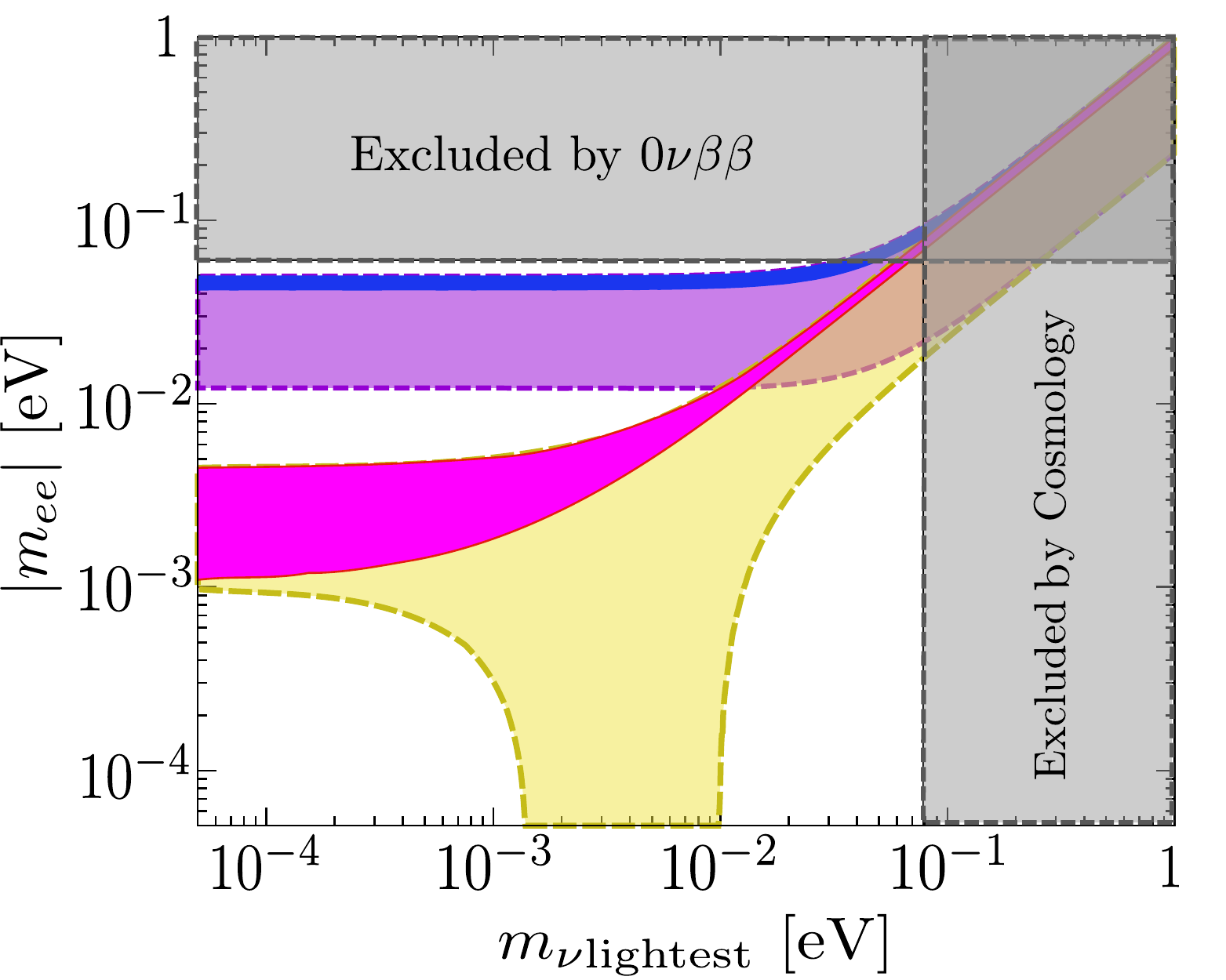}}
			\subfigure{ \includegraphics[width=8.5cm, height=7cm]{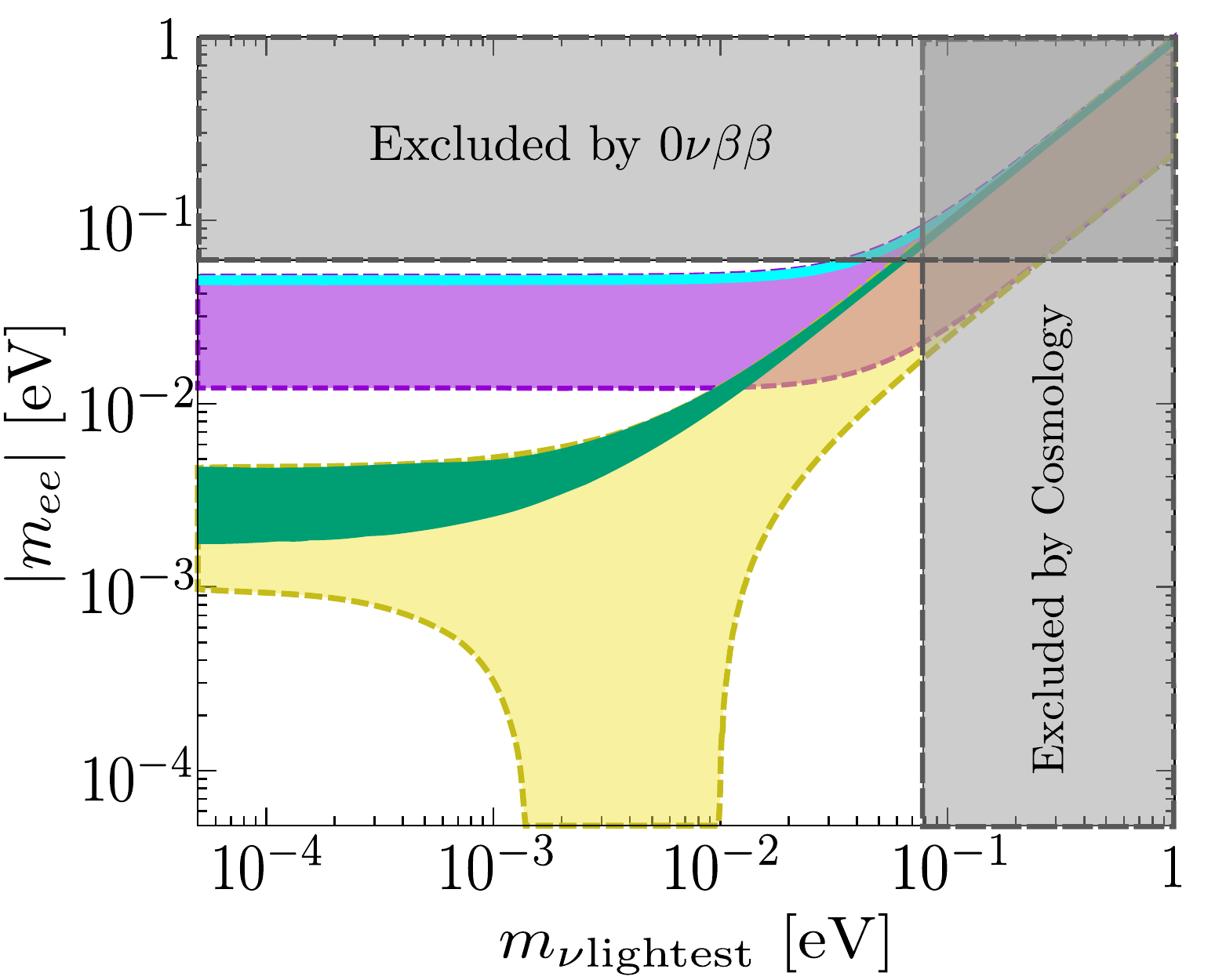}}
		\end{tabular}
		\caption{ Plots show the allowed regions for the magnitude of Majorana effective mass $|m_{ee}|$.
			Respectively, for an inverted and normal neutrino mass hierarchy, the yellow and purple stripes are obtained from the current experimental data on neutrino 
			oscillations at 3$\sigma$~\cite{global-fit-2021}. 
			In the upper left panel, the magenta area is for a normal hierarchy while the blue area is for an inverted hierarchy, 
			both areas are obtained from ${\bf M}_{\ell}^{1}$ and ${\bf M}_{\ell}^{5}$. 
			In the upper right panel, the green area is for a normal hierarchy while the cyan area is for an inverted hierarchy, both areas are obtained from  
			${\bf M}_{\ell}^{2}$ and ${\bf M}_{\ell}^{4}$.
			From KamLAND-ZEN~\cite{KamLAND-Zen:2016pfg} and EXO-200~\cite{EXO:2017poz} we have the following upper limit $|m_{ee}| < 0.061$, which correspond to the 
			horizontal grey band, whereas vertical grey band corresponds to results reported by Planck collaboration~\cite{Planck:2015fie}.}
		\label{Fig:Mee:Type-II}
	\end{center}
\end{figure}
Consequently, to reproduce the values for the leptonic flavor mixing angles, 
at $3\sigma$ obtained form the global fit eq.~(\ref{Ec:exp-mixing-angles}), for a normal and inverted hierarchy,  the free parameter $\delta_{\ell}$ should be in the 
following numerical interval:
\begin{equation}\label{Eq:Type-II:val-delta-l}
 \begin{array}{ll}\vspace{2mm} 
  \delta_{\ell} \, \in \, \left[ 1.054, 1.058 \right] & \textrm{for} \, \,{\bf M}_{\ell}^{1} \, \textrm{and} \, {\bf M}_{\ell}^{5},\\
  \delta_{\ell} \, \in \, \left[ 1.0013, 1.0052 \right] & \textrm{for} \,\, {\bf M}_{\ell}^{2} \, \textrm{and} \, {\bf M}_{\ell}^{4}.
 \end{array}
\end{equation}
In this equivalent class, from expressions in Appendix~\ref{Appendix-Para-Type-II} and figure~\ref{Fig:Angulos-Clase-II-phi-a-Phi-c} we have:
\begin{enumerate}
  \item For the mass matrices ${\bf M}_{\ell}^{1}$ and ${\bf M}_{\ell}^{5}$, the reactor, solar and atmospheric mixing angles have a weak dependence on the phase 
   factors $\phi_{a}$. However, the atmospheric and reactor angles have a weak dependence on the phase factor $\phi_{c}$. 
   On the other hand, to reproduce the current experimental data, 
   at $3\sigma$ eq.(\ref{Ec:exp-mixing-angles}), from the right panels in figure~\ref{Fig:Angulos-Clase-II-phi-a-Phi-c} for the $\theta_{12}$ angle we see that: 
   \begin{equation}\label{Eq:Type-II:val-phi-c}
    \begin{array}{ll}\vspace{2mm} 
     \phi_{c} \, \in \, \left[ -140^{\circ}, 140^{\circ} \right] & \textrm{for} \, {\bf M}_{\ell}^{1} ,\\
     |\phi_{c}| \, \in \, \left[ 38^{\circ}, 180^{\circ} \right] & \textrm{for} \, {\bf M}_{\ell}^{5} .
    \end{array}
   \end{equation}
 \item For the mass matrices ${\bf M}_{\ell}^{2}$ and ${\bf M}_{\ell}^{4}$, the reactor and atmospheric mixing angles do not have an explicit dependence on phase 
  factor $\phi_{c}$. The solar angle has a weak dependence on phase factors $\phi_{a}$ and $\phi_{c}$. While the reactor angle has a weak dependence on phase factor 
  $\phi_{a}$. On the other hand, to reproduce the current experimental data, at $3\sigma$ eq.(\ref{Ec:exp-mixing-angles}), from the right panels in 
  figure~\ref{Fig:Angulos-Clase-II-phi-a-Phi-c} for the $\theta_{23}$ angle we obtain: 
   \begin{equation}\label{Eq:Type-II:val-phi-a}
    \begin{array}{ll}\vspace{2mm} 
     |\phi_{a}| \, \in \, \left[ 73^{\circ}, 180^{\circ} \right] & \textrm{for} \, {\bf M}_{\ell}^{2} \, \textrm{and} \, {\bf M}_{\ell}^{4}.
    \end{array}
   \end{equation}  
\end{enumerate}

In figure~\ref{Fig:Mee:Type-II} we show the allowed regions for the magnitude of the Majorana effective mass $|m_{ee}|$, eq.~(\ref{ec:mee-0}), which were obtained in a 
model-independent context where the neutrino mass matrix has the form given in eq.~(\ref{Eq:M_TBM}), while the charged lepton matrix is represented for an element of 
the equivalent class with two texture zeros type-II. eq.~(\ref{Eq:EQ-Type-I}). Each one of these regions was obtained by considering the values given in 
eqs.~(\ref{Eq:Type-II:val-delta-l})-(\ref{Eq:Type-II:val-phi-a}) for the free parameter $\delta_{\ell}$ and the associated to the CP violation phases 
$\phi_{a}$ and $\phi_{c}$. 
%
\subsection{Equivalent class with two texture zeros type-III}
%
The equivalent class for Hermitian matrices with two texture zeros type-III have the form~\cite{Canales:2012dr}: 
\begin{equation}\label{Eq:EQ-Type-III} 
 \begin{array}{ccc}
  {\bf M}_{\ell}^{0} = 
  \left( \begin{array}{ccc}
    0           & a_{\ell}     & e_{\ell} \\
   a_{\ell}^{*} & 0            & c_{\ell} \\
   e_{\ell}^{*} & c_{\ell}^{*} & d_{\ell}
  \end{array} \right), &
  {\bf M}_{\ell}^{1} = 
  \left( \begin{array}{ccc}
    0           & a_{\ell}^{*}  & c_{\ell} \\
   a_{\ell}     & 0             & e_{\ell} \\
   c_{\ell}^{*} & e_{\ell}^{*}  & d_{\ell}
  \end{array} \right), &
  {\bf M}_{\ell}^{2} = 
  \left( \begin{array}{ccc}
   d_{\ell} & c_{\ell}^{*}  & e_{\ell}^{*}  \\
   c_{\ell} & 0             & a_{\ell}^{*}  \\
   e_{\ell} & a_{\ell}      & 0
  \end{array} \right), \\
  {\bf M}_{\ell}^{3} = 
  \left( \begin{array}{ccc}
   0            & e_{\ell}  & a_{\ell}  \\
   e_{\ell}^{*} & d_{\ell}  & c_{\ell}^{*}  \\
   a_{\ell}^{*} & c_{\ell}  & 0
  \end{array} \right), &
  {\bf M}_{\ell}^{4} = 
  \left( \begin{array}{ccc}
   d_{\ell} & e_{\ell}^{*}  & c_{\ell}^{*}  \\
   e_{\ell} & 0             & a_{\ell}  \\
   c_{\ell} & a_{\ell}^{*}  & 0
  \end{array} \right), &
  {\bf M}_{\ell}^{5} = 
  \left( \begin{array}{ccc}
   0            & c_{\ell} & a_{\ell}^{*}  \\
   c_{\ell}^{*} & d_{\ell} & e_{\ell}^{*}  \\
   a_{\ell}     & e_{\ell} & 0
  \end{array} \right),
 \end{array}
\end{equation}
\begin{figure}[!htbp]
 \begin{center}
  \begin{tabular}{cc} 
  \subfigure{ \includegraphics[width=7cm, height=5cm]{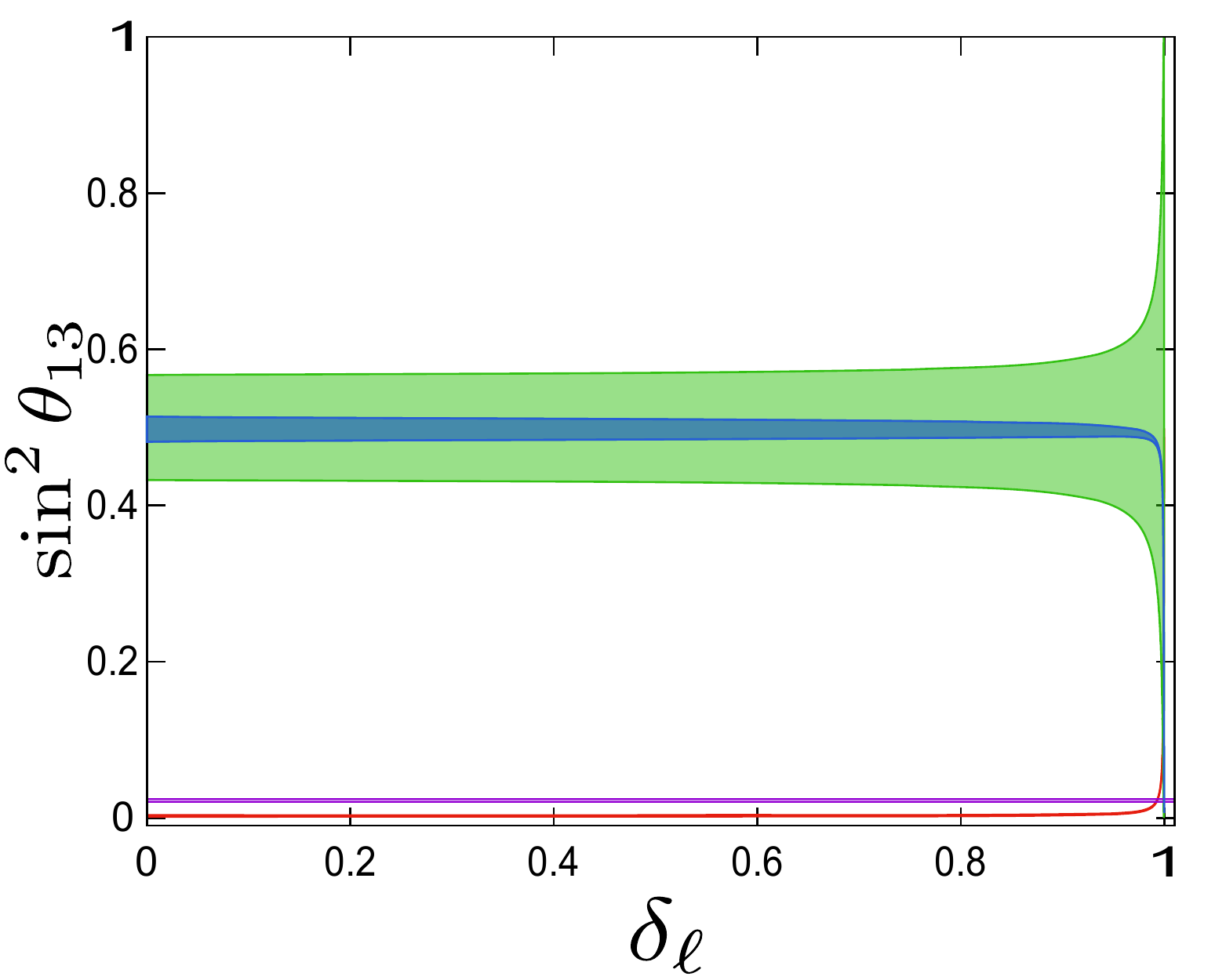} }
  \subfigure{ \includegraphics[width=7cm, height=5cm]{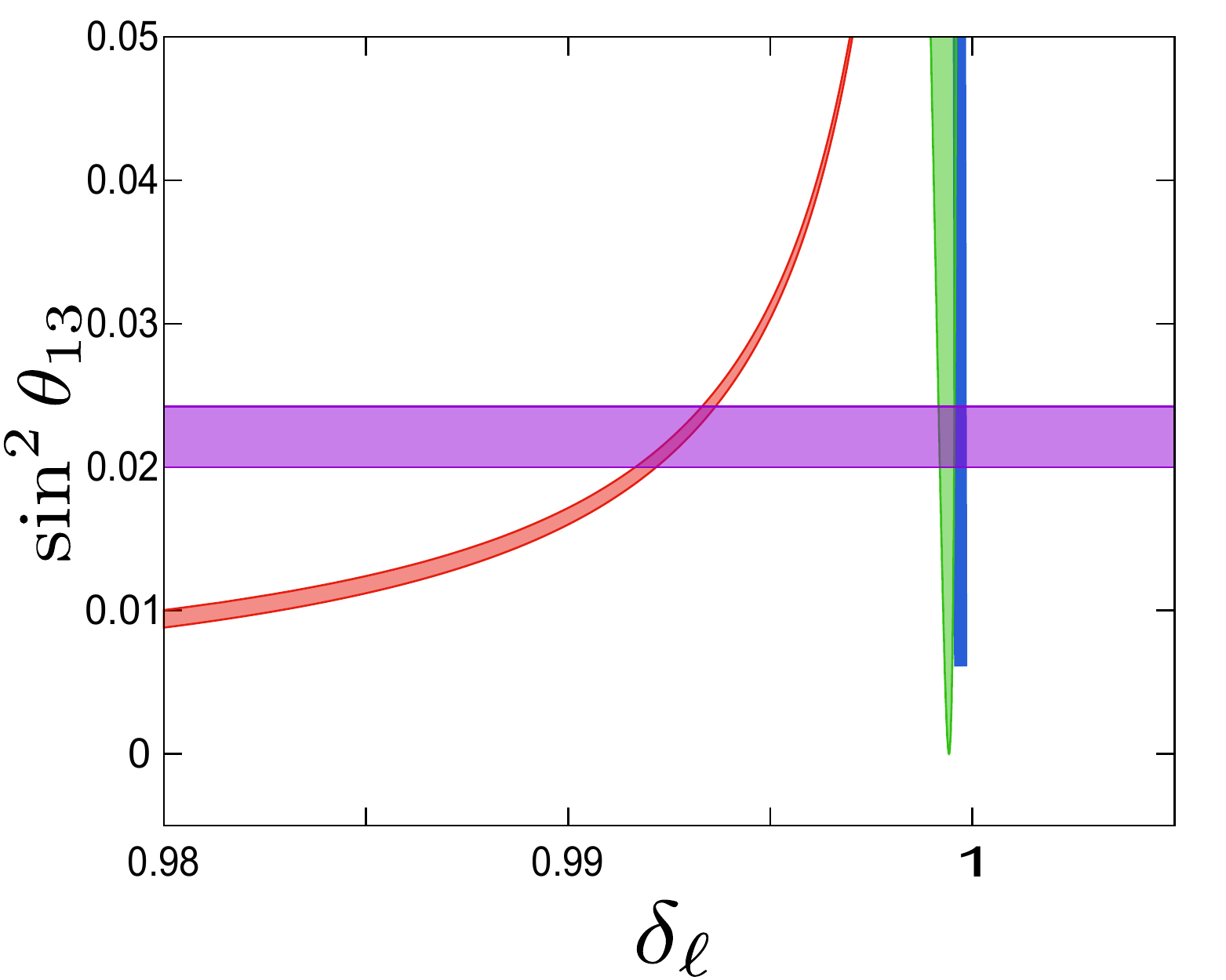} } \\
  \subfigure{ \includegraphics[width=7cm, height=5cm]{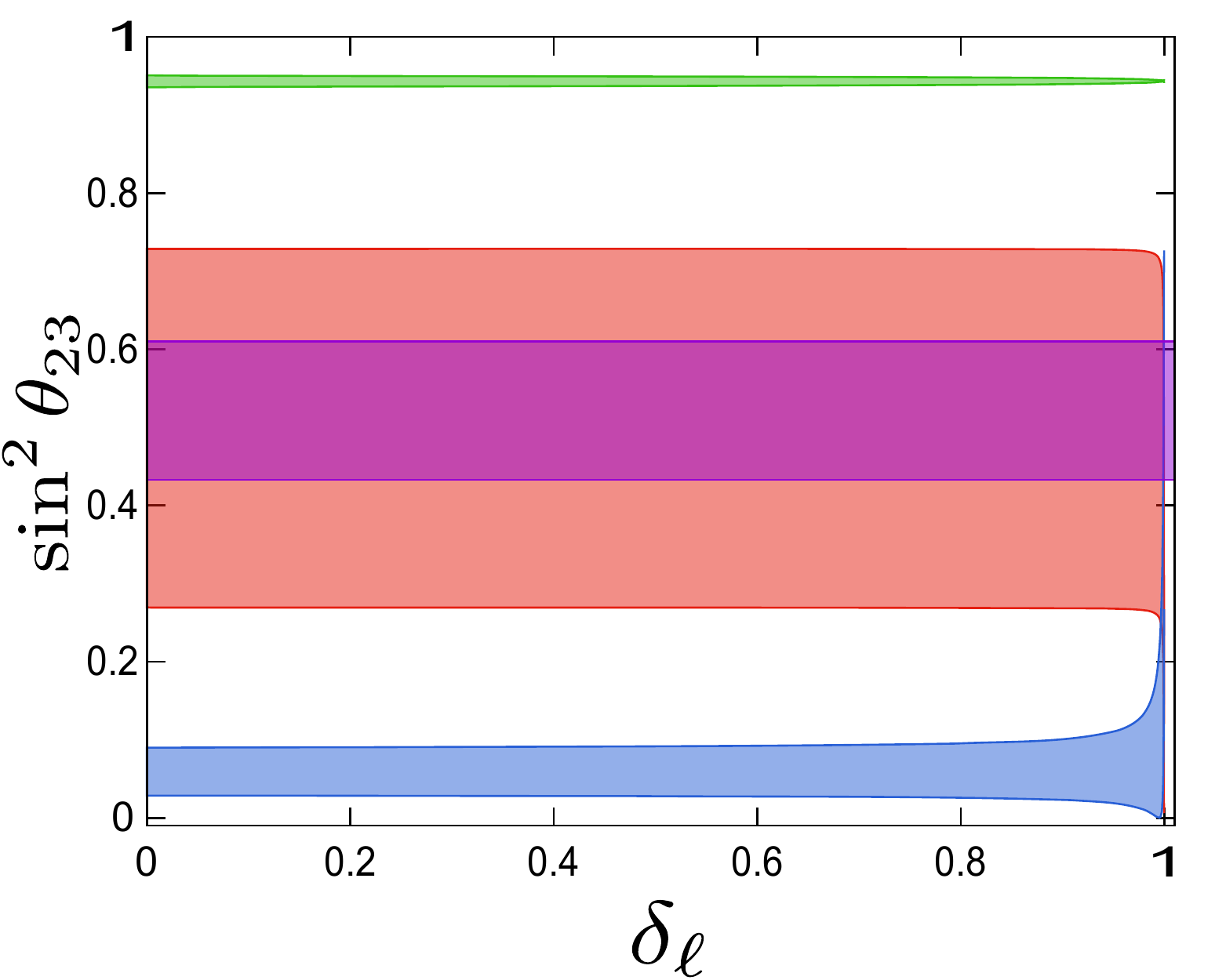} }
  \subfigure{ \includegraphics[width=7cm, height=5cm]{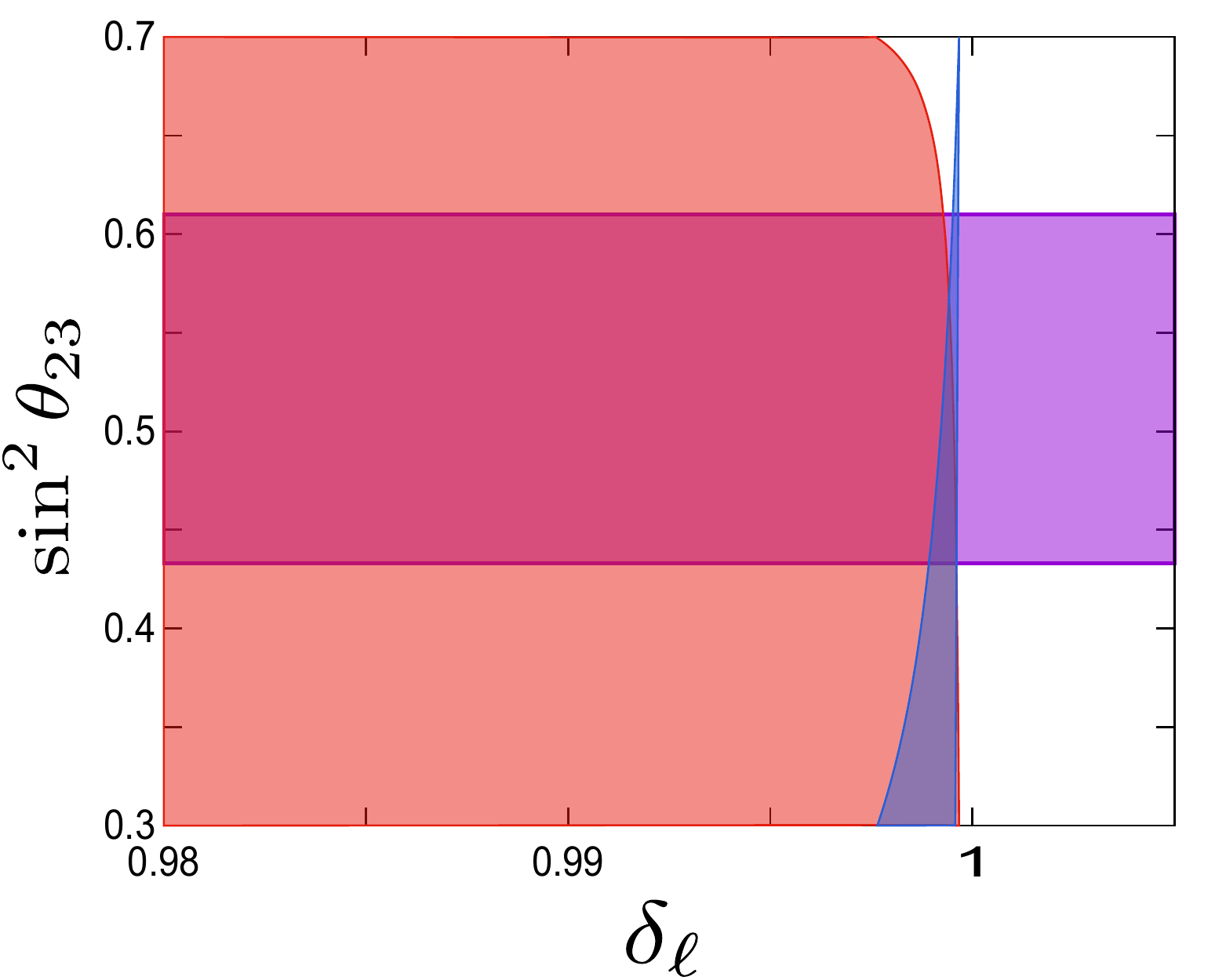} } \\
  \subfigure{ \includegraphics[width=7cm, height=5cm]{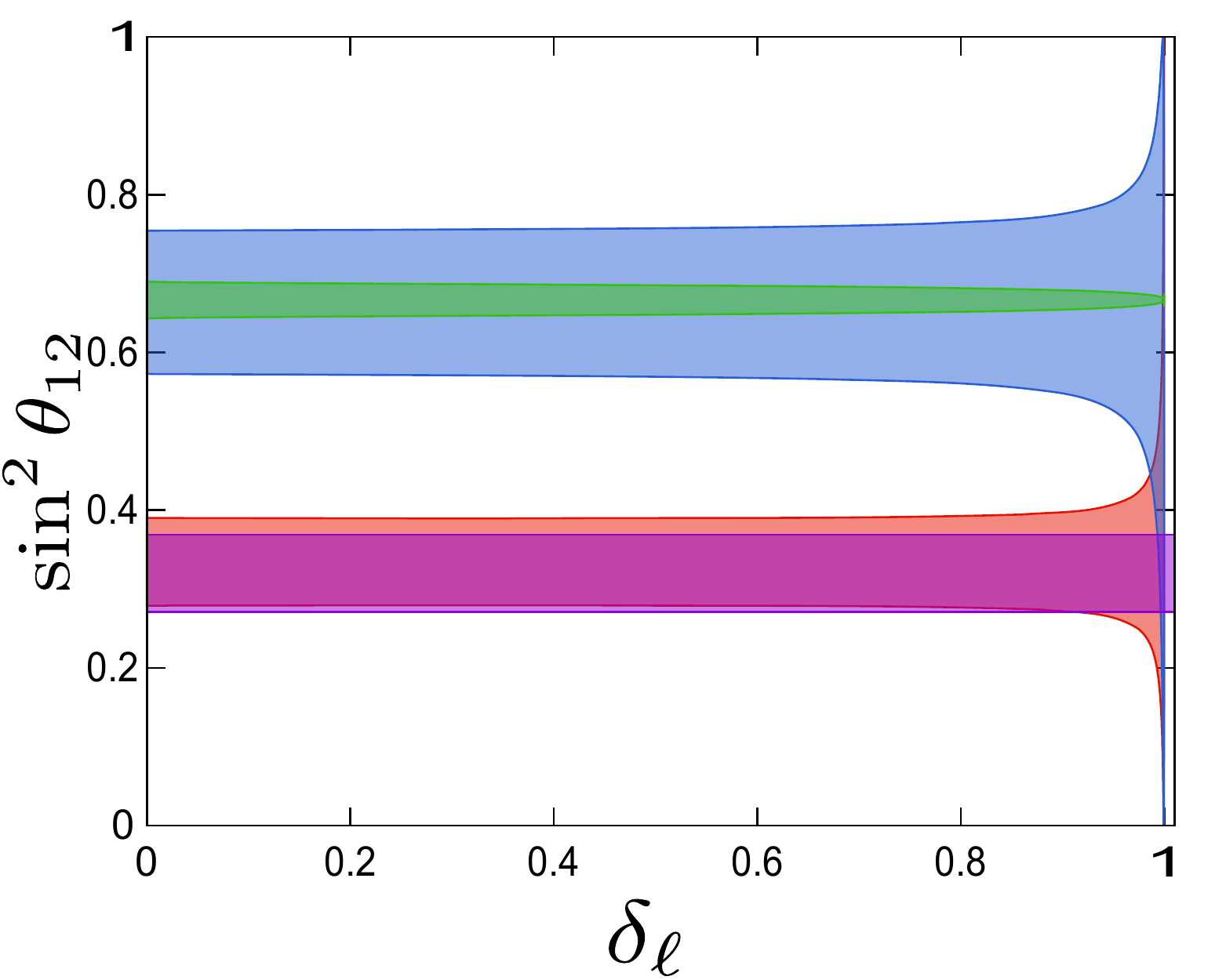} } 
  \subfigure{ \includegraphics[width=7cm, height=5cm]{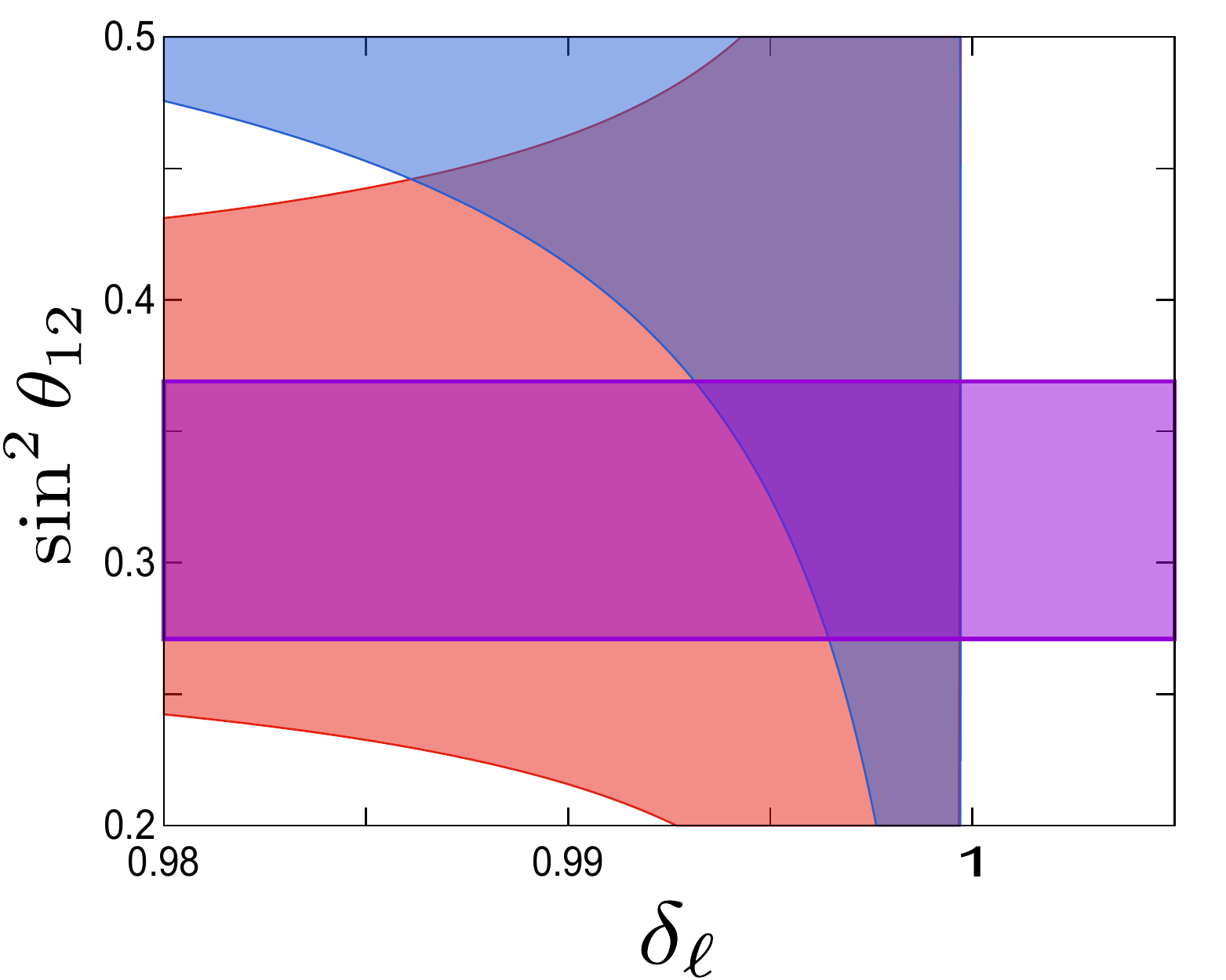} }
  \end{tabular}
  \caption{The allowed regions for the reactor (upper panels), atmospheric (middle panels), and solar (lower panels) mixing angles and free parameter $\delta_{\ell}$. 
   The purple stripe corresponds to the values at 3$\sigma$ for the reactor, atmospheric and solar mixing angles obtained from the global fit, for normal and inverted 
   hierarchy~\cite{global-fit-2021}.
   In these panels, the red area is for ${\bf M}_{\ell}^{0}$ and ${\bf M}_{\ell}^{3}$, blue area is for ${\bf M}_{\ell}^{1}$ and ${\bf M}_{\ell}^{5}$, 
   the green area is for ${\bf M}_{\ell}^{2}$ and ${\bf M}_{\ell}^{4}$. 
   The right panels show a zoom-in of the region where the theoretical expressions for the mixing angles simultaneously reproduce the current experimental data.}
   \label{Fig:Angulos-Clase-IV-delta-l}
 \end{center}
\end{figure}
where  
\begin{equation}
 \begin{array}{l}\vspace{2mm}
  a_{\ell} = \sqrt{ \frac{1}{ 1 + \tan^{2} \beta_{\ell} } \frac{ \widetilde{m}_{e} \widetilde{m}_{\mu} }{ 1 - \delta_{\ell} } } e^{ i \phi_{a} }, \;
  c_{\ell} = \left( \frac{ 1 - \tan^{2} \beta_{\ell} }{ 1 + \tan^{2} \beta_{\ell} } \sqrt{ \frac{ f_{\ell 1} f_{\ell 2} f_{\ell 3} }{ 1 - \delta_{\ell} } }
   +
   \left[ \texttt{s}_{3} - 2 \left( 1 - \delta_{\ell} \right) + \texttt{s}_{1} \widetilde{m}_{e} + \texttt{s}_{2} \widetilde{m}_{\mu} \right]
    \frac{ \tan \beta_{\ell} }{ 1 + \tan^{2} \beta_{\ell} } \right) e^{ i \phi_{c} }, \\ \vspace{2mm}
  e_{\ell} = 
   \sqrt{ 
    \frac{ \tan^{2} \beta_{\ell} }{ 1 + \tan^{2} \beta_{\ell} }
    \frac{ \widetilde{m}_{e} \widetilde{m}_{\mu} 
    }{ 1 - \delta_{\ell}  } } e^{ i \phi_{e} }, \;
  d_{\ell} = 
    \frac{ 2 \tan \beta_{\ell} }{ 1 + \tan^{2} \beta_{\ell} } \sqrt{  \frac{ f_{\ell 1} f_{\ell 2} f_{\ell 3} }{ 1 - \delta_{\ell} } } 
    + \left( \texttt{s}_{3} + \texttt{s}_{1} \widetilde{m}_{e} + \texttt{s}_{2} \widetilde{m}_{\mu} \right) \frac{ \tan^{2} \beta_{\ell} }{ 1 + \tan^{2} \beta_{\ell} } 
    + \left( 1 - \delta_{\ell} \right) \frac{ 1 - \tan^{2} \beta_{\ell} }{ 1 + \tan^{2} \beta_{\ell} } ,  \\ \vspace{2mm}
  \tan \beta_{\ell \pm} = 
   \sqrt{  \frac{ f_{\ell 1} f_{\ell 2} f_{\ell 3} }{ \left( 1 - \delta_{\ell} \right)^{3} } } 
   \left(
    1 \pm 
    \sqrt{
     1 - 
     \frac{ 
      \left( \texttt{s}_{3} - \left( 1 - \delta_{\ell} \right) + \texttt{s}_{1} \widetilde{m}_{e} + \texttt{s}_{2} \widetilde{m}_{\mu} \right)
      \left( 1 - \delta_{\ell} \right)^{2}
     }{  
      f_{\ell 1} f_{\ell 2} f_{\ell 3} 
     }    
    }
   \right) ,
 \end{array}
\label{eq:tanbetal}
\end{equation}
\begin{figure}[!htbp]
 \begin{center}
  \begin{tabular}{cc} 
  \subfigure{ \includegraphics[width=7cm, height=5cm]{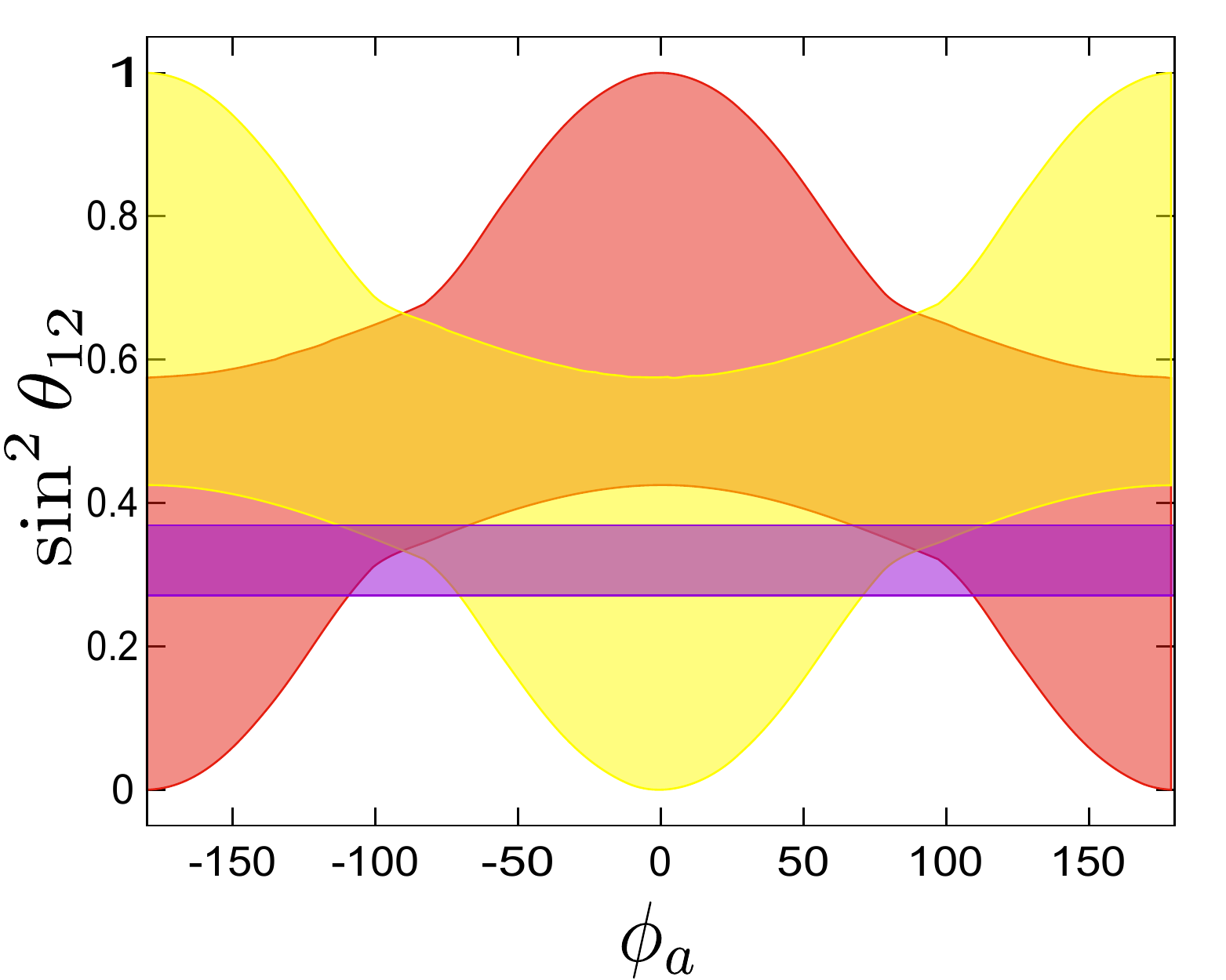} }
  \subfigure{ \includegraphics[width=7cm, height=5cm]{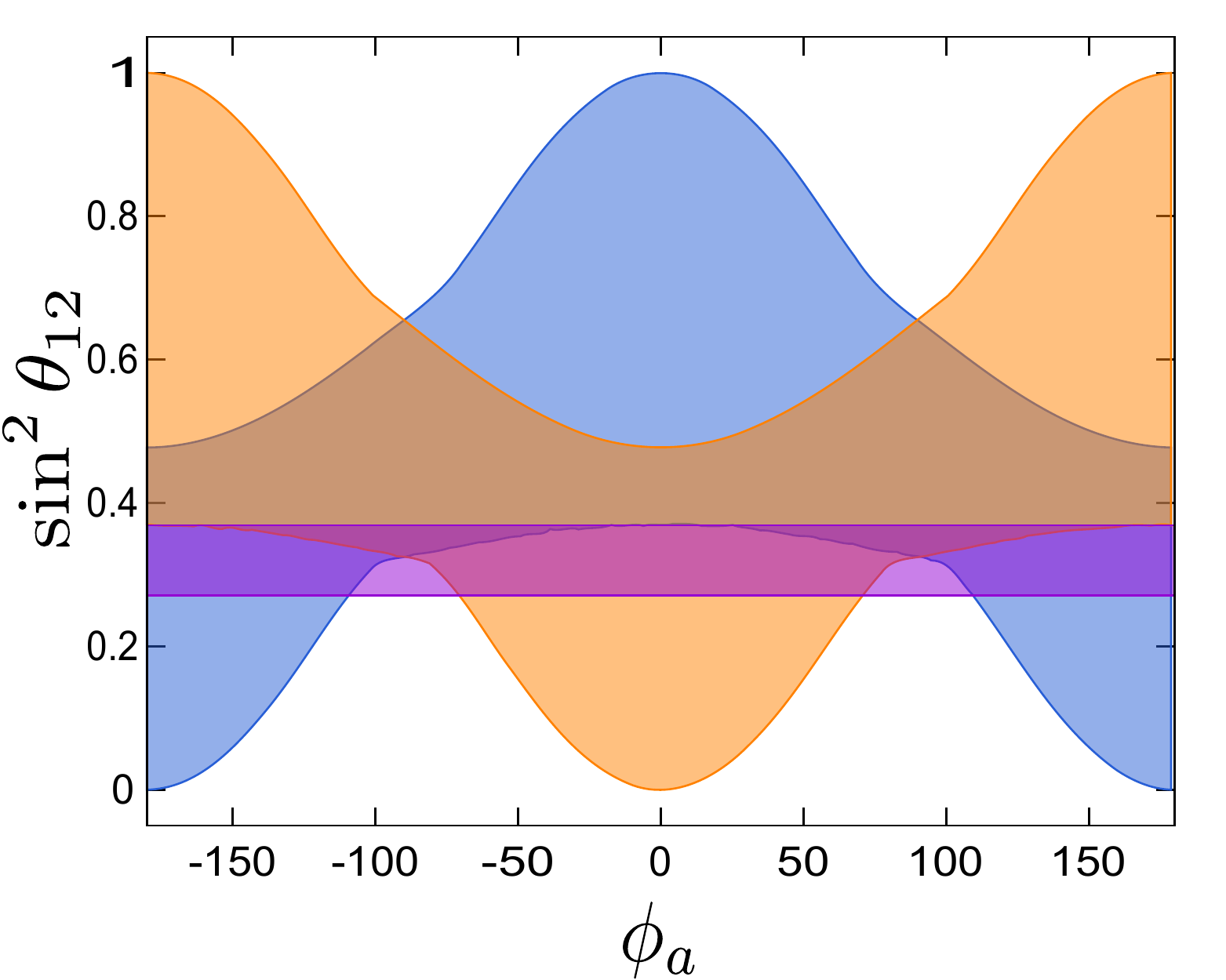} } \\
  \subfigure{ \includegraphics[width=7cm, height=5cm]{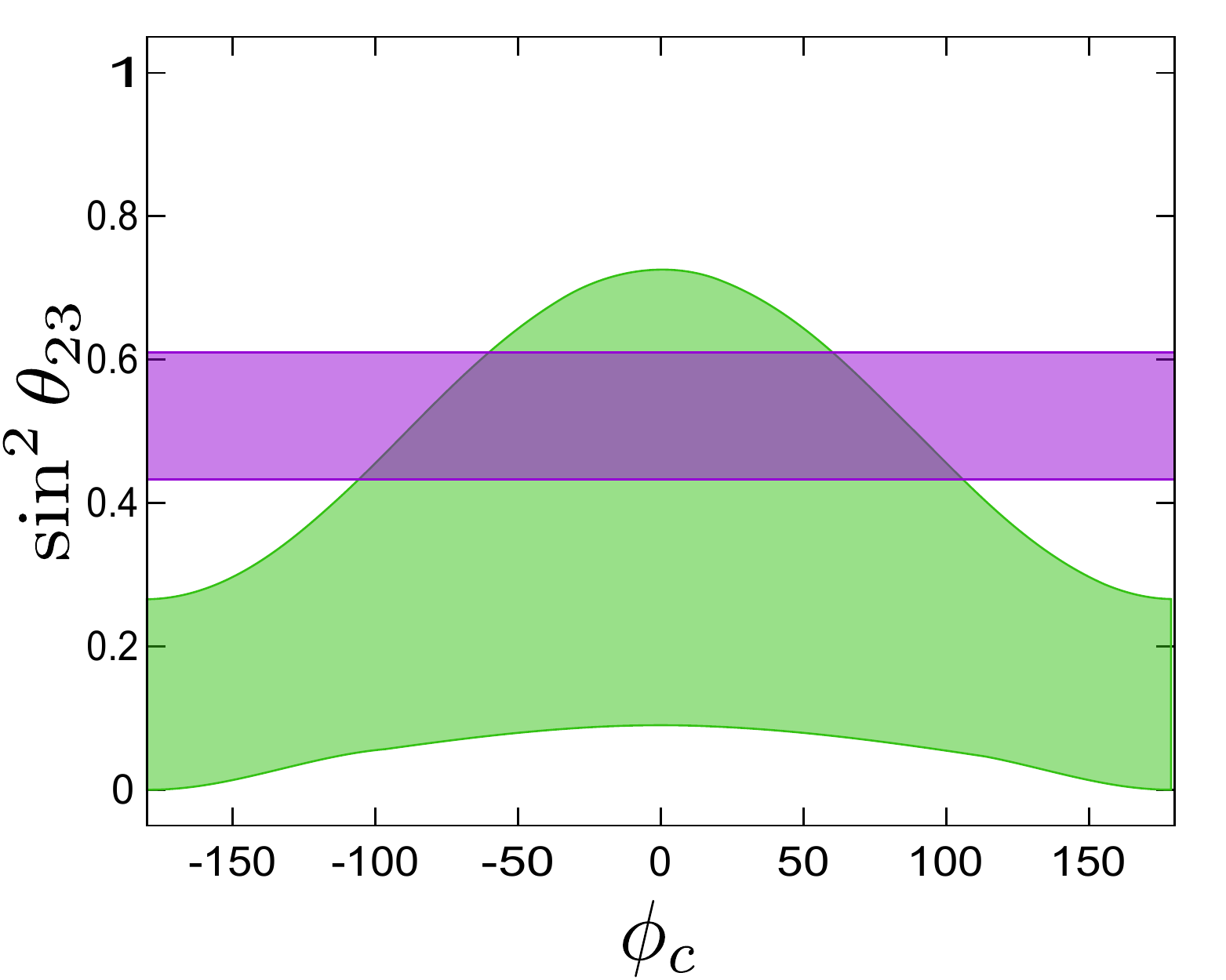} } 
  \end{tabular}
  \caption{The allowed regions for the solar (upper panels) and atmospheric (lower panel) mixing angles and free parameters $\phi_{a}$ and $\phi_{c}$.
   The purple stripe corresponds to the values at 3$\sigma$ for the solar and atmospheric mixing angles obtained from the global fit, for normal and inverted 
   hierarchy~\cite{global-fit-2021}.
   In the upper panels, the red area is for ${\bf M}_{\ell}^{0}$, yellow area is for ${\bf M}_{\ell}^{3}$, blue area is for ${\bf M}_{\ell}^{1}$, 
   orange area is for ${\bf M}_{\ell}^{5}$. In the lower panel, the green area is for ${\bf M}_{\ell}^{0}$ and ${\bf M}_{\ell}^{3}$.
   }\label{Fig:Angulos-Clase-III-phi-a-Phi-c} 
 \end{center}
\end{figure}
with
$f_{\ell 1} = 1 - \texttt{s}_{1} \widetilde{m}_{e} - \delta_{\ell}$, 
$f_{\ell 2} = \texttt{s}_{3} \left( 1 -  \texttt{s}_{2} \widetilde{m}_{\mu} - \delta_{\ell} \right)$, 
$f_{\ell 3} = 1 + \texttt{s}_{3} \left( \delta_{\ell} - 1 \right)$,
$\widetilde{m}_{e} = \frac{ m_{e} }{ m_{\tau} }$, $\widetilde{m}_{\mu} = \frac{ m_{\mu} }{ m_{\tau} }$, 
$\phi_{a} = \arg \left \{ a_{\ell} \right \}$, $\phi_{c} = \arg \left \{ c_{\ell} \right \}$, and $\phi_{e} = \arg \left \{ e_{\ell} \right \}$. 
In this case, the diagonal matrix of phase factors is
${\bf P}_{\ell} = \mathrm{diag} \left( 1,  e^{ i \phi_{a} } ,e^{ i \left( \phi_{a} + \phi_{c} \right) }  \right) $, while the phase factors satisfy the relation 
$\phi_{e} = \phi_{a} + \phi_{c}$. The real orthogonal matrix ${\bf O}_{\ell}$ is constructed with the help of the general eigenvectors given in 
eq.~(\ref{Eq:General_Eigenvector}), which are the  eigenvectors of the charged lepton mass matrix. The explicit form of  
${\bf O}_{\ell} = \left( | M_{1} \rangle, | M_{2} \rangle, | M_{3} \rangle \right)$  is
\begin{equation}
 {\bf O}_{\ell} = {\bf R}^{\top} {\cal O}_{\ell}
\end{equation}
where
\begin{equation}
 \begin{array}{ll}
  {\bf R} = 
  \left( \begin{array}{ccc}\vspace{2mm}
   1 & 0                                                                 &       0 \\ \vspace{2mm}
   0 &  \frac{ 1 }{ \sqrt{ 1 + \tan^{2} \beta_{\ell} } }                 & \frac{ \tan \beta_{\ell} }{ \sqrt{ 1 + \tan^{2} \beta_{\ell} } }  \\ \vspace{2mm}
   0 & -\frac{ \tan \beta_{\ell} }{ \sqrt{ 1 + \tan^{2} \beta_{\ell} } } &  \frac{ 1 }{ \sqrt{ 1 + \tan^{2} \beta_{\ell} } }
  \end{array} \right) 
  \quad \textrm{and} \quad
  {\cal O}_{\ell} = 
  \left( \begin{array}{ccc}\vspace{2mm}
   \texttt{s}_{1} \sqrt{ \frac{ \widetilde{m}_{\mu} f_{\ell 1} }{ D_{\ell 1} } } &
   \texttt{s}_{2} \sqrt{ \frac{ \widetilde{m}_{e}   f_{\ell 2} }{ D_{\ell 2} } } &
   \texttt{s}_{3} \sqrt{ \frac{ \widetilde{m}_{e} \widetilde{m}_{\mu} f_{\ell 3} }{ D_{\ell 3} } } \\ \vspace{2mm}
   \sqrt{ \frac{ \widetilde{m}_{e}   \left( 1 - \delta_{\ell} \right) f_{\ell 1} }{ D_{\ell 1} } } &
   \sqrt{ \frac{ \widetilde{m}_{\mu} \left( 1 - \delta_{\ell} \right) f_{\ell 2} }{ D_{\ell 2} } } &
   \sqrt{ \frac{ \left( 1 - \delta_{\ell} \right) f_{\ell 3} }{ D_{\ell 3} } } \\ \vspace{2mm}
  -\sqrt{ \frac{ \widetilde{m}_{e} f_{\ell 2} f_{\ell 3} }{ D_{\ell 1} } } &
   \texttt{s}_{1} \texttt{s}_{2} 
   \sqrt{ \frac{ \widetilde{m}_{\mu} f_{\ell 1} f_{\ell 3} }{ D_{\ell 2} } } &
   \texttt{s}_{3} \sqrt{ \frac{ f_{\ell 1} f_{\ell 2} }{ D_{\ell 3} } }
  \end{array} \right),
 \end{array}
\end{equation}
with 
 \begin{equation}
  \begin{array}{l} \vspace{2mm}
   D_{\ell 1} =       
    \left( 1 - \delta_{\ell} \right) \left( \widetilde{m}_{\mu} + \texttt{s}_{3} \widetilde{m}_{e} \right) \left( 1 + \texttt{s}_{2} \widetilde{m}_{e} \right), \quad
   D_{\ell 2} =       
    \left( 1 - \delta_{\ell} \right) \left( \widetilde{m}_{\mu} + \texttt{s}_{3} \widetilde{m}_{e} \right) \left( 1 + \texttt{s}_{1} \widetilde{m}_{\mu} \right), 
    \\ \vspace{2mm}  
   D_{\ell 3} =       
    \left( 1 - \delta_{\ell} \right) \left( 1 + \texttt{s}_{2} \widetilde{m}_{e} \right) \left( 1 + \texttt{s}_{1} \widetilde{m}_{\mu} \right) .  
  \end{array}
 \end{equation}  
\begin{figure}[!htbp]
 \begin{center}
  \begin{tabular}{cc}
  \subfigure{ \includegraphics[width=8.5cm, height=7cm]{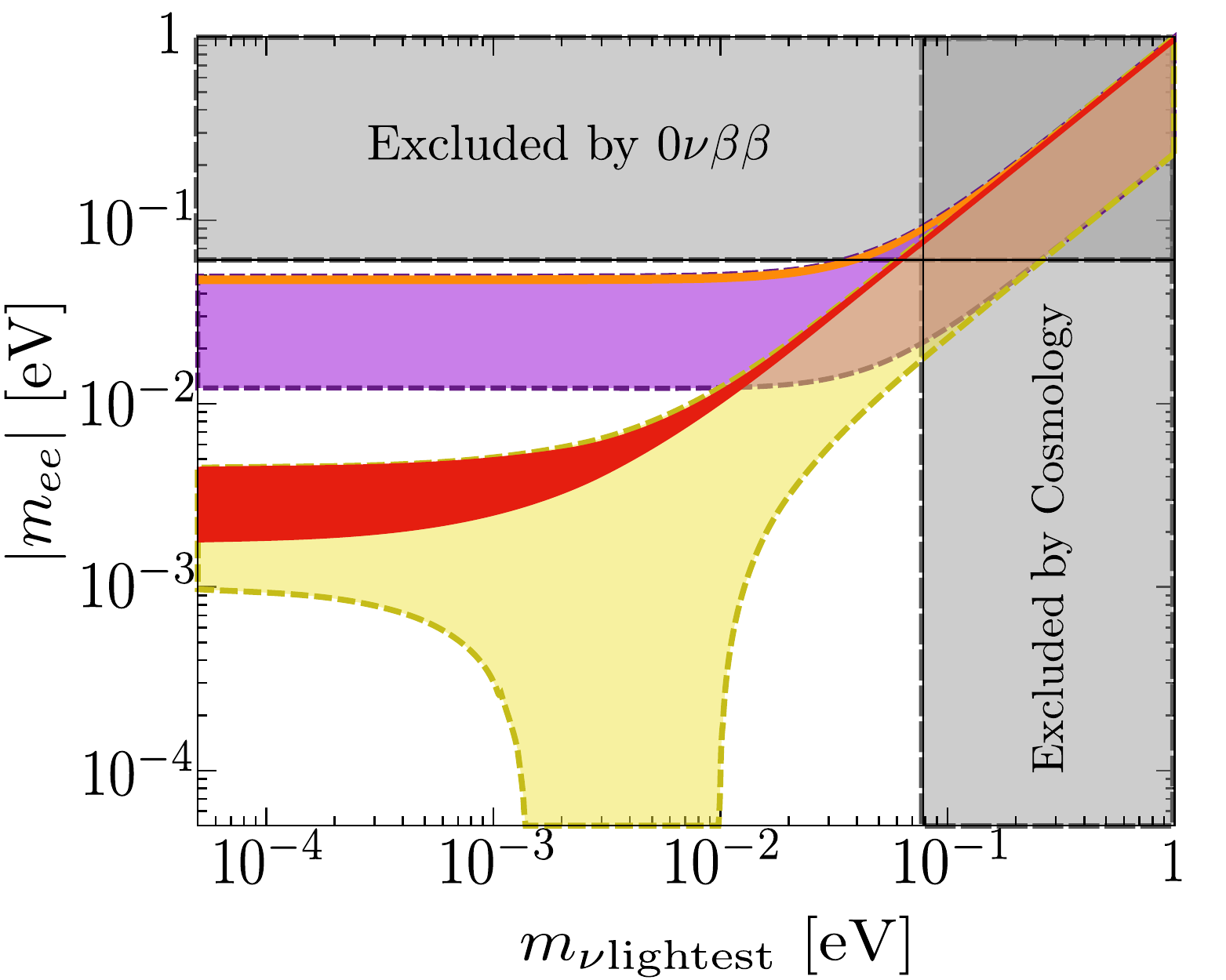}}
  \subfigure{ \includegraphics[width=8.5cm, height=7cm]{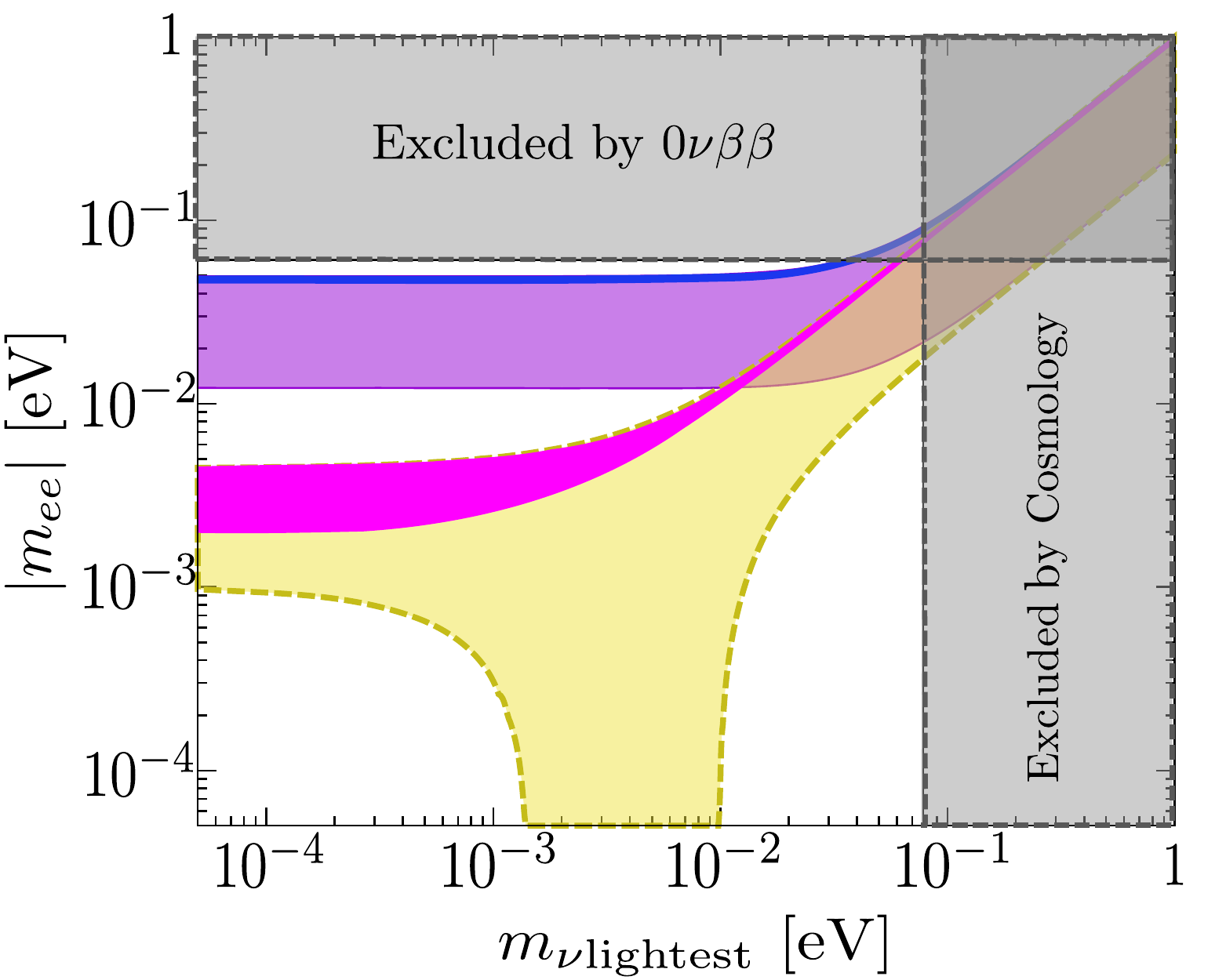}}
  \end{tabular}
  \caption{ These panels  show the allowed regions for the magnitude of Majorana effective mass $|m_{ee}|$.
   Respectively, for an inverted and normal neutrino mass hierarchy, the yellow and purple stripes are obtained from the current experimental data on neutrino 
   oscillations at 3$\sigma$~\cite{global-fit-2021}. 
   In the upper right panel, the magenta area is for a normal hierarchy while the blue area is for an inverted hierarchy, 
   both areas are obtained from ${\bf M}_{\ell}^{1}$ and ${\bf M}_{\ell}^{5}$. 
   In the upper left panel, the red area is for a normal hierarchy while the orange area is for an inverted hierarchy, both areas are obtained from  
   ${\bf M}_{\ell}^{2}$ and ${\bf M}_{\ell}^{4}$.
   From KamLAND-ZEN~\cite{KamLAND-Zen:2016pfg} and EXO-200~\cite{EXO:2017poz} we have the following upper limit $|m_{ee}| < 0.061$, which correspond to the 
   horizontal grey band, whereas vertical grey band corresponds to results reported by Planck collaboration~\cite{Planck:2015fie}. }\label{Fig:Mee:Type-III}
 \end{center}
\end{figure}
The parameter $\delta_{\ell}$ must satisfy the conditions 
\begin{equation}\label{eq:conditionsdelta-type-III}
 \begin{array}{ll}
  \textrm{\bf A.} \quad
  0 < \delta_{\ell} < 1 - \widetilde{m}_{\mu}, & \mathrm{for } \qquad m_{e} = - m_{e} \quad 
   \left( \texttt{s}_{1} = -1, \; \texttt{s}_{2} = +1, \; \texttt{s}_{3} = +1 \right). \\
  \textrm{\bf B.} \quad
  0 < \delta_{\ell} < 1 - \widetilde{m}_{e},\quad \delta_{\ell} \neq \widetilde{m}_{\mu} - \widetilde{m}_{e}, & \mathrm{for } \qquad m_{\mu} = - m_{\mu} \quad 
   \left( \texttt{s}_{1} = +1, \; \texttt{s}_{2} = -1, \; \texttt{s}_{3} = +1 \right). \\ 
  \textrm{\bf C.} \quad
  1 - \widetilde{m}_{\mu} < \delta_{\ell} < 1 - \widetilde{m}_{e}, & \mathrm{for } \qquad  m_{\tau} = - m_{\tau} \quad 
   \left( \texttt{s}_{1} = +1, \; \texttt{s}_{2} = +1, \; \texttt{s}_{3} = -1 \right).
 \end{array}
\end{equation}
In this case, the flavor mixing angles in eq.~(\ref{Eq:Mix-Angle-0}) have the form:
\begin{equation}
 \begin{array}{l}
  \sin^{2} \theta_{12} = 
   \frac{1}{3} \frac{ \widetilde{m}_{e} }{ \widetilde{m}_{\mu} } \varepsilon_{12} , \qquad
  \sin^{2} \theta_{23} = 
   \frac{1}{2} 
   \frac{
    \left( 1 + \texttt{s}_{2} \widetilde{m}_{e}   \right)
   }{
    \left( 1 + \texttt{s}_{1} \widetilde{m}_{\mu} \right) 
   } \varepsilon_{23}, \qquad
  \sin^{2} \theta_{13} = 
   \frac{1}{2} \frac{ \widetilde{m}_{e} }{ \widetilde{m}_{\mu} } \varepsilon_{13}.
 \end{array}
\end{equation}
The explicit form of the $\varepsilon_{ij}$ parameters is given in the Appendix~\ref{Appendix-Para-Type-III}. 
From the allowed regions of flavor mixing angles shown in figure~\ref{Fig:Angulos-Clase-IV-delta-l}, 
and computed taken into account the condition $\textrm{\bf B}$~\eqref{eq:conditionsdelta-type-III} and for $\tan \beta_{\ell +}$ \eqref{eq:tanbetal}, 
we obtain that in this equivalent class, the charged lepton mass 
matrices ${\bf M}_{\ell}^{2}$ and  ${\bf M}_{\ell}^{4}$ cannot correctly reproduce current experimental data of the solar and atmospheric mixing angles. 
So to reproduce the values for the leptonic flavor mixing angles, at $3\sigma$ obtained form the global fit eq.~(\ref{Ec:exp-mixing-angles}), for a normal and inverted 
hierarchy,  the free parameter $\delta_{\ell}$ should be in the following numerical interval:
\begin{equation}\label{Eq:Type-III:val-delta-l}
 \begin{array}{ll}\vspace{2mm} 
  \delta_{\ell} \, \in \, \left[0.9916, 0.9936 \right] & \textrm{for} \, {\bf M}_{\ell}^{0} \, \textrm{and} \, {\bf M}_{\ell}^{3},\\
  \delta_{\ell} \, \approx 0.9997 & \textrm{for} \, {\bf M}_{\ell}^{1} \, \textrm{and} \, {\bf M}_{\ell}^{5}.
 \end{array}
\end{equation}
In this equivalent class, from expressions in Appendix~\ref{Appendix-Para-Type-III} and figure~\ref{Fig:Angulos-Clase-III-phi-a-Phi-c} we have:
\begin{enumerate}
 \item For the mass matrices ${\bf M}_{\ell}^{0}$ and ${\bf M}_{\ell}^{3}$, the reactor and atmospheric mixing angles do not have an explicit dependence on phase factor 
  $\phi_{a}$. While the solar and reactor mixing angles have a weak dependence on phase factors $\phi_{c}$. On the other hand, to reproduce the current experimental 
  data, at $3\sigma$ eq.(\ref{Ec:exp-mixing-angles}), from the  upper left and lower panels in figure~\ref{Fig:Angulos-Clase-III-phi-a-Phi-c} 
  for the $\theta_{12}$ and $\theta_{23}$ angles we obtain that:
 
   \begin{equation}\label{Eq:Type-III:val-phi-c}
    \begin{array}{ll}\vspace{2mm} 
     | \phi_{a} | \, \in \, \left[ 67^{\circ}, 180^{\circ} \right] & \textrm{for} \, \, {\bf M}_{\ell}^{0} ,\\ 
     \phi_{a} \, \in \, \left[ -113^{\circ}, 113^{\circ} \right] & \textrm{for} \, \, {\bf M}_{\ell}^{3} ,\\
     \phi_{c} \, \in \, \left[ -106^{\circ}, 106^{\circ} \right] & \textrm{for} \,\, {\bf M}_{\ell}^{0} \, \textrm{and} \, {\bf M}_{\ell}^{3}.
    \end{array}
   \end{equation}
 \item For the mass matrices ${\bf M}_{\ell}^{1}$ and ${\bf M}_{\ell}^{5}$, the reactor and atmospheric mixing angles have a weak dependence on phase factors 
  $\phi_{a}$ and $\phi_{c}$. While the solar mixing angle has a weak dependence on phase factor  $\phi_{c}$. On the other hand, to reproduce the current experimental 
  data, at $3\sigma$ eq.(\ref{Ec:exp-mixing-angles}), from the  upper right panel in figure~\ref{Fig:Angulos-Clase-III-phi-a-Phi-c} for the $\theta_{12}$ angle we 
  obtain that:
   \begin{equation}\label{Eq:Type-III:val-phi-a}
    \begin{array}{ll}\vspace{2mm} 
     | \phi_{a} | \, \in \, \left[ 10^{\circ}, 180^{\circ} \right] & \textrm{for}\, \, {\bf M}_{\ell}^{1} ,\\ 
     \phi_{a} \, \in \, \left[ -170^{\circ}, 170^{\circ} \right] & \textrm{for} \,\, {\bf M}_{\ell}^{5} .
    \end{array}
   \end{equation}
\end{enumerate}

In figure~\ref{Fig:Mee:Type-III} we show the allowed regions for the magnitude of the Majorana effective mass $|m_{ee}|$, ec.~(\ref{ec:mee-0}), which were obtained in a 
model-independent context where the neutrino mass matrix has the form given in eq.~(\ref{Eq:M_TBM}), while the charged lepton matrix is represented for an element of 
the equivalent class with two texture zeros type-III. eq.~(\ref{Eq:EQ-Type-III}). Each one of these regions was obtained by considering the values given in 
eqs.~(\ref{Eq:Type-III:val-delta-l})-(\ref{Eq:Type-III:val-phi-a}) for the free parameter $\delta_{\ell}$ and the associated to the CP violation phases 
$\phi_{a}$ and $\phi_{c}$.
%
%
	\begin{figure}[t]
	\begin{center}
		\begin{tabular}{cc}
			\subfigure{ \includegraphics[width=7cm, height=5cm]{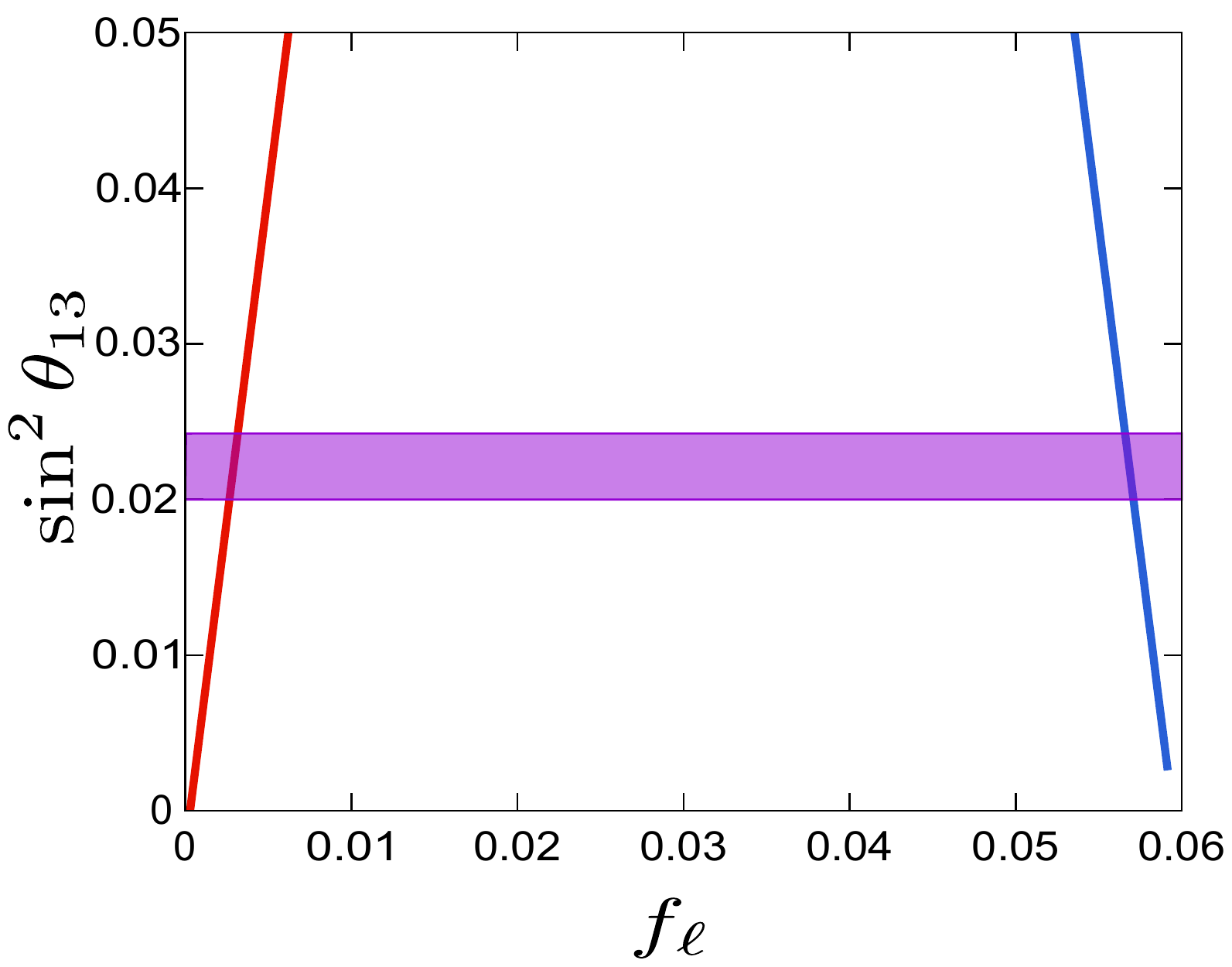}} 
			\subfigure{ \includegraphics[width=7cm, height=5cm]{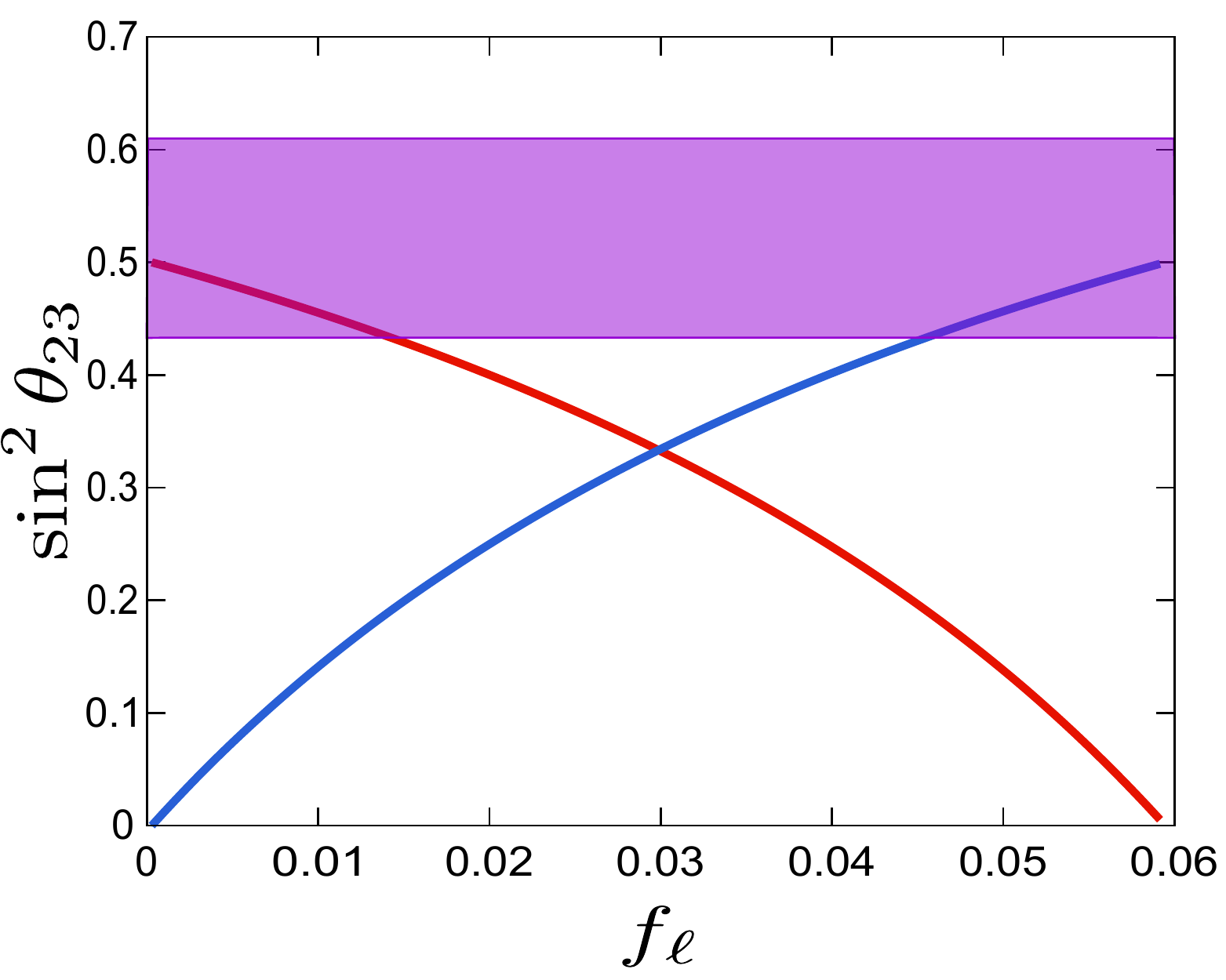}}  \\ 
			\subfigure{ \includegraphics[width=7cm, height=5cm]{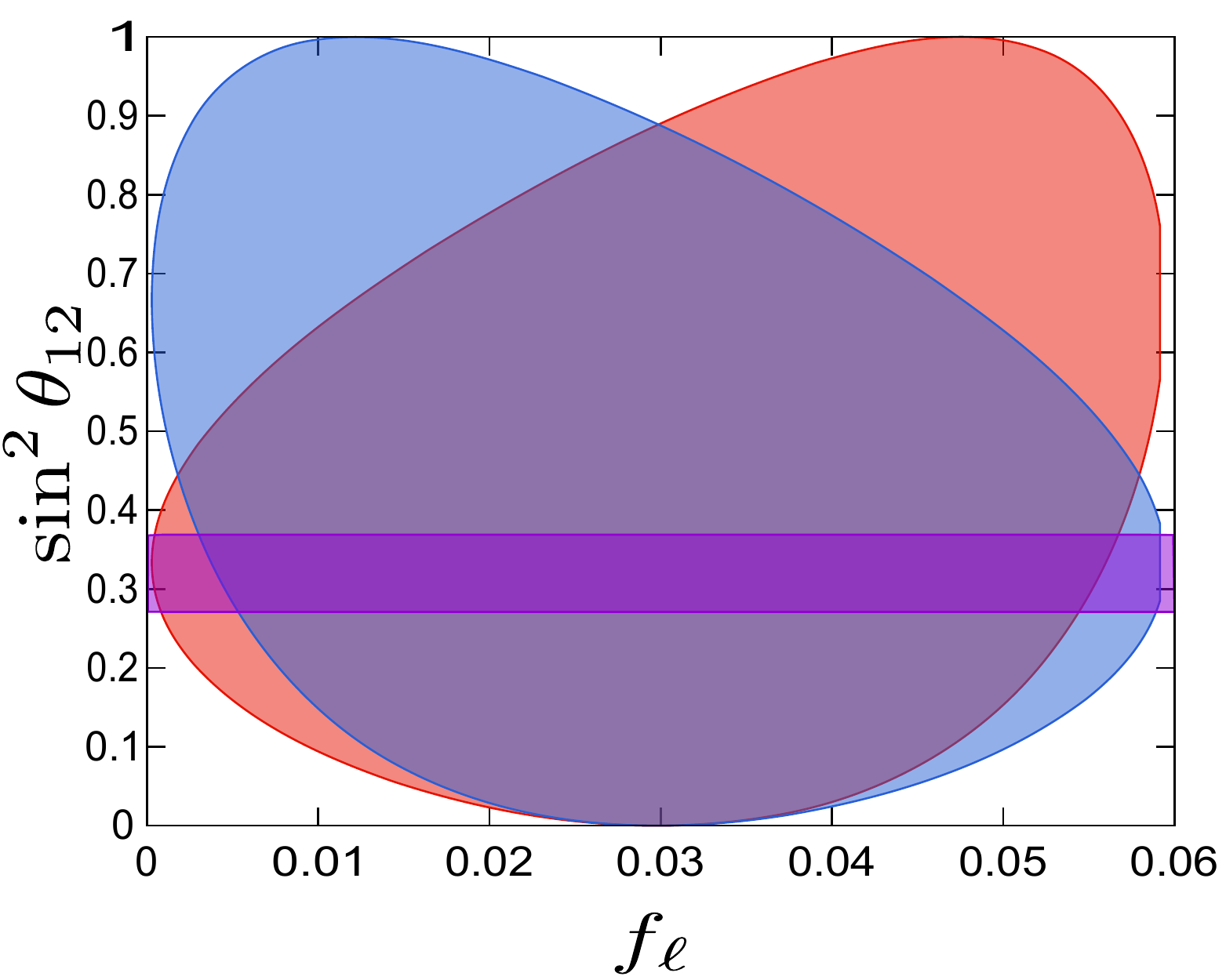}}
		\end{tabular}
		\caption{The allowed regions for the reactor, atmospheric (upper panels), and solar (lower panels) mixing angles and free parameter $f_{\ell}$. 
			The purple stripe corresponds to the values at 3$\sigma$ for the reactor, atmospheric and solar mixing angles obtained from the global fit, for normal and inverted 
			hierarchy~\cite{global-fit-2021}.
			In these panels, the red area is for ${\bf M}_{\ell}^{0}$ and ${\bf M}_{\ell}^{3}$, and blue area is for ${\bf M}_{\ell}^{1}$ and ${\bf M}_{\ell}^{5}$.
		}\label{Fig:Angulos-Clase-IV-fl}
	\end{center}
\end{figure}
%
\subsection{Equivalent class with two texture zeros type-IV}
%
The equivalent class for Hermitian matrices with two texture zeros type-IV have the form~\cite{Canales:2012dr}: 
\begin{equation}\label{Eq:EQ-Type-IV}
 \begin{array}{ccc}
  {\bf M}_{\ell}^{0} = 
  \left( \begin{array}{ccc}
   f_{\ell}     & a_{\ell}  & 0        \\
   a_{\ell}^{*} & b_{\ell}  & 0        \\
   0            & 0         & d_{\ell}
  \end{array} \right), &
  {\bf M}_{\ell}^{1} = 
  \left( \begin{array}{ccc}
   b_{\ell}     & a_{\ell}  & 0        \\
   a_{\ell}^{*} & f_{\ell}  & 0        \\
   0            & 0         & d_{\ell}
  \end{array} \right), &
  {\bf M}_{\ell}^{2} = 
  \left( \begin{array}{ccc}
   d_{\ell} & 0         & 0             \\
   0        & b_{\ell}  & a_{\ell}^{*}  \\
   0        & a_{\ell}  & f_{\ell}
  \end{array} \right), \\
  {\bf M}_{\ell}^{3} = 
  \left( \begin{array}{ccc}
   f_{\ell}      & 0         & a_{\ell} \\
   0             & d_{\ell}  & 0        \\
   a_{\ell}^{*}  & 0         & b_{\ell}
  \end{array} \right), &
  {\bf M}_{\ell}^{4} = 
  \left( \begin{array}{ccc}
   d_{\ell} & 0             & 0  \\
   0        & f_{\ell}      & a_{\ell}  \\
   0        & a_{\ell}^{*}  & b_{\ell}
  \end{array} \right), &
  {\bf M}_{\ell}^{5} = 
  \left( \begin{array}{ccc}
   b_{\ell}     & 0        & a_{\ell}^{*}  \\
   0            & d_{\ell} & 0             \\
   a_{\ell}     & 0        & f_{\ell}
  \end{array} \right),
 \end{array}
\end{equation}
where  $d_{\ell} = 1$
\begin{equation}
 \begin{array}{l}
  a_{\ell} = \sqrt{ \left( \widetilde{f}_{\ell} - \widetilde{m}_{e} \right) \left( \widetilde{m}_{\mu} - \widetilde{f}_{\ell} \right) } 
   \, e^{i \phi_{a} }, \quad
  b_{\ell} = \widetilde{m}_{e} + \widetilde{m}_{\mu} - \widetilde{f}_{\ell},
 \end{array}
\end{equation}
where $\widetilde{m}_{e} = \frac{ m_{e} }{ m_{\tau} }$, $\widetilde{m}_{\mu} = \frac{ m_{\mu} }{ m_{\tau} }$, 
$\widetilde{f}_{\ell} = \frac{ f_{\ell} }{ m_{\tau} }$, and 
$\phi_{a} = \arg \left \{ a_{\ell} \right \}$. 
In this case, the diagonal matrix of phase factors is
${\bf P}_{\ell} = \mathrm{diag} \left( 1,  e^{ i \phi_{a} } ,e^{ i \phi_{a} }  \right) $. 
The real orthogonal matrix ${\bf O}_{\ell}$ is
\begin{equation}
 \begin{array}{l}\vspace{2mm}
  {\bf O}_{\ell} =
  \left( \begin{array}{ccc} \vspace{2mm}
  \sqrt{ \frac{ \widetilde{m}_{\mu} - \widetilde{f}_{\ell} }{ \widetilde{m}_{\mu} - \widetilde{m}_{e} } } &
  \sqrt{ \frac{ \widetilde{f}_{\ell} - \widetilde{m}_{e} }{ \widetilde{m}_{\mu} - \widetilde{m}_{e} } }   &
  0 \\ \vspace{2mm}
  - \sqrt{ \frac{ \widetilde{f}_{\ell} - \widetilde{m}_{e} }{ \widetilde{m}_{\mu} - \widetilde{m}_{e} } }  &
  \sqrt{ \frac{ \widetilde{m}_{\mu} - \widetilde{f}_{\ell} }{ \widetilde{m}_{\mu} - \widetilde{m}_{e} } } & 
  0 \\ 
  0 & 0 & 1
  \end{array} \right).
 \end{array}
\end{equation}
%
%
\begin{figure}[t]
	\begin{center}
		\begin{tabular}{cc}
			\subfigure{ \includegraphics[width=8.5cm, height=7cm]{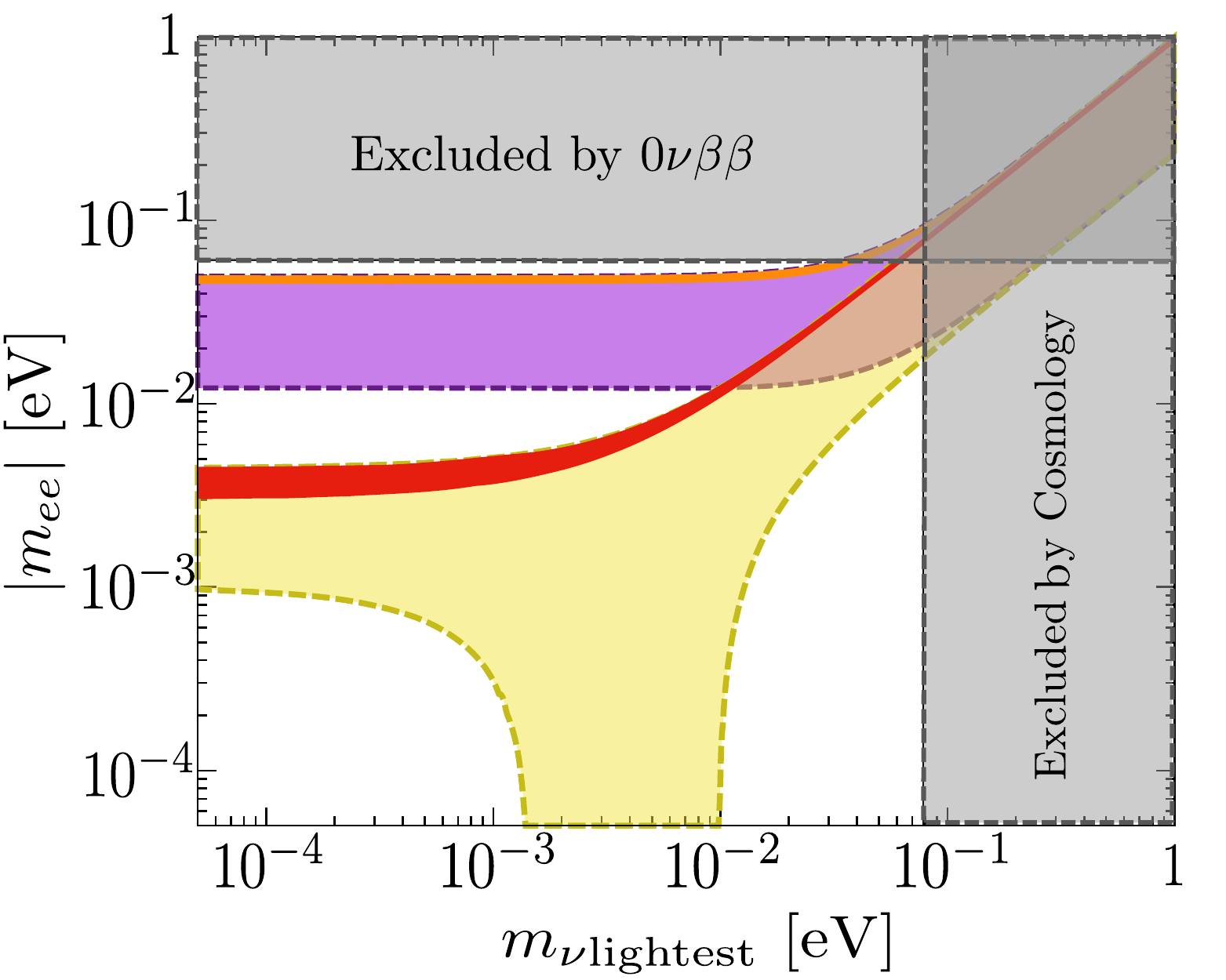}}
			\subfigure{ \includegraphics[width=8.5cm, height=7cm]{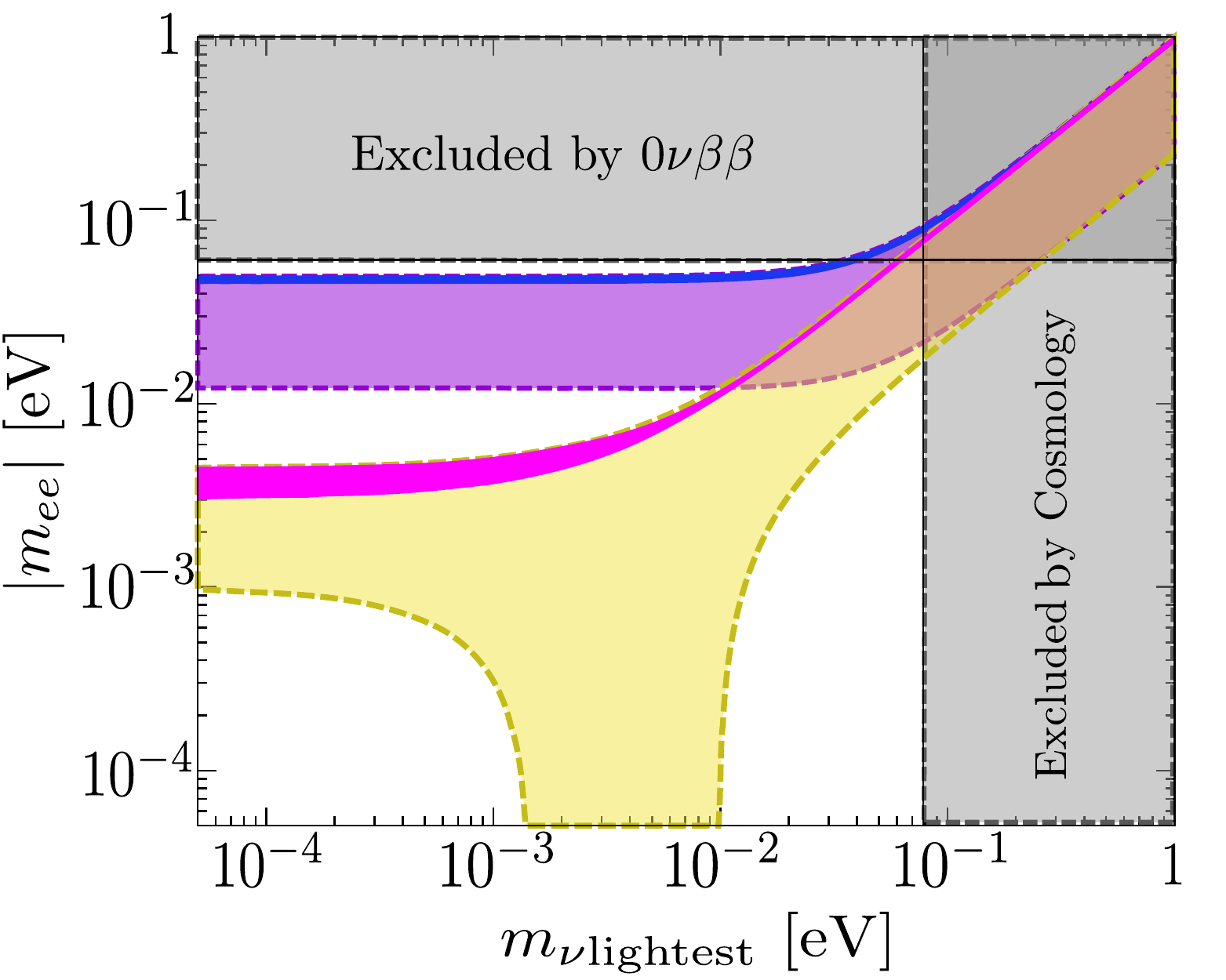}}
		\end{tabular}
		\caption{ In these panels, we show the allowed regions for the magnitude of Majorana effective mass $|m_{ee}|$.
			Respectively, for an inverted and normal neutrino mass hierarchy, the yellow and purple stripes are obtained from the current experimental data on neutrino 
			oscillations at 3$\sigma$~\cite{global-fit-2021}. 
			In the upper right panel, the magenta area is for a normal hierarchy while the blue area is for an inverted hierarchy, 
			both areas are obtained from ${\bf M}_{\ell}^{1}$ and ${\bf M}_{\ell}^{5}$. 
			In the upper left panel, the red area is for a normal hierarchy while the orange area is for an inverted hierarchy, both areas are obtained from  
			${\bf M}_{\ell}^{2}$ and ${\bf M}_{\ell}^{4}$.
			From KamLAND-ZEN~\cite{KamLAND-Zen:2016pfg} and EXO-200~\cite{EXO:2017poz} we have the following upper limit $|m_{ee}| < 0.061$, which correspond to the 
			horizontal grey band, whereas vertical grey band corresponds to results reported by Planck collaboration~\cite{Planck:2015fie}.}
		\label{Fig:Mee:Type-IV}
	\end{center}
\end{figure}
The parameter $\widetilde{f}_{\ell}$ must satisfy the condition $\widetilde{m}_{\mu} > \widetilde{f}_{\ell} > \widetilde{m}_{e}$. 
In this case, the flavor mixing angles in eq.~(\ref{Eq:Mix-Angle-0}) have the form:
\begin{equation}
 \begin{array}{l}
  \sin^{2} \theta_{12} = \frac{1}{3} \varepsilon_{12}, \qquad
  \sin^{2} \theta_{23} = \frac{1}{2} \varepsilon_{23}, \qquad
  \sin^{2} \theta_{13} = \frac{1}{2} \frac{ \widetilde{m}_{e} }{ \widetilde{m}_{\mu} } \varepsilon_{13}.
 \end{array}
\end{equation}
The explicit form of the $\varepsilon_{ij}$ parameters is given in the Appendix~\ref{Appendix-Para-Type-IV}. In this equivalent class, the charged lepton mass matrices 
${\bf M}_{\ell}^{2}$ and  ${\bf M}_{\ell}^{4}$ cannot correctly reproduce current experimental data of the solar and atmospheric mixing angles, since these mass 
matrices generate the same theoretical expression for $\varepsilon_{12}=\varepsilon_{23}$. 
In figure~\ref{Fig:Angulos-Clase-IV-fl} we show the allowed regions for the reactor, atmospheric, and solar mixing angles and the free parameter $f_{\ell}$. From this 
regions we obtain that to reproduce the values for the leptonic flavor mixing angles, at $3\sigma$, obtained form the global fit, for a normal and inveterted hierarchy, 
the free parameter $f_{\ell}$ should be in the following numerical interval 
\begin{equation}\label{Eq:Type-IV:val-f-l}
 \begin{array}{l}\vspace{2mm}
  f_{\ell} \, \in \, \left[ 2.66 \times 10^{-3} , 3.16 \times 10^{-3} \right] 
   \quad \textrm{for} \, {\bf M}_{\ell}^{0} \, \textrm{and} \, {\bf M}_{\ell}^{3}. \\
  f_{\ell} \, \in \, \left[ 5.66 \times 10^{-2} , 5.71 \times 10^{-2} \right] 
   \quad \textrm{for} \, {\bf M}_{\ell}^{1} \, \textrm{and} \, {\bf M}_{\ell}^{5}.
 \end{array}
\end{equation}
For the mass matrices ${\bf M}_{\ell}^{0}$, ${\bf M}_{\ell}^{1}$, ${\bf M}_{\ell}^{3}$ and ${\bf M}_{\ell}^{5}$, the theoretical expressions for reactor and atmospheric 
mixing angles do not explicitly have a dependence on the phase factor $\phi_{a}$, while the solar angle has a weak dependence on the same phase factor. 
 
In figure~\ref{Fig:Mee:Type-IV} we show the allowed regions for the magnitude of the Majorana effective mass $|m_{ee}|$, eq.~(\ref{ec:mee-0}), which were obtained in a 
model-independent context where the neutrino mass matrix has the form given in eq.~(\ref{Eq:M_TBM}), while the charged lepton matrix is represented for an element of 
the equivalent class with two texture zeros type-IV, eq.~(\ref{Eq:EQ-Type-IV}). Each one of these regions was obtained by considering the values given in 
eqs.~(\ref{Eq:Type-IV:val-f-l}) for the free parameter $f_{\ell}$.
%
\section{Summary}\label{sec:summary}
%
In a model-independent theoretical framework, we present a generalization of TBM leptonic flavor mixing pattern. 
In this modification to the TBM pattern, the unitary matrix that diagonalizes to the neutrino mass matrix is represented by means of TBM  flavor mixing pattern, 
eq.~(\ref{Eq:Utbm-0}), whereas the charged lepton mass matrix is represented by one of the elements of the equivalence classes with two texture zeros, 
eqs.~(\ref{Eq:EQ-Type-I}),~(\ref{Eq:EQ-Type-II}),~(\ref{Eq:EQ-Type-III}), and~(\ref{Eq:EQ-Type-IV}). 
For these four equivalent classes, we show a deviation from the TBM 
pattern in terms of the charged lepton masses as well as the theoretical expressions and their parameter space for the mixing angles. 
Furthermore, from the theoretical expressions of $\varepsilon_{ij}$ in Appendix~\ref{sec:mixingangles} we have for each type of equivalent class the ${\bf M}^0_{\ell}$ 
and ${\bf M}^3_{\ell}$ matrices generate  the same expressions for $\epsilon_{23}$ and $\epsilon_{13}$; similarly for ${\bf M}^2_{\ell}$ and ${\bf M}^4_{\ell}$, as well 
as ${\bf M}^1_{\ell}$ and ${\bf M}^5_{\ell}$. Finally, we present the phenomenological implications of numerical values of the ``Majorana-like'' phase factors on the neutrinoless double-beta decay. 

From the analysis performed for the four types of equivalence classes, we had that:
For the equivalent class type-I, on the one hand, it is easy to conclude that all charged lepton mass matrices ${\bf M}^{i}_{\ell}$ ($i=0,\dots, 5$) are able to reproduce the current experimental values of reactor, solar 
and atmospheric angles. But, the numerical interval of the free parameter $\delta_{\ell}$, for the ${\bf M}_{\ell}^{1}$, ${\bf M}_{\ell}^{2}$, 
${\bf M}_{\ell}^{4}$ and ${\bf M}_{\ell}^{5}$, is too small eq.~(\ref{Eq:Type-I:val-delta-l}). 
On the other hand, for all mass matrices ${\bf M}^{i}_{\ell}$ ($i=0,\dots, 5$), the solar and reactor mixing angles have a weak dependence on phases $\phi_{a}$ and 
$\phi_{c}$; whereas for ${\bf M}_{\ell}^{1}$ and ${\bf M}_{\ell}^{5}$, the atmospheric mixing angle has a weak dependence on $\phi_{a}$ and $\phi_{c}$. 
However, to reproduce the current experimental data for  the  mixing angle $\theta_{23}$, eq.~(\ref{Ec:exp-mixing-angles}), for ${\bf M}_{\ell}^{0}$ and 
${\bf M}_{\ell}^{3}$, the  phase $\phi_{c}$ is in the numerical interval (\ref{Eq:Type-I:val-phi-c});  for  ${\bf M}_{\ell}^{2}$ and 
${\bf M}_{\ell}^{4}$, $\phi_{a}$  runs over the numerical interval (\ref{Eq:Type-I:val-phi-a}).

In case of  the equivalent class type-II,  the charged lepton mass matrices ${\bf M}_{\ell}^{0}$ and ${\bf M}_{\ell}^{3}$ cannot simultaneously reproduce the current 
experimental data on the neutrino oscillations. But the remaining mass matrices, ${\bf M}_{\ell}^{1}$, ${\bf M}_{\ell}^{2}$, ${\bf M}_{\ell}^{4}$ and 
${\bf M}_{\ell}^{5}$, reproduce the experimental values of the three leptonic mixing angles, where the free parameter $\delta_{\ell}$ has the numerical 
interval given in  eq.~(\ref{Eq:Type-II:val-delta-l}). Moreover, for these four mass matrices, the reactor mixing angle has a weak dependence on the phases $\phi_{a}$ and $\phi_{c}$. 
The solar mixing angle has a weak dependence on the phases $\phi_{a}$ and $\phi_{c}$ for the ${\bf M}_{\ell}^{2}$ and ${\bf M}_{\ell}^{4}$, whereas to 
reproduce the current experimental data for  $\theta_{12}$, eq.~(\ref{Ec:exp-mixing-angles}); for ${\bf M}_{\ell}^{1}$ and  ${\bf M}_{\ell}^{5}$ ,
 the  phase $\phi_{c}$ is in the numerical interval given in eq.~(\ref{Eq:Type-II:val-phi-c}). The atmospheric mixing angle has a weak dependence on the phases 
$\phi_{a}$ and $\phi_{c}$ for the ${\bf M}_{\ell}^{1}$ and ${\bf M}_{\ell}^{5}$. Nevertheless, to reproduce the current experimental data for  $\theta_{23}$, 
for ${\bf M}_{\ell}^{2}$ and  ${\bf M}_{\ell}^{4}$, the  phase $\phi_{a}$ run over numerical interval given in eq.~(\ref{Eq:Type-II:val-phi-a}).

For the equivalent class type-III,  the charged lepton mass matrices ${\bf M}_{\ell}^{2}$ and ${\bf M}_{\ell}^{4}$ cannot simultaneously reproduce the current 
experimental data on the neutrino oscillations. However, for the numerical interval of the free parameter $\delta_{\ell}$ given in eq.~(\ref{Eq:Type-III:val-delta-l}), 
 ${\bf M}_{\ell}^{0}$, ${\bf M}_{\ell}^{1}$, ${\bf M}_{\ell}^{3}$ and  ${\bf M}_{\ell}^{5}$,  correctly reproduce the experimental data of the three 
leptonic mixing angles. For the four previous matrices, the reactor mixing angle has a weak dependence on phases $\phi_{a}$ and  $\phi_{c}$. The atmospheric mixing 
angle has a weak dependence on the phases $\phi_{a}$ and  $\phi_{c}$ for the ${\bf M}_{\ell}^{2}$ and ${\bf M}_{\ell}^{4}$, while to reproduce the 
experimental data for $\theta_{23}$, for ${\bf M}_{\ell}^{1}$ and ${\bf M}_{\ell}^{5}$, the phase $\phi_{a}$ is in the numerical interval given in 
eq.~(\ref{Eq:Type-III:val-phi-c}). Furthermore, to reproduce the current experimental data for  the solar mixing angle for the four charged lepton mass matrices, the phase factors  $\phi_{a}$ and  
$\phi_{c}$ are in the numerical intervals given in eqs.~(\ref{Eq:Type-III:val-phi-c}) and~(\ref{Eq:Type-III:val-phi-a}). 

Finally, of the equivalent class type-IV we concluded that the charged lepton mass matrices ${\bf M}_{\ell}^{2}$ and  ${\bf M}_{\ell}^{4}$ cannot simultaneously reproduce the current 
experimental data on the neutrino oscillations, since for these mass matrices $\varepsilon_{12}=\varepsilon_{23}$. 
However, the remaining four mass matrices reproduce the experimental data of the three lepton mixing angles, where  the numerical interval of the free parameter 
$f_{\ell}$ is given in eq.~(\ref{Eq:Type-IV:val-f-l}). And for this case, the atmospheric, reactor, and solar mixing angles have a weak dependence on phase factor $\phi_{a}$.

%
\appendix
%
\section{Permutation group}\label{sec:permutationgroup}
The permutations of symmetry group $S_{3}$ can be represented on the reducible 
triplet as~\cite{Georgi:1999wka,Ishimori:2010au}: 
\begin{equation}\label{eq:A-1}
 \begin{array}{ccc} \vspace{2mm}
  {\bf T}_{0} = 
   \left( \begin{array}{ccc}
    1 & 0 & 0 \\
    0 & 1 & 0 \\
    0 & 0 & 1
   \end{array} \right), &
  {\bf T}_{1} =
   \left( \begin{array}{ccc}
    0 & 1 & 0 \\
    1 & 0 & 0 \\
    0 & 0 & 1
   \end{array}  \right), &
  {\bf T}_{2} =
   \left( \begin{array}{ccc}
    0 & 0 & 1 \\
    0 & 1 & 0 \\
    1 & 0 & 0
   \end{array}  \right) , \\
 {\bf T}_{3} =
   \left( \begin{array}{ccc}
    1 & 0 & 0 \\
    0 & 0 & 1 \\
    0 & 1 & 0
   \end{array}  \right) , &
 {\bf T}_{4} = 
   \left( \begin{array}{ccc}
    0 & 1 & 0 \\
    0 & 0 & 1 \\
    1 & 0 & 0
   \end{array}  \right) , & 
 {\bf T}_{5} = 
   \left( \begin{array}{ccc} 
    0 & 0 & 1 \\
    1 & 0 & 0 \\
    0 & 1 & 0
   \end{array}  \right) .   
 \end{array}
\end{equation}
%
%
\section{General Eigenvector for a $3 \times 3$ complex matrix}\label{sec:generleig}
%
 The general shape of a $3 \times 3$ matrix is:
 \begin{equation}
  {\bf M} = 
  \left( \begin{array}{ccc}
   m_{11} & m_{12} & m_{13} \\
   m_{21} & m_{22} & m_{23} \\
   m_{31} & m_{32} & m_{33}  
  \end{array}   \right), 
 \end{equation}
 where all elements of ${\bf M}$ are complex. The three eigenvalues, 
 $\left| {\bf M}_{k} \right.\rangle$, of the ${\bf M}$ matrix 
  have the form~\cite{SimetryS3Dr_Felix}:
 \begin{equation}\label{Eq:General_Eigenvector}
  \left| M_{k} \right. \rangle = \frac{1}{ N_{k} } 
  \left( \begin{array}{c}
   a_{k} \\
   b_{k} \\
   c_{k}
  \end{array}   \right)   
  = \frac{1}{ N_{k} }
  \left( \begin{array}{c}
   \left( \lambda_{k} - m_{22} \right) m_{13} + m_{12} m_{23} \\
   \left( \lambda_{k} - m_{11} \right) m_{23} + m_{21} m_{13} \\
   \left( \lambda_{k} - m_{11} \right) \left( \lambda_{k} - m_{22} \right) - m_{12} m_{21}
  \end{array}   \right).
 \end{equation}
 In this expression the $\lambda_{k}$, with $k=1,2,3$, correspond to the eigenvalues of 
 the ${\bf M}$ matrix, and the 
 $N_{k} \equiv \sqrt{ \langle {\bf M}_{k} \left| {\bf M}_{k} \right.\rangle }$  are the 
 normalization constants. 
 Now, it is easy to show that the $\left| {\bf M}_{k} \right.\rangle$ are the eigenvectors of 
 the ${\bf M}$ matrix,  we only need to consider the eigenvalues equation 
 ${\bf M} \, \left| {\bf M}_{k} \right. \rangle = 
 \lambda_{k} \left| {\bf M}_{k} \right.\rangle$ and 
 the explicit form of the characteristic polynomial which is given by the expression
 $\textrm{det} \left \{ \lambda_{k} \mathbb{I}_{3 \times 3} - {\bf M}  \right\} = 0$. 
 The explicit form of the eigenvalues equation is 
 \begin{equation}\label{Eq:A3}
  \left( \begin{array}{ccc}
   m_{11} & m_{12} & m_{13} \\
   m_{21} & m_{22} & m_{23} \\
   m_{31} & m_{32} & m_{33}  
  \end{array}   \right)
  \left( \begin{array}{c}
   a_{k} \\
   b_{k} \\
   c_{k}
  \end{array}  \right) =
  \lambda_{k} 
  \left( \begin{array}{c}
   a_{k} \\
   b_{k} \\
   c_{k}
  \end{array}  \right)  .
 \end{equation}
 The first row of the right-handed side of eq.~(\ref{Eq:A3}) is
 \begin{equation}\label{Eq:A4}
  \begin{array}{rl}
   m_{11} a_{k} + m_{12} b_{k} + m_{13} c_{k} = &  
       m_{11} \left( \left( \lambda_{k} - m_{22} \right) m_{13} + m_{12} m_{23} \right) 
    +  m_{12} \left( \left( \lambda_{k} - m_{11} \right) m_{23} + m_{21} m_{13} \right) \\ &
    +  m_{13} \left( \left( \lambda_{k} - m_{11} \right) \left( \lambda_{k} - m_{22} \right) 
     - m_{12} m_{21} \right), \\ 
   = &
   \lambda_{k} \left( \left( \lambda_{k} - m_{22} \right) m_{13} + m_{12} m_{23}  \right), \\
   = & \lambda_{k} a_{k}.
  \end{array}
 \end{equation}
 The second row of the right-handed side of eq.~(\ref{Eq:A3}) is 
 \begin{equation}\label{Eq:A5}
  \begin{array}{rl}
   m_{21} a_{k} + m_{22} b_{k} + m_{23} c_{k} = &  
       m_{21} \left( \left( \lambda_{k} - m_{22} \right) m_{13} + m_{12} m_{23} \right) 
    +  m_{22} \left( \left( \lambda_{k} - m_{11} \right) m_{23} + m_{21} m_{13} \right) \\ &
    +  m_{23} \left( \left( \lambda_{k} - m_{11} \right) \left( \lambda_{k} - m_{22} \right) 
     - m_{12} m_{21} \right), \\ 
   = &
   \lambda_{k} \left( \left( \lambda_{k} - m_{11} \right) m_{23} + m_{21} m_{13} \right) , \\
   = & \lambda_{k} b_{k}.
  \end{array}
 \end{equation} 
 The third row of the right-handed side of eq.~(\ref{Eq:A3}) is 
 \begin{equation}\label{Eq:A6}
  \begin{array}{rl}
   m_{31} a_{k} + m_{32} b_{k} + m_{33} c_{k} = &  
       m_{31} \left( \left( \lambda_{k} - m_{22} \right) m_{13} + m_{12} m_{23} \right) 
    +  m_{32} \left( \left( \lambda_{k} - m_{11} \right) m_{23} + m_{21} m_{13} \right) \\ &
    +  m_{33} \left( \left( \lambda_{k} - m_{11} \right) \left( \lambda_{k} - m_{22} \right) 
     - m_{12} m_{21} \right), \\ 
   = & \left( \lambda_{k} - m_{22} \right) m_{13} m_{31} + m_{31} m_{12} m_{23} 
   + m_{32} m_{21} m_{13} + \left( \lambda_{k} - m_{11} \right) m_{32} m_{23} \\ &
   + \left( \lambda_{k} - m_{22} \right) \left( \lambda_{k} - m_{11} \right) m_{33}
   - m_{33} m_{12} m_{21} .
  \end{array}
 \end{equation}  
 With help of the characteristic polynomial in terms of ${\bf M}$ matrix invariants
 (trace and determinant)~\cite{SimetryS3Dr_Felix} 
 \begin{equation}
  \lambda_{i}^{3} - \textrm{Tr} \left\{ {\bf M} \right\} \lambda_{i}^{2} 
 - \chi\left\{ {\bf M} \right\} \lambda_{i} - \textrm{det} \left\{ {\bf M} \right\} = 0,
 \end{equation}
 where $\chi\left\{ {\bf M} \right\}$ is a function of the trace with the following explicit 
 form:
 \begin{equation}
  \chi \left\{ {\bf M} \right\} \equiv 
   - \textrm{Tr} \left\{ \textrm{adj} \left\{\bf M \right\} \right\} = 
   \frac{1}{2} \left( \textrm{Tr} \left\{ {\bf M}^{2} \right\} 
    - \textrm{Tr} \left\{ {\bf M} \right\}^{2} 
 \right) .
 \end{equation}   
 We obtain that
 \begin{equation}\label{Eq:A9}
  \begin{array}{l}
  \left( \lambda_{k} - m_{22} \right) m_{13} m_{31} + m_{31} m_{12} m_{23} 
   + m_{32} m_{21} m_{13} + \left( \lambda_{k} - m_{11} \right) m_{32} m_{23} 
   + \left( \lambda_{k} - m_{22} \right) \left( \lambda_{k} - m_{11} \right) m_{33} \\
   - m_{33} m_{12} m_{21} = 
   \lambda_{k} 
   \left( \left( \lambda_{k} - m_{11} \right) \left( \lambda_{k} - m_{22} \right) 
    - m_{12} m_{21} \right) = \lambda_{k} c_{k} . 
  \end{array}
 \end{equation}
 From eq.~(\ref{Eq:A9}) the expression in eq.~(\ref{Eq:A6}) takes the form:
 \begin{equation}\label{Eq:A10}
  \begin{array}{rl}
   m_{31} a_{k} + m_{32} b_{k} + m_{33} c_{k} = & 
   \lambda_{k} 
   \left( \left( \lambda_{k} - m_{11} \right) \left( \lambda_{k} - m_{22} \right) 
    - m_{12} m_{21} \right) = \lambda_{k} c_{k}.
  \end{array}
 \end{equation}   
 Now, with help of Eqs.~(\ref{Eq:A4}), (\ref{Eq:A5}) and~(\ref{Eq:A10}) we can conclude that
 the vectors $\left| {\bf M}_{k} \right.\rangle$ are eigenvectors of ${\bf M}$ matrix. 
%
\section{Mixing angles parameters}\label{sec:mixingangles}
%
%
\subsection{Parameters of equivalent class with two texture zeros type-I}\label{Appendix-Para-Type-I}
%
For the mass matrices ${\bf M}_{\ell}^{0}$ and ${\bf M}_{\ell}^{3}$,
\begin{equation}
  \begin{array}{l}\vspace{2mm}
   \varepsilon_{23} = 
    \frac{
     \left( 1 + \texttt{s}_{3} \left( \delta_{\ell} - 1 \right) \right) f_{\ell 1} + \left( 1 - \delta_{\ell} \right) f_{\ell 2} 
     + \texttt{s}_{1} \texttt{s}_{2} 
     2 \sqrt{ \left( 1 + \texttt{s}_{3} \left( \delta_{\ell} - 1 \right) \right) \left( 1 - \delta_{\ell} \right) f_{\ell 1} f_{\ell 2} } 
     \cos \phi_{c}
    }{
     \left( 1 - \delta_{\ell} \right) \left(1 + \texttt{s}_{3} \frac{ \widetilde{m}_{e} }{ \widetilde{m}_{\mu} }  \right)   
     \left( 1 + \texttt{s}_{2} \widetilde{m}_{e} \right)
     +
     \frac{1}{2} \frac{ \widetilde{m}_{e} }{ \widetilde{m}_{\mu} }
     \left(
      f_{\ell 2} \left( 1 + \texttt{s}_{3} \left( \delta_{\ell} - 1 \right) \right) + f_{\ell 1} \left( 1 - \delta_{\ell} \right) 
      - 2 \sqrt{ \left( 1 + \texttt{s}_{3} \left( \delta_{\ell} - 1 \right) \right) \left( 1 - \delta_{\ell} \right)  f_{\ell 1} f_{\ell 2} }
      \cos \phi_{c}
     \right)
    } , \\ \vspace{2mm}
   \varepsilon_{13} = 
    \frac{
     f_{\ell 2} \left( 1 + \texttt{s}_{3} \left( \delta_{\ell} - 1 \right) \right) 
     + f_{\ell 1} \left( 1 - \delta_{\ell} \right) 
     - 2 \sqrt{ \left( 1 - \delta_{\ell} \right) \left( 1 + \texttt{s}_{3} \left( \delta_{\ell} - 1 \right) \right) f_{\ell 1} f_{\ell 2} } \cos \phi_{c}
    }{
     \left( 1 - \delta_{\ell} \right)
     \left(1 + \texttt{s}_{3} \frac{ \widetilde{m}_{e} }{ \widetilde{m}_{\mu} }  \right) \left( 1 + \texttt{s}_{2} \widetilde{m}_{e} \right)
    }  , \\ \vspace{2mm}
   \varepsilon_{12} = 
   \frac{
    f_{\ell 2} \left( 1 + \texttt{s}_{3} \left( \delta_{\ell} - 1 \right) \right) +
    f_{\ell 1} \frac{ \widetilde{m}_{\mu} }{ \widetilde{m}_{e} } +
    f_{\ell 1} \left( 1 - \delta_{\ell} \right) +
    2 \sqrt{ \left( 1 + \texttt{s}_{3} \left( \delta_{\ell} - 1 \right) \right) \left( 1 - \delta_{\ell} \right) f_{\ell 1} f_{\ell 2} } c_{c} 
    +(-1)^{i} \texttt{s}_{1} 2 \left(
     \sqrt{ \frac{ \widetilde{m}_{\mu} }{ \widetilde{m}_{e} } \left( 1 + \texttt{s}_{3} \left( \delta_{\ell} - 1 \right) \right) f_{\ell 1} f_{\ell 2} } 
     c_{ac} 
    +
    f_{\ell 1} \sqrt{ \frac{ \widetilde{m}_{\mu} }{ \widetilde{m}_{e} } \left( 1 - \delta_{\ell} \right) }  c_{a} \right)
   }{
    \left( 1 - \delta_{\ell} \right) \left( 1 + \texttt{s}_{3} \frac{ \widetilde{m}_{e} }{ \widetilde{m}_{\mu} } \right)
    \left( 1 + \texttt{s}_{2} \widetilde{m}_{e} \right) 
    -
    \frac{1}{2} \frac{ \widetilde{m}_{e} }{ \widetilde{m}_{\mu} } f_{\ell 2} \left( 1 + \texttt{s}_{3} \left( \delta_{\ell} - 1 \right) \right)
    +
    \left(1 - \delta_{\ell}\right) f_{\ell 1}
    -
    2 \sqrt{ \left(1 - \delta_{\ell} \right) \left( 1 + \texttt{s}_{3} \left( \delta_{\ell} - 1 \right) \right) f_{\ell 1} f_{\ell 2} }
    c_{c}
   } ,
 \end{array}
\end{equation}
where $i = 0$ for ${\bf M}_{\ell}^{0}$, $i=1$ for ${\bf M}_{\ell}^{3}$, $c_{a} = \cos \phi_{a}$, $c_{c} = \cos \phi_{c}$, and 
$c_{ac} = \cos \left( \phi_{a} + \phi_{c} \right)$. 
For ${\bf M}_{\ell}^{1}$ and ${\bf M}_{\ell}^{5}$ 
\begin{equation}
  \begin{array}{l} \vspace{2mm}
   \varepsilon_{23} = 
    \frac{ 
     \frac{ \widetilde{m}_{e} }{ \widetilde{m}_{\mu} } f_{\ell 2} + \left( 1 + \texttt{s}_{3} \left( \delta_{\ell} - 1 \right) \right) f_{\ell 1}
     + \texttt{s}_{1} 2 
     \sqrt{ \frac{ \widetilde{m}_{e} }{ \widetilde{m}_{\mu} }  \left( 1 + \texttt{s}_{3} \left( \delta_{\ell} - 1 \right) \right) f_{\ell 1} f_{\ell 2} }
     \cos \left( \phi_{a} + \phi_{c} \right)
    }{ 
     \left( 1 - \delta_{\ell} \right)
     \left( 1 + \texttt{s}_{3} \frac{ \widetilde{m}_{e} }{ \widetilde{m}_{\mu} }  \right) \left( 1 + \texttt{s}_{2} \widetilde{m}_{e} \right) 
     -
     \frac{1}{2} \frac{ \widetilde{m}_{e} }{ \widetilde{m}_{\mu} }
     \left( 
     f_{\ell 2} \left( 1 + \texttt{s}_{3} \left( \delta_{\ell} - 1 \right) \right)   
     + \frac{ \widetilde{m}_{\mu} }{ \widetilde{m}_{e} } f_{\ell 1}  
     - \texttt{s}_{1} 2 \sqrt{ \frac{ \widetilde{m}_{\mu} }{ \widetilde{m}_{e} } 
     \left( 1 + \texttt{s}_{3} \left( \delta_{\ell} - 1 \right) \right) f_{\ell 1} f_{\ell 2}  } \cos \left( \phi_{a} + \phi_{c} \right) \right)
    }
   ,\\ \vspace{2mm}
   \varepsilon_{13} = 
    \frac{
     f_{\ell 2} \left( 1 + \texttt{s}_{3} \left( \delta_{\ell} - 1 \right) \right)   
     + \frac{ \widetilde{m}_{\mu} }{ \widetilde{m}_{e} } f_{\ell 1}  
     - \texttt{s}_{1} 2 \sqrt{ \frac{ \widetilde{m}_{\mu} }{ \widetilde{m}_{e} } 
     \left( 1 + \texttt{s}_{3} \left( \delta_{\ell} - 1 \right) \right) f_{\ell 1} f_{\ell 2}  } \cos \left( \phi_{a} + \phi_{c} \right)
    }{
     \left( 1 - \delta_{\ell} \right)
     \left( 1 + \texttt{s}_{3} \frac{ \widetilde{m}_{e} }{ \widetilde{m}_{\mu} }  \right) \left( 1 + \texttt{s}_{2} \widetilde{m}_{e} \right)
    } 
   ,\\ \vspace{2mm}
   \varepsilon_{12} =  
   \frac{
    f_{\ell 2} \left( 1 + \texttt{s}_{3} \left( \delta_{\ell} - 1 \right) \right) +
    f_{\ell 1} \frac{ \widetilde{m}_{\mu} }{ \widetilde{m}_{e} } +
    f_{\ell 1} \left( 1 - \delta_{\ell} \right) +
    2 \texttt{s}_{1} 
     \sqrt{ \frac{ \widetilde{m}_{\mu} }{ \widetilde{m}_{e} } \left( 1 + \texttt{s}_{3} \left( \delta_{\ell} - 1 \right) \right) f_{\ell 1} f_{\ell 2} } 
     c_{ac} 
     + (-1)^{i} 2 \left(
     \sqrt{ \left( 1 + \texttt{s}_{3} \left( \delta_{\ell} - 1 \right) \right) \left( 1 - \delta_{\ell} \right) f_{\ell 1} f_{\ell 2} } c_{c}  +
     \texttt{s}_{1} f_{\ell 1} \sqrt{ \frac{ \widetilde{m}_{\mu} }{ \widetilde{m}_{e} } \left( 1 - \delta_{\ell} \right) }  c_{a} 
    \right)
   }{
    \left( 1 - \delta_{\ell} \right) \left( 1 + \texttt{s}_{3} \frac{ \widetilde{m}_{e} }{ \widetilde{m}_{\mu} } \right)
    \left( 1 + \texttt{s}_{2} \widetilde{m}_{e} \right) 
    -
    \frac{1}{2} \frac{ \widetilde{m}_{e} }{ \widetilde{m}_{\mu} } f_{\ell 2} \left( 1 + \texttt{s}_{3} \left( \delta_{\ell} - 1 \right) \right)
    +
    f_{\ell 1} \frac{ \widetilde{m}_{\mu} }{ \widetilde{m}_{e} }
    -
    2 \texttt{s}_{1} \sqrt{ \frac{ \widetilde{m}_{\mu} }{ \widetilde{m}_{e} } \left( 1 + \texttt{s}_{3} \left( \delta_{\ell} - 1 \right) \right) f_{\ell 1} f_{\ell 2} }
    c_{ac}
   } ,
   \end{array}
\end{equation}
where $i = 0$ for ${\bf M}_{\ell}^{1}$, $i=1$ for ${\bf M}_{\ell}^{5}$, $c_{a} = \cos \phi_{a}$, $c_{c} = \cos \phi_{c}$, and 
$c_{ac} = \cos \left( \phi_{a} + \phi_{c} \right)$. 
For ${\bf M}_{\ell}^{2}$ and ${\bf M}_{\ell}^{4}$ 
\begin{equation}
 \begin{array}{l}\vspace{2mm}
  \varepsilon_{23} = 
   \frac{
    f_{\ell 2} \left( 1 - \delta_{\ell} + \frac{ \widetilde{m}_{e} }{ \widetilde{m}_{\mu} } 
    + \texttt{s}_{2} 2 \sqrt{ \frac{ \widetilde{m}_{e} }{ \widetilde{m}_{\mu} } \left( 1 - \delta_{\ell} \right) }  \cos \phi_{a} \right)
   }{
    \left( 1 - \delta_{\ell} \right)
    \left( 1 + \texttt{s}_{3} \frac{ \widetilde{m}_{e} }{ \widetilde{m}_{\mu} }  \right) \left( 1 + \texttt{s}_{2} \widetilde{m}_{e} \right)
    - \frac{1}{2} \frac{ \widetilde{m}_{e} }{ \widetilde{m}_{\mu} } f_{\ell 1} \left( 1 - \delta_{\ell}  + \frac{ \widetilde{m}_{\mu} }{ \widetilde{m}_{e} }  
    + \texttt{s}_{1} 2 \sqrt{  \frac{ \widetilde{m}_{\mu} }{ \widetilde{m}_{e} } \left( 1 - \delta_{\ell} \right) } \cos \phi_{a} \right)
   } ,\;
  \varepsilon_{13} = 
   \frac{ 
    f_{\ell 1} \left( 1 - \delta_{\ell}  + \frac{ \widetilde{m}_{\mu} }{ \widetilde{m}_{e} }  
    + \texttt{s}_{1} 2 \sqrt{  \frac{ \widetilde{m}_{\mu} }{ \widetilde{m}_{e} } \left( 1 - \delta_{\ell} \right) } \cos \phi_{a} \right)    
   }{
    \left( 1 - \delta_{\ell} \right)
    \left( 1 + \texttt{s}_{3} \frac{ \widetilde{m}_{e} }{ \widetilde{m}_{\mu} }  \right) \left( 1 + \texttt{s}_{2} \widetilde{m}_{e} \right)
   }, \\ \vspace{2mm}
   \varepsilon_{12} = 
   \frac{
    f_{\ell 2} \left( 1 + \texttt{s}_{3} \left( \delta_{\ell} - 1 \right) \right) +
    f_{\ell 1} \frac{ \widetilde{m}_{\mu} }{ \widetilde{m}_{e} } +
    f_{\ell 1} \left( 1 - \delta_{\ell} \right) -
    2 \texttt{s}_{1} f_{\ell 1} \sqrt{ \frac{ \widetilde{m}_{\mu} }{ \widetilde{m}_{e} } \left( 1 - \delta_{\ell} \right) }  c_{a} 
    + (-1)^{i} 2 \left(
     \sqrt{ \left( 1 + \texttt{s}_{3} \left( \delta_{\ell} - 1 \right) \right) \left( 1 - \delta_{\ell} \right) f_{\ell 1} f_{\ell 2} } c_{c} -
     \texttt{s}_{1} 
     \sqrt{ \frac{ \widetilde{m}_{\mu} }{ \widetilde{m}_{e} } \left( 1 + \texttt{s}_{3} \left( \delta_{\ell} - 1 \right) \right) f_{\ell 1} f_{\ell 2} } 
     c_{ac} \right)
   }{
    \left( 1 - \delta_{\ell} \right) \left( 1 + \texttt{s}_{3} \frac{ \widetilde{m}_{e} }{ \widetilde{m}_{\mu} } \right)
    \left( 1 + \texttt{s}_{2} \widetilde{m}_{e} \right) 
    -
    \frac{ \widetilde{m}_{e} }{ \widetilde{m}_{\mu} } \frac{f_{\ell 1}}{2}  
    \left(  1 - \delta_{\ell}  + \frac{ \widetilde{m}_{\mu} }{ \widetilde{m}_{e} }  
    + \texttt{s}_{1} 2 \sqrt{  \frac{ \widetilde{m}_{\mu} }{ \widetilde{m}_{e} } \left( 1 - \delta_{\ell} \right) } \cos \phi_{a} 
    \right)
   },
  \end{array} 
\end{equation}
where $i = 0$ for ${\bf M}_{\ell}^{2}$, $i=1$ for ${\bf M}_{\ell}^{4}$, $c_{a} = \cos \phi_{a}$, $c_{c} = \cos \phi_{c}$, and 
$c_{ac} = \cos \left( \phi_{a} + \phi_{c} \right)$. 
%
\subsection{Parameters of equivalent class with two texture zeros type-II}\label{Appendix-Para-Type-II}
%
For the mass matrices ${\bf M}_{\ell}^{0}$ and ${\bf M}_{\ell}^{3}$,
\begin{equation}
 \begin{array}{l}\vspace{2mm}
  \varepsilon_{23} = 
   \frac{
    f_{\ell 1} \delta_{\ell} + f_{\ell 2} \widetilde{\mu}_{\ell} - 2 \sqrt{ \widetilde{\mu}_{\ell} \delta_{\ell} f_{\ell 1} f_{\ell 2} } \cos \phi_{c}
   }{
    \widetilde{\mu}_{\ell} \left( 1 + \frac{ \widetilde{m}_{e} }{ \widetilde{m}_{\mu} } \right) \left(1 - \widetilde{m}_{e} \right)
    - \frac{1}{2} \frac{ \widetilde{\sigma}_{\ell 1} }{ \widetilde{m}_{\mu} } 
    \left( f_{\ell 2} \delta_{\ell} + \widetilde{\mu}_{\ell} f_{\ell 1} + 2 \sqrt{ \widetilde{\mu}_{\ell} \delta_{\ell} f_{\ell 1} f_{\ell 2} } \cos \phi_{c} \right)
   }, \;
  \varepsilon_{13} =
   \frac{
    f_{\ell 2} \delta_{\ell} +
    \widetilde{\mu}_{\ell} f_{\ell 1} +
    2 \sqrt{ \widetilde{\mu}_{\ell} \delta_{\ell} f_{\ell 1} f_{\ell 2} } \cos \phi_{c}
   }{
    \widetilde{\mu}_{\ell} \left( 1 + \frac{ \widetilde{m}_{e} }{ \widetilde{m}_{\mu} } \right) \left(1 - \widetilde{m}_{e} \right)
   }, \\\vspace{2mm}
  \varepsilon_{12} = 
   \frac{ 
    f_{\ell 2} \delta_{\ell}
    +
    f_{\ell 1} \widetilde{\mu}_{\ell} 
    +
    f_{\ell 1} \frac{ \widetilde{\sigma}_{\ell 2} \widetilde{\sigma}_{\ell 3} }{ \widetilde{\sigma}_{\ell 1} } 
    - 
    2 \sqrt{ \widetilde{\mu}_{\ell} \delta_{\ell} f_{\ell 1} f_{\ell 2} } \cos \phi_{c}
    + (-1)^{i} 2
    \sqrt{ \frac{ \widetilde{\sigma}_{\ell 2} \widetilde{\sigma}_{\ell 3} }{ \widetilde{\sigma}_{\ell 1} }  \delta_{\ell} f_{\ell 1} f_{\ell 2}  }
    \cos \left(  \phi_{a} + \phi_{c} \right)
    + (-1)^{(i+1)} 2
    f_{\ell 1}  \sqrt{ \frac{ \widetilde{\sigma}_{\ell 2} \widetilde{\sigma}_{\ell 3} }{ \widetilde{\sigma}_{\ell 1} } \widetilde{\mu}_{\ell}   } 
    \cos \phi_{a}
   }{
    \widetilde{\mu}_{\ell}
    \left( 1 - \widetilde{m}_{e} \right)
    \left( 1 + \frac{ \widetilde{m}_{e} }{ \widetilde{m}_{\mu} } \right)
    - 
    \frac{ \widetilde{\sigma}_{\ell 1} }{ 2 \widetilde{m}_{\mu} } 
    \left( f_{\ell 2} \delta_{\ell} +  f_{\ell 1} \widetilde{\mu}_{\ell} 
    + 2 \sqrt{ \widetilde{\mu}_{\ell} \delta_{\ell} f_{\ell 1} f_{\ell 2} } \cos \phi_{c}  \right)   },
 \end{array}
\end{equation}
where $i = 0$ for ${\bf M}_{\ell}^{0}$, $i=1$ for ${\bf M}_{\ell}^{3}$. 
For the mass matrices ${\bf M}_{\ell}^{1}$ and ${\bf M}_{\ell}^{5}$,
\begin{equation}
 \begin{array}{l}\vspace{2mm}
  \varepsilon_{23} =
   \frac{ 
    \frac{ \widetilde{\sigma}_{\ell 1} \widetilde{\sigma}_{\ell 3}  }{ \widetilde{\sigma}_{\ell 2} } f_{\ell 2}  +  
    \delta_{\ell} f_{\ell 1} + 
    2 \sqrt{ 
     \frac{ \widetilde{\sigma}_{\ell 1} \widetilde{\sigma}_{\ell 3}  }{ \widetilde{\sigma}_{\ell 2} }  
     \delta_{\ell} f_{\ell 1} f_{\ell 2}
    }  \cos \left( \phi_{a} + \phi_{c} \right)
   }{
    \widetilde{\mu}_{\ell} \left( 1 + \frac{ \widetilde{m}_{e} }{ \widetilde{m}_{\mu} } \right) \left( 1 - \widetilde{m}_{e} \right) -
    \frac{1}{2} \frac{ \widetilde{\sigma}_{\ell 1} }{ \widetilde{m}_{\mu} }
    \left( f_{\ell 2} \delta_{\ell} +
    \frac{ \widetilde{\sigma}_{\ell 2} \widetilde{\sigma}_{\ell 3} }{ \widetilde{\sigma}_{\ell 1} } f_{\ell 1} -
    2 \sqrt{ \frac{ \widetilde{\sigma}_{\ell 2} \widetilde{\sigma}_{\ell 3} }{ \widetilde{\sigma}_{\ell 1} } \delta_{\ell} f_{\ell 1} f_{\ell 2} } 
    \cos \left( \phi_{a} + \phi_{c} \right) \right)
   }, \;
  \varepsilon_{13} =
   \frac{
    f_{\ell 2} \delta_{\ell} +
    \frac{ \widetilde{\sigma}_{\ell 2} \widetilde{\sigma}_{\ell 3} }{ \widetilde{\sigma}_{\ell 1} } f_{\ell 1} -
    2 \sqrt{ \frac{ \widetilde{\sigma}_{\ell 2} \widetilde{\sigma}_{\ell 3} }{ \widetilde{\sigma}_{\ell 1} } \delta_{\ell} f_{\ell 1} f_{\ell 2} } 
    \cos \left( \phi_{a} + \phi_{c} \right)
   }{
    \widetilde{\mu}_{\ell} \left( 1 + \frac{ \widetilde{m}_{e} }{ \widetilde{m}_{\mu} } \right) \left(1 - \widetilde{m}_{e} \right)
   }, \\\vspace{2mm}
  \varepsilon_{12} = 
   \frac{ 
    f_{\ell 2} \delta_{\ell}
    +
    f_{\ell 1} \widetilde{\mu}_{\ell} 
    +
    f_{\ell 1} \frac{ \widetilde{\sigma}_{\ell 2} \widetilde{\sigma}_{\ell 3} }{ \widetilde{\sigma}_{\ell 1} } 
    +(-1)^{(i+1)}
    2 \sqrt{ \widetilde{\mu}_{\ell} \delta_{\ell} f_{\ell 1} f_{\ell 2} } \cos \phi_{c}
    +  2
    \sqrt{ \frac{ \widetilde{\sigma}_{\ell 2} \widetilde{\sigma}_{\ell 3} }{ \widetilde{\sigma}_{\ell 1} }  \delta_{\ell} f_{\ell 1} f_{\ell 2}  }
    \cos \left(  \phi_{a} + \phi_{c} \right)
    + (-1)^{(i+1)} 2
    f_{\ell 1}  \sqrt{ \frac{ \widetilde{\sigma}_{\ell 2} \widetilde{\sigma}_{\ell 3} }{ \widetilde{\sigma}_{\ell 1} } \widetilde{\mu}_{\ell}   } 
    \cos \phi_{a}
   }{
    \widetilde{\mu}_{\ell}
    \left( 1 - \widetilde{m}_{e} \right)
    \left( 1 + \frac{ \widetilde{m}_{e} }{ \widetilde{m}_{\mu} } \right)
    - 
    \frac{ \widetilde{\sigma}_{\ell 1} }{ 2 \widetilde{m}_{\mu} } 
    \left( 
     f_{\ell 2} \delta_{\ell} 
     +  
     f_{\ell 1} \frac{ \widetilde{\sigma}_{\ell 2} \widetilde{\sigma}_{\ell 3} }{ \widetilde{\sigma}_{\ell 1} }
     -
     2 \sqrt{ \frac{ \widetilde{\sigma}_{\ell 2} \widetilde{\sigma}_{\ell 3} }{ \widetilde{\sigma}_{\ell 1} } 
      \delta_{\ell} f_{\ell 1} f_{\ell 2} } 
      \cos \left(  \phi_{a} + \phi_{c} \right)  \right)   },
 \end{array}
\end{equation}
where $i = 0$ for ${\bf M}_{\ell}^{1}$, $i=1$ for ${\bf M}_{\ell}^{5}$. 
For the mass matrices ${\bf M}_{\ell}^{2}$ and ${\bf M}_{\ell}^{4}$,
\begin{equation}
 \begin{array}{l}\vspace{2mm}
  \varepsilon_{23} = 
   \frac{ 
    f_{\ell 2} \left(
    \widetilde{\mu}_{\ell} 
    +
    \frac{ \widetilde{\sigma}_{\ell 1} \widetilde{\sigma}_{\ell 3} }{ \widetilde{\sigma}_{\ell 2} }
    -
    2 \sqrt{ \widetilde{\mu}_{\ell} \frac{ \widetilde{\sigma}_{\ell 1} \widetilde{\sigma}_{\ell 3} }{ \widetilde{\sigma}_{\ell 2} } }
     \cos \phi_{a} 
    \right) 
   }{
   \widetilde{\mu}_{\ell} \left( 1 + \frac{ \widetilde{m}_{e} }{ \widetilde{m}_{\mu} } \right) \left( 1 - \widetilde{m}_{e} \right) 
   - 
   \frac{1}{2} \frac{ \widetilde{\sigma}_{\ell 1} }{ \widetilde{m}_{\mu} }
    f_{\ell 1} 
    \left( 
     \widetilde{\mu}_{\ell} + \frac{ \widetilde{\sigma}_{\ell 2} \widetilde{\sigma}_{\ell 3} }{ \widetilde{\sigma}_{\ell 1} } 
     - 2 \sqrt{ \widetilde{\mu}_{\ell} \frac{ \widetilde{\sigma}_{\ell 2} \widetilde{\sigma}_{\ell 3} }{ \widetilde{\sigma}_{\ell 1} } }
     \cos \phi_{a}  
    \right)
   }, \;
  \varepsilon_{13} =
   \frac{
    f_{\ell 1} 
    \left( 
     \widetilde{\mu}_{\ell} + \frac{ \widetilde{\sigma}_{\ell 2} \widetilde{\sigma}_{\ell 3} }{ \widetilde{\sigma}_{\ell 1} } 
     - 2 \sqrt{ \widetilde{\mu}_{\ell} \frac{ \widetilde{\sigma}_{\ell 2} \widetilde{\sigma}_{\ell 3} }{ \widetilde{\sigma}_{\ell 1} } }
     \cos \phi_{a}  
    \right)
   }{
    \widetilde{\mu}_{\ell} \left( 1 + \frac{ \widetilde{m}_{e} }{ \widetilde{m}_{\mu} } \right) \left(1 - \widetilde{m}_{e} \right)
   }, \\\vspace{2mm}
  \varepsilon_{12} =
   \frac{ 
    f_{\ell 2} \delta_{\ell}
    +
    f_{\ell 1} \widetilde{\mu}_{\ell} 
    +
    f_{\ell 1} \frac{ \widetilde{\sigma}_{\ell 2} \widetilde{\sigma}_{\ell 3} }{ \widetilde{\sigma}_{\ell 1} } 
    +(-1)^{i}
    2 \sqrt{ \widetilde{\mu}_{\ell} \delta_{\ell} f_{\ell 1} f_{\ell 2} } \cos \phi_{c}
    +  (-1)^{(i+1)}2
    \sqrt{ \frac{ \widetilde{\sigma}_{\ell 2} \widetilde{\sigma}_{\ell 3} }{ \widetilde{\sigma}_{\ell 1} }  \delta_{\ell} f_{\ell 1} f_{\ell 2}  }
    \cos \left(  \phi_{a} + \phi_{c} \right)
    +  2
    f_{\ell 1}  \sqrt{ \frac{ \widetilde{\sigma}_{\ell 2} \widetilde{\sigma}_{\ell 3} }{ \widetilde{\sigma}_{\ell 1} } \widetilde{\mu}_{\ell}   } 
    \cos \phi_{a}
   }{
    \widetilde{\mu}_{\ell}
    \left( 1 - \widetilde{m}_{e} \right)
    \left( 1 + \frac{ \widetilde{m}_{e} }{ \widetilde{m}_{\mu} } \right)
    - 
    \frac{ \widetilde{\sigma}_{\ell 1} }{ 2 \widetilde{m}_{\mu} } f_{\ell 1}
    \left( 
     \widetilde{\mu}_{\ell} 
     +  
     \frac{ \widetilde{\sigma}_{\ell 2} \widetilde{\sigma}_{\ell 3} }{ \widetilde{\sigma}_{\ell 1} }
     -
     2 \sqrt{ \frac{ \widetilde{\sigma}_{\ell 2} \widetilde{\sigma}_{\ell 3} }{ \widetilde{\sigma}_{\ell 1} } 
      \widetilde{\mu}_{\ell} } 
      \cos  \right)   }.
 \end{array}
\end{equation}
%
\subsection{Parameters of equivalent class with two texture zeros type-III}\label{Appendix-Para-Type-III}
%
For the mass matrices ${\bf M}_{\ell}^{0}$ and ${\bf M}_{\ell}^{3}$,
\begin{equation}
 \begin{array}{l} \vspace{2mm}
  \varepsilon_{23} =
   \frac{
   	\left( 1 + t_{\beta}^{2} \right) \left( f_{\ell 2} \left( 1 - \delta_{\ell} \right) + f_{\ell 1} f_{\ell 3} \right)
   	+
   	2 \left( 
   	 f_{\ell 2} t_{\beta} \left( 1 - \delta_{\ell} \right) - f_{\ell 1} f_{\ell 3} t_{\beta} 
   	 +
   	 \texttt{s}_{1} \texttt{s}_{2} 	\left( 1 - t_{\beta}^{2} \right)
   	 \sqrt{ f_{\ell 1} f_{\ell 2} f_{\ell 3} \left( 1 - \delta_{\ell} \right) }
   	\right) \cos \phi_{c}
   }{
    \left( 1 + t_{\beta}^{2} \right) \left( 1 - \delta_{\ell} \right) \left( 1 + \texttt{s}_{3} \frac{ \widetilde{m}_{e} }{ \widetilde{m}_{\mu} } \right)
    \left( 1 + \texttt{s}_{2} \widetilde{m}_{e} \right)
    -
    \frac{1}{2} \frac{ \widetilde{m}_{e} }{ \widetilde{m}_{\mu} } 
    \left(
    \left( 1 + t_{\beta}^{2} \right) \left( f_{\ell 1} \left( 1 - \delta_{\ell} \right) + f_{\ell 2} f_{\ell 3} \right) 
    +
    2 \left( 
     t_{\beta} \left( f_{\ell 1} \left( 1 - \delta_{\ell} \right) - f_{\ell 2} f_{\ell 3} \right) 
     - 
     \sqrt{ \left( 1 - \delta_{\ell} \right) f_{\ell 1} f_{\ell 2} f_{\ell 3}  }
    \right) \cos \phi_{c}     
    \right)
   }  ,\\  \vspace{2mm}
  \varepsilon_{13} = 
   \frac{
    \left( 1 + t_{\beta}^{2} \right) \left( f_{\ell 1} \left( 1 - \delta_{\ell} \right) + f_{\ell 2} f_{\ell 3} \right) 
    +
    2 \left( 
     t_{\beta} \left( f_{\ell 1} \left( 1 - \delta_{\ell} \right) - f_{\ell 2} f_{\ell 3} \right) 
     - 
     \sqrt{ \left( 1 - \delta_{\ell} \right) f_{\ell 1} f_{\ell 2} f_{\ell 3}  }
    \right) \cos \phi_{c}
   }{
    \left( 1 + t_{\beta}^{2} \right) \left( 1 - \delta_{\ell} \right) \left( 1 + \texttt{s}_{3} \frac{ \widetilde{m}_{e} }{ \widetilde{m}_{\mu} } \right)
    \left( 1 + \texttt{s}_{2} \widetilde{m}_{e} \right)
   },\\ \vspace{2mm}
  \varepsilon_{12} =  
   \frac{ 
    \left( \sqrt{ f_{\ell 2} f_{\ell 3} } t_{\beta} + \sqrt{ f_{\ell 1} \left( 1 - \delta_{\ell} \right) } \right)^{2}   
    + 
    \left( \sqrt{ f_{\ell 2} f_{\ell 3} } - t_{\beta} \sqrt{ f_{\ell 1} \left( 1 - \delta_{\ell} \right) } \right)^{2}   
    + 
    2 \left( \sqrt{ f_{\ell 2} f_{\ell 3} } t_{\beta} + \sqrt{ f_{\ell 1} \left( 1 - \delta_{\ell} \right) } \right)
    \left( \sqrt{ f_{\ell 2} f_{\ell 3} } - t_{\beta} \sqrt{ f_{\ell 1} \left( 1 - \delta_{\ell} \right) } \right)
    c_{c}
   }{ 
    \left( 1 + t_{\beta}^{2} \right) \left( 1 - \delta_{\ell} \right) \left( 1 + \texttt{s}_{3} \frac{ \widetilde{m}_{e} }{ \widetilde{m}_{\mu} } \right)
    \left( 1 + \texttt{s}_{2} \widetilde{m}_{e} \right)
    -
    N\varepsilon_{13} 
   } 
   \times \\ \qquad \times
   \frac{
    + \frac{ \widetilde{m}_{\mu} }{ \widetilde{m}_{e} } \left( 1 + t_{\beta}^{2} \right) f_{\ell 1}
    + (-1)^{i} 2 \texttt{s}_{1} \sqrt{ \frac{ \widetilde{m}_{\mu} }{ \widetilde{m}_{e} } \left( 1 + t_{\beta}^{2} \right) f_{\ell 1} } 
     \left( \sqrt{ f_{\ell 2} f_{\ell 3} } t_{\beta} + \sqrt{ f_{\ell 1} \left( 1 - \delta_{\ell} \right) }\right) c_{a}
    + (-1)^{i} 2 \texttt{s}_{1} \sqrt{ \frac{ \widetilde{m}_{\mu} }{ \widetilde{m}_{e} } \left( 1 + t_{\beta}^{2} \right) f_{\ell 1} } 
     \left( \sqrt{ f_{\ell 2} f_{\ell 3} } - t_{\beta} \sqrt{ f_{\ell 1} \left( 1 - \delta_{\ell} \right) } \right) c_{ac}
   }{
   
   } ,
 \end{array}
\end{equation}
where $i = 0$ for ${\bf M}_{\ell}^{0}$, $i=1$ for ${\bf M}_{\ell}^{3}$, $c_{a} = \cos \phi_{a}$, $c_{c} = \cos \phi_{c}$,
$c_{ac} = \cos \left( \phi_{a} + \phi_{c} \right)$, and $t_{\beta} = \tan \beta$, 
\begin{equation}
 \begin{array}{l}
  N\varepsilon_{13} = 
    \left( 1 + t_{\beta}^{2} \right) \left( f_{\ell 1} \left( 1 - \delta_{\ell} \right) + f_{\ell 2} f_{\ell 3} \right) 
    +
    2 \left( 
     t_{\beta} \left( f_{\ell 1} \left( 1 - \delta_{\ell} \right) - f_{\ell 2} f_{\ell 3} \right) 
     - 
     \sqrt{ \left( 1 - \delta_{\ell} \right) f_{\ell 1} f_{\ell 2} f_{\ell 3}  }
    \right) \cos \phi_{c}.
 \end{array}
\end{equation} 
For the mass matrices ${\bf M}_{\ell}^{1}$ and ${\bf M}_{\ell}^{5}$, 
\begin{equation}
 \begin{array}{l}\vspace{2mm}
  \varepsilon_{23} = 
   \frac{ 
    \frac{ \widetilde{m}_{e} }{ \widetilde{m}_{\mu} } f_{\ell 2} \left( 1 + t_{\beta}^{2} \right)
    +
    f_{\ell 1} f_{\ell 3}
    +
    f_{\ell 2} t_{\beta}^{2} \left( 1 - \delta_{\ell} \right)
    +
    \texttt{s}_{1} \texttt{s}_{2} t_{\beta} \sqrt{ f_{\ell 1} f_{\ell 2} f_{\ell 3}  \left( 1 - \delta_{\ell} \right) } 
    + 
    2 \sqrt{ 1 + t_{\beta}^{2} } \sqrt{ \frac{ \widetilde{m}_{e} }{ \widetilde{m}_{\mu} }}
    \left(
     \texttt{s}_{1} \sqrt{ f_{\ell 1} f_{\ell 2} f_{\ell 3} }
     +
     \texttt{s}_{2} t_{\beta} f_{\ell 2} \sqrt{1 - \delta_{\ell} }
    \right)\cos \left( \phi_{a} + \phi_{c} \right)
   }{ 
    \left( 1 + t_{\beta}^{2} \right) \left( 1 - \delta_{\ell} \right) \left( 1 + \texttt{s}_{3} \frac{ \widetilde{m}_{e} }{ \widetilde{m}_{\mu} } \right)
    \left( 1 + \texttt{s}_{2} \widetilde{m}_{e} \right) 
    - N\varepsilon_{13} 
   } , \\ \vspace{2mm}
  \varepsilon_{13} = 
   \frac{
    \left( 1 + t_{\beta}^{2} \right) \frac{ \widetilde{m}_{\mu} }{ \widetilde{m}_{e} } f_{\ell 1} 
    +
    f_{\ell 2} f_{\ell 3}  + f_{\ell 1} t_{\beta}^{2} \left( 1 - \delta_{\ell} \right) 
    -
    2 t_{\beta} \sqrt{ f_{\ell 1} f_{\ell 2} f_{\ell 3}  \left( 1 - \delta_{\ell} \right) }  
    -
    2 \texttt{s}_{1} \sqrt{ 1 + t_{\beta}^{2} } \sqrt{  \frac{ \widetilde{m}_{\mu} }{  \widetilde{m}_{e} } }
    \left(
     \sqrt{ f_{\ell 1} f_{\ell 2} f_{\ell 3} } 
     -
     t_{\beta} f_{\ell 1} \sqrt{ \left( 1 - \delta_{\ell} \right) }
    \right) \cos \left( \phi_{a} + \phi_{c} \right)
   }{
    \left( 1 + t_{\beta}^{2} \right) \left( 1 - \delta_{\ell} \right) \left( 1 + \texttt{s}_{3} \frac{ \widetilde{m}_{e} }{ \widetilde{m}_{\mu} } \right)
    \left( 1 + \texttt{s}_{2} \widetilde{m}_{e} \right) 
   } ,\\ \vspace{2mm}
  \varepsilon_{12} =  
   \frac{ 
    \left( \sqrt{ f_{\ell 2} f_{\ell 3} } t_{\beta} + \sqrt{ f_{\ell 1} \left( 1 - \delta_{\ell} \right) } \right)^{2}
    +
    \left( \sqrt{ f_{\ell 2} f_{\ell 3} } - t_{\beta} \sqrt{ f_{\ell 1} \left( 1 - \delta_{\ell} \right) } \right)^{2}
    + (-1)^{j} 2 c_{c}
    \left( \sqrt{ f_{\ell 2} f_{\ell 3} } t_{\beta} + \sqrt{ f_{\ell 1} \left( 1 - \delta_{\ell} \right) } \right)
    \left( \sqrt{ f_{\ell 2} f_{\ell 3} } - t_{\beta} \sqrt{ f_{\ell 1} \left( 1 - \delta_{\ell} \right) } \right)
   }{
    \left( 1 + t_{\beta}^{2} \right) \left( 1 - \delta_{\ell} \right) \left( 1 + \texttt{s}_{3} \frac{ \widetilde{m}_{e} }{ \widetilde{m}_{\mu} } \right)
    \left( 1 + \texttt{s}_{2} \widetilde{m}_{e} \right) 
    - N\varepsilon_{13}    
   } 
   \times \\ \qquad \times
   \frac{
    + \frac{ \widetilde{m}_{\mu} }{ \widetilde{m}_{e} } \left( 1 + t_{\beta} \right) f_{\ell 1}
    + (-1)^{j} 2 \texttt{s}_{1}
     \sqrt{ \frac{ \widetilde{m}_{\mu} }{ \widetilde{m}_{e} } \left( 1 + t_{\beta} \right) f_{\ell 1} }
     \left( \sqrt{ f_{\ell 2} f_{\ell 3} } t_{\beta} + \sqrt{ f_{\ell 1} \left( 1 - \delta_{\ell} \right) } \right) 
     c_{a}
    + 2 \texttt{s}_{1}  
    \sqrt{ \frac{ \widetilde{m}_{\mu} }{ \widetilde{m}_{e} } \left( 1 + t_{\beta} \right) f_{\ell 1} }
    \left( \sqrt{ f_{\ell 2} f_{\ell 3} } - t_{\beta} \sqrt{ f_{\ell 1} \left( 1 - \delta_{\ell} \right) } \right)
    c_{ac}
   }{
   },
 \end{array}
\end{equation} 
where $i = 0$ for ${\bf M}_{\ell}^{1}$, $i=1$ for ${\bf M}_{\ell}^{5}$, $c_{a} = \cos \phi_{a}$, $c_{c} = \cos \phi_{c}$,
$c_{ac} = \cos \left( \phi_{a} + \phi_{c} \right)$, $t_{\beta} = \tan \beta$, and 
\begin{equation}
 \begin{array}{ll} \vspace{2mm}
  N\varepsilon_{13} = &
  \frac{1}{2} \frac{ \widetilde{m}_{e} }{ \widetilde{m}_{\mu} } 
    \left(
    \left( 1 + t_{\beta}^{2} \right) \frac{ \widetilde{m}_{\mu} }{ \widetilde{m}_{e} } f_{\ell 1} 
    +
    f_{\ell 2} f_{\ell 3}  + f_{\ell 1} t_{\beta}^{2} \left( 1 - \delta_{\ell} \right) 
    -
    2 t_{\beta} \sqrt{ f_{\ell 1} f_{\ell 2} f_{\ell 3}  \left( 1 - \delta_{\ell} \right) } \right. \\
    &
    \left.-
    2 \texttt{s}_{1} \sqrt{ 1 + t_{\beta}^{2} } \sqrt{  \frac{ \widetilde{m}_{\mu} }{  \widetilde{m}_{e} } }
    \left(
     \sqrt{ f_{\ell 1} f_{\ell 2} f_{\ell 3} } 
     -
      t_{\beta} f_{\ell 1} \sqrt{ \left( 1 - \delta_{\ell} \right) }
    \right) \cos \left( \phi_{a} + \phi_{c} \right)
    \right) .
 \end{array}
\end{equation}
For ${\bf M}_{\ell}^{2}$ and ${\bf M}_{\ell}^{4}$ 
\begin{equation}
 \begin{array}{l}\vspace{2mm}
  \varepsilon_{13} =
  \frac{ 
   \left( 1 + t_{\beta}^{2} \right) \frac{ \widetilde{m}_{\mu} }{ \widetilde{m}_{e} } f_{\ell 1} 
    +
   f_{\ell 2} f_{\ell 3}  t_{\beta}^{2} 
   +
   f_{\ell 1} \left( 1 - \delta_{\ell} \right) 
   +
   2 t_{\beta} \sqrt{ \left( 1 - \delta_{\ell} \right) f_{\ell 1} f_{\ell 2} f_{\ell 3} }
   +
   2 \texttt{s}_{1} \sqrt{ 1 + t_{\beta}^{2} } \sqrt{  \frac{ \widetilde{m}_{\mu} }{  \widetilde{m}_{e} } }
    \left(
     t_{\beta} \sqrt{ f_{\ell 1} f_{\ell 2} f_{\ell 3} } 
     +
     f_{\ell 1} \sqrt{ \left( 1 - \delta_{\ell} \right) }
    \right) \cos \phi_{a} 
   }{
    \left( 1 + t_{\beta}^{2} \right) \left( 1 - \delta_{\ell} \right) \left( 1 + \texttt{s}_{3} \frac{ \widetilde{m}_{e} }{ \widetilde{m}_{\mu} } \right)
    \left( 1 + \texttt{s}_{2} \widetilde{m}_{e} \right) 
   }, \\ \vspace{2mm}
  \varepsilon_{23} = 
  \frac{ 
   f_{\ell 2} \left( \frac{ \widetilde{m}_{e} }{  \widetilde{m}_{\mu} } \left( 1 + t_{\beta}^{2} \right) + 1 - \delta_{\ell} \right) 
   +
   f_{\ell 1} f_{\ell 3} t_{\beta}^{2} 
   -
   2 \texttt{s}_{1} \texttt{s}_{2} t_{\beta} 
   \sqrt{ \frac{ \widetilde{m}_{e} }{ \widetilde{m}_{\mu} } f_{\ell 1} f_{\ell 2} f_{\ell 3} \left( 1 - \delta_{\ell} \right) } 
   +
   2 \sqrt{ \frac{ \widetilde{m}_{e} }{ \widetilde{m}_{\mu} } \left( 1 + t_{\beta}^{2} \right) } \left(
    \texttt{s}_{2} f_{\ell 2} \sqrt{ 1 - \delta_{\ell} }
    -
    \texttt{s}_{1} t_{\beta} \sqrt{ f_{\ell 1} f_{\ell 2} f_{\ell 3} }
   \right) \cos \phi_{a}
  }{
    \left( 1 + t_{\beta}^{2} \right) \left( 1 - \delta_{\ell} \right) \left( 1 + \texttt{s}_{3} \frac{ \widetilde{m}_{e} }{ \widetilde{m}_{\mu} } \right)
    \left( 1 + \texttt{s}_{2} \widetilde{m}_{e} \right)  
    -
    N\varepsilon_{13} 
  },\\ \vspace{2mm}
  \varepsilon_{12} =  
   \frac{
    \left( \sqrt{ f_{\ell 2} f_{\ell 3} } t_{\beta} + \sqrt{ f_{\ell 1} \left( 1 - \delta_{\ell} \right) } \right)^{2}
    +   
    \left( \sqrt{ f_{\ell 2} f_{\ell 3} } - t_{\beta} \sqrt{ f_{\ell 1} \left( 1 - \delta_{\ell} \right) } \right)^{2}
    + (-1)^{i+1} 2
    \left( \sqrt{ f_{\ell 2} f_{\ell 3} } t_{\beta} + \sqrt{ f_{\ell 1} \left( 1 - \delta_{\ell} \right) } \right)
    \left( \sqrt{ f_{\ell 2} f_{\ell 3} } - t_{\beta} \sqrt{ f_{\ell 1} \left( 1 - \delta_{\ell} \right) } \right)
    c_{c}
   }{
    \left( 1 + t_{\beta}^{2} \right) \left( 1 - \delta_{\ell} \right) \left( 1 + \texttt{s}_{3} \frac{ \widetilde{m}_{e} }{ \widetilde{m}_{\mu} } \right)
    \left( 1 + \texttt{s}_{2} \widetilde{m}_{e} \right)  
    -
    N\varepsilon_{13}    
   } 
   \times \\ \qquad \times
   \frac{
    + \frac{ \widetilde{m}_{\mu} }{ \widetilde{m}_{e} } \left( 1 + t_{\beta} \right) f_{\ell 1}
    - 2 \texttt{s}_{1} 
     \sqrt{ \frac{ \widetilde{m}_{\mu} }{ \widetilde{m}_{e} } \left( 1 + t_{\beta} \right) f_{\ell 1} }
     \left( \sqrt{ f_{\ell 2} f_{\ell 3} } t_{\beta} + \sqrt{ f_{\ell 1} \left( 1 - \delta_{\ell} \right) } \right)
     c_{a}
    -2 \texttt{s}_{1} 
     \sqrt{ \frac{ \widetilde{m}_{\mu} }{ \widetilde{m}_{e} } \left( 1 + t_{\beta} \right) f_{\ell 1} }
     \left( \sqrt{ f_{\ell 2} f_{\ell 3} } - t_{\beta} \sqrt{ f_{\ell 1} \left( 1 - \delta_{\ell} \right) } \right)
     c_{ac}
   }{
   
   },
 \end{array}
\end{equation}
where $i = 0$ for ${\bf M}_{\ell}^{2}$, $i=1$ for ${\bf M}_{\ell}^{4}$, $c_{a} = \cos \phi_{a}$, $c_{c} = \cos \phi_{c}$,
$c_{ac} = \cos \left( \phi_{a} + \phi_{c} \right)$, $t_{\beta} = \tan \beta$, and 
\begin{equation}
 \begin{array}{rl}
  N\varepsilon_{13} = & 
   \frac{1}{2} \frac{ \widetilde{m}_{e} }{ \widetilde{m}_{\mu} } \left( \left( 1 + t_{\beta}^{2} \right) \frac{ \widetilde{m}_{\mu} }{ \widetilde{m}_{e} } f_{\ell 1} 
    +
   f_{\ell 2} f_{\ell 3}  t_{\beta}^{2} 
   +
   f_{\ell 1} \left( 1 - \delta_{\ell} \right) 
   +
   2 t_{\beta} \sqrt{ \left( 1 - \delta_{\ell} \right) f_{\ell 1} f_{\ell 2} f_{\ell 3} } 
   \right. \\ 
   & + \left.
   2 \texttt{s}_{1} \sqrt{ 1 + t_{\beta}^{2} } \sqrt{  \frac{ \widetilde{m}_{\mu} }{  \widetilde{m}_{e} } }
    \left(
     t_{\beta} \sqrt{ f_{\ell 1} f_{\ell 2} f_{\ell 3} } 
     +
     f_{\ell 1} \sqrt{ \left( 1 - \delta_{\ell} \right) }
    \right) \cos \phi_{a} \right) .  
 \end{array}
\end{equation}
%
\subsection{Parameters of equivalent class with two texture zeros type-IV}\label{Appendix-Para-Type-IV}
%
For the mass matrices ${\bf M}_{\ell}^{0}$ and ${\bf M}_{\ell}^{3}$,
\begin{equation}
 \begin{array}{l}
  \varepsilon_{13} =
   \frac{ \widetilde{m}_{\mu} }{ \widetilde{m}_{e} }
   \frac{ \widetilde{f}_{\ell} - \widetilde{m}_{e} }{ \widetilde{m}_{\mu} - \widetilde{m}_{e} }, \qquad
  \varepsilon_{23} =  
   \frac{ 
    \widetilde{m}_{\mu} - \widetilde{f}_{\ell} 
   }{
     \widetilde{m}_{\mu} - \frac{1}{2} \widetilde{m}_{e} - \frac{1}{2} \widetilde{f}_{\ell} 
   } , \qquad
  \varepsilon_{12} = 
   \frac{
    \widetilde{m}_{\mu} - \widetilde{m}_{e} 
    +(-1)^{i+1} 2 \sqrt{ \left( \widetilde{f}_{\ell} - \widetilde{m}_{e} \right) \left( \widetilde{m}_{\mu} - \widetilde{f}_{\ell} \right) } \cos \phi_{a}
   }{
    \widetilde{m}_{\mu}  - \frac{ \widetilde{m}_{e} }{2} - \frac{ \widetilde{f}_{\ell} }{2}
   } .
 \end{array}
\end{equation}
For the mass matrices ${\bf M}_{\ell}^{1}$ and ${\bf M}_{\ell}^{5}$, 
\begin{equation}
 \begin{array}{l}
  \varepsilon_{13} =
   \frac{ \widetilde{m}_{\mu} }{ \widetilde{m}_{e} }
   \frac{ \widetilde{m}_{\mu} - \widetilde{f}_{\ell} }{ \widetilde{m}_{\mu} - \widetilde{m}_{e} }, \qquad
  \varepsilon_{23} =
   \frac{ 
    \widetilde{f}_{\ell} - \widetilde{m}_{e}
   }{
    \frac{1}{2} \widetilde{m}_{\mu} - \widetilde{m}_{e} + \frac{1}{2} \widetilde{f}_{\ell} 
   }, \qquad
  \varepsilon_{12} = 
   \frac{
    \widetilde{m}_{\mu} - \widetilde{m}_{e} 
    +(-1)^{i+1} 2 \sqrt{ \left( \widetilde{f}_{\ell} - \widetilde{m}_{e} \right) \left( \widetilde{m}_{\mu} - \widetilde{f}_{\ell} \right) } \cos \phi_{a}
   }{
    \frac{ \widetilde{m}_{\mu} }{2} - \widetilde{m}_{e}  -  \frac{ \widetilde{f}_{\ell} }{2}
   } 
 \end{array}
\end{equation}
For ${\bf M}_{\ell}^{2}$ and ${\bf M}_{\ell}^{4}$ 
\begin{equation}
 \begin{array}{l}\vspace{2mm}
  \varepsilon_{13} =
   \frac{ \widetilde{m}_{\mu} }{ \widetilde{m}_{e} }
   \left( 
    1 - 2 
    \frac{ 
     \sqrt{ \left( \widetilde{f}_{\ell} - \widetilde{m}_{e} \right) \left( \widetilde{m}_{\mu} - \widetilde{f}_{\ell} \right) } 
    }{
     \widetilde{m}_{\mu} - \widetilde{m}_{e} 
    }  \cos \phi_{a} 
   \right), \qquad
  \varepsilon_{23} = \varepsilon_{12} =
   \frac{ 
    \widetilde{m}_{\mu} - \widetilde{m}_{e} 
    + 2 \sqrt{ \left( \widetilde{f}_{\ell} - \widetilde{m}_{e} \right) \left( \widetilde{m}_{\mu} - \widetilde{f}_{\ell} \right) } 
    \cos \phi_{a}
   }{
    \frac{1}{2} \left( \widetilde{m}_{\mu} - \widetilde{m}_{e} \right) 
    + \sqrt{ \left( \widetilde{f}_{\ell} - \widetilde{m}_{e} \right) \left( \widetilde{m}_{\mu} - \widetilde{f}_{\ell} \right) } 
    \cos \phi_{a}
   }.
 \end{array}
\end{equation}

\begin{acknowledgments}
This work has been partially supported by \textit{CONACYT-SNI (M\'exico)}. 
\end{acknowledgments}


\end{document}